\documentclass[useAMS, usenatbib]{mn2e}
\usepackage{graphicx}
\usepackage{longtable}
\usepackage{multicol}
\usepackage{lscape}
\usepackage{subfig}
\usepackage{natbib}
\usepackage{threeparttable}

\title[On the nature of the red, 2MASS selected AGN in the local Universe I: an optical spectroscopic study]{On the nature of the red, 2MASS selected AGN in the local Universe I: an optical spectroscopic study}
\author[M. Rose, C. N. Tadhunter, J. Holt, J. Rodr\'{i}guez Zaur\'{i}n]{M. Rose$^{1}$\thanks{E-mail:m.rose@sheffield.ac.uk (MR); c.tadhunter@sheffield.ac.uk (CNT)}, C. N. Tadhunter$^{1}$\footnotemark[1], J. Holt$^{2}$ and J. Rodr\'{i}guez Zaur\'{i}n$^{3}$\\
$^{1}$Department of Physics and Astronomy, University of Sheffield, Sheffield S3 7RH\\ 
$^{2}$Leiden Observatory, Leiden University, P.O. Box 9513, 2300 RA Leiden, Netherlands\\
$^{3}$Instituto de Astrofisica de Canarias, E38205 - La Laguna (Tenerife), Spain}
\begin{document}

\date{}

\pagerange{\pageref{firstpage}--\pageref{lastpage}} \pubyear{2011}

\maketitle

\label{firstpage}

\begin{abstract}

\noindent We present optical spectra for a representative sample of 27 nearby (z $<$ 0.2) 2MASS-selected AGN with red near-IR colours (J-K$_S$ $\ga$ 2.0). The spectra were taken with the ISIS spectrograph on the WHT with the aim of determining the nature of the red 2MASS AGN, in particular whether they are young quasars obscured by their natal cocoon of gas and dust. We compare our findings with those obtained for comparison samples of PG quasars and unobscured type 1 AGN.

The spectra show a remarkable variety, including moderately reddened type 1 objects  (45\%), type 1 objects that appear similar to traditional UV/optical selected AGN (11\%), narrow-line Seyfert 1 AGN (15\%), type 2 AGN (22\%) and HII/composite objects (7\%). The high Balmer decrements that we measure in many of the type 1 objects are consistent with their red J-K$_S$ colours being due to moderate levels of dust extinction (0.2 $<$ E(B-V) $<$ 1.2). However, we measure only modest velocity shifts and widths for the broader [OIII]$\lambda$5007 emission line components that are similar to those measured in the comparison samples.  This suggests that the outflows in the red 2MASS objects are not unusual compared with those of optical/UV selected AGN of similar luminosity. In addition, the Eddington ratios for the 2MASS sample are relatively modest.

Overall, based on their optical spectra, we find no clear evidence that the population of red, 2MASS selected AGN at low redshifts represents young quasars. Most plausibly, these objects are normal type 1 AGN that are moderately obscured by material in the outer layers of the circum-nuclear tori or in the disks of the host galaxies.

\end{abstract}

\begin{keywords}
galaxies:active --- galaxies:Seyfert --- quasars:emission lines --- quasars:general 
\end{keywords}

\section{Introduction}

Most of our understanding of Active Galactic Nuclei (AGN) is based on samples of AGN detected in UV/optical surveys. Indeed, key advances such as measuring high cosmological redshifts \citep{hazard} and developing the orientation-based unified schemes (see \citealt{antonucci}) are based on the study of AGN at UV/optical wavelengths. However, there has been increasing evidence that a large fraction of AGN have not been detected in UV/optical surveys because they are obscured by large columns of dust ( \citealt{webster}; \citealt{cutri}). Surveys at infrared and radio wavelengths suggest that up to 80\% of all AGN may have been missed in current UV/optical surveys (see \citealt{low}; \citealt{webster}; \citealt{francis}). 

A key result of the Two Micron All Sky Survey (2MASS) was the discovery of a population of AGN that appear redder than their traditional optical/UV selected counterparts at infrared wavelengths. The red 2MASS AGN were selected to have J-K$_S$ $\ga$ 2.0, distinguishing them from most optical/UV selected AGN that have bluer near-IR colours (J-K$_S$ $<$ 2.0; \citealt{cutri}), and K-band magnitudes in the range 11.0 $<$ K$_S$ $<$ 14.9 mag. The latter criterion was chosen so that bright (K$_S$ $<$ 11.0) Galactic AGB and carbon stars were excluded, along with `normal' galaxies in the redshift range 0.4 $<$ z $<$ 0.5 (K$_S$ $>$ 14.9; \citealt{cutri}). Spectroscopic follow up of the objects which fulfill these criteria revealed that the majority of the candidates have spectra resembling AGN \citep{cutri}. More recent surveys have determined that these \lq red quasars' do indeed represent a large fraction of the overall population of quasars in the local Universe ($\sim$15--60$\%$; \citealt{glikman07}; \citealt{glikman12}). 

Because the red quasars potentially make up a large fraction of the AGN population, it is important that we understand their relationship to the unobscured population: are they young, dust obscured versions of \lq typical quasars\rq\ (\citealt{hutchings03}; \citealt{georgakakis}; \citealt{glikman07}; \citealt{glikman12})? Or can their properties be explained by a specific viewing angle in the orientation-based unified schemes \citep{wilkes}? Or do they perhaps represent a new population altogether? Are these objects all luminous quasars, or are some less luminous AGN? 

Past studies of red 2MASS quasars have found reddenings in the range 0.1 $<$ E(B-V) $<$ 3.2 magnitudes based on Balmer decrements and fits to the continuum shape (e.g. \citealt{glikman07}; \citealt{georgakakis}; \citealt{kuraab}; \citealt{glikman12}; \citealt{canalizo12}), although it is important to recognise that some have negligible reddening \citep{glikman07}. In addition, imaging studies of the host galaxies find evidence for disturbed morphologies in up to 70\% of cases (\citealt{hutchings03}; \citealt{marble03}; \citealt{urrutia}), which may support the idea that host galaxies have undergone recent interactions with other galaxies. Such interactions could drive material into the central region of the galaxies, enshrouding the SMBH in a cocoons of dust \citep{hopkins}, making them difficult to detect in traditional UV/optical surveys \citep{cutri}.

Many previous studies of \lq red quasars' have focussed on samples covering a wide range of redshift (0 $<$ z $<$ 3: \citealt{glikman07}; \citealt{georgakakis}; \citealt{glikman12}). The problem with such samples is it is difficult to investigate the intrinsic diversity of the red quasar population in the face of possible evolutionary and/or luminosity-dependent effects. Therefore it is important to study a sample of 2MASS-selected AGN that covers a more limited range in redshift.

This paper reports observations of an RA-limited sample of 27 2MASS selected quasars (J-K$_S$ $\ga$ 2.0) with a narrow range of redshifts: 0.09 $<$ z $<$ 0.20. We aim to determine how the red 2MASS AGN fit into our picture of active galaxies in general, and how important they are to our understanding of AGN. In this, the first of two papers (Paper I), we report a spectroscopic study of this sample, focussing on the AGN properties, emission line outflows and dust extinction. The spectral properties are compared to control samples of optical/UV selected quasars and AGN. Paper II in the series will focus on the near- and mid-IR properties of the sample, and will include a comprehensive comparison with several complete samples of active galaxies.

 The cosmological parameters used throughout this paper are adopted from WMAP: $H_\circ$ = 71 km s$^{-1}$, $\Omega_M$ = 0.27 and $\Omega_\Lambda$ = 0.73 \citep{spergel}.  

\section{Observations and data reduction}

\begin{figure*}
\centering
\includegraphics[scale=0.25]{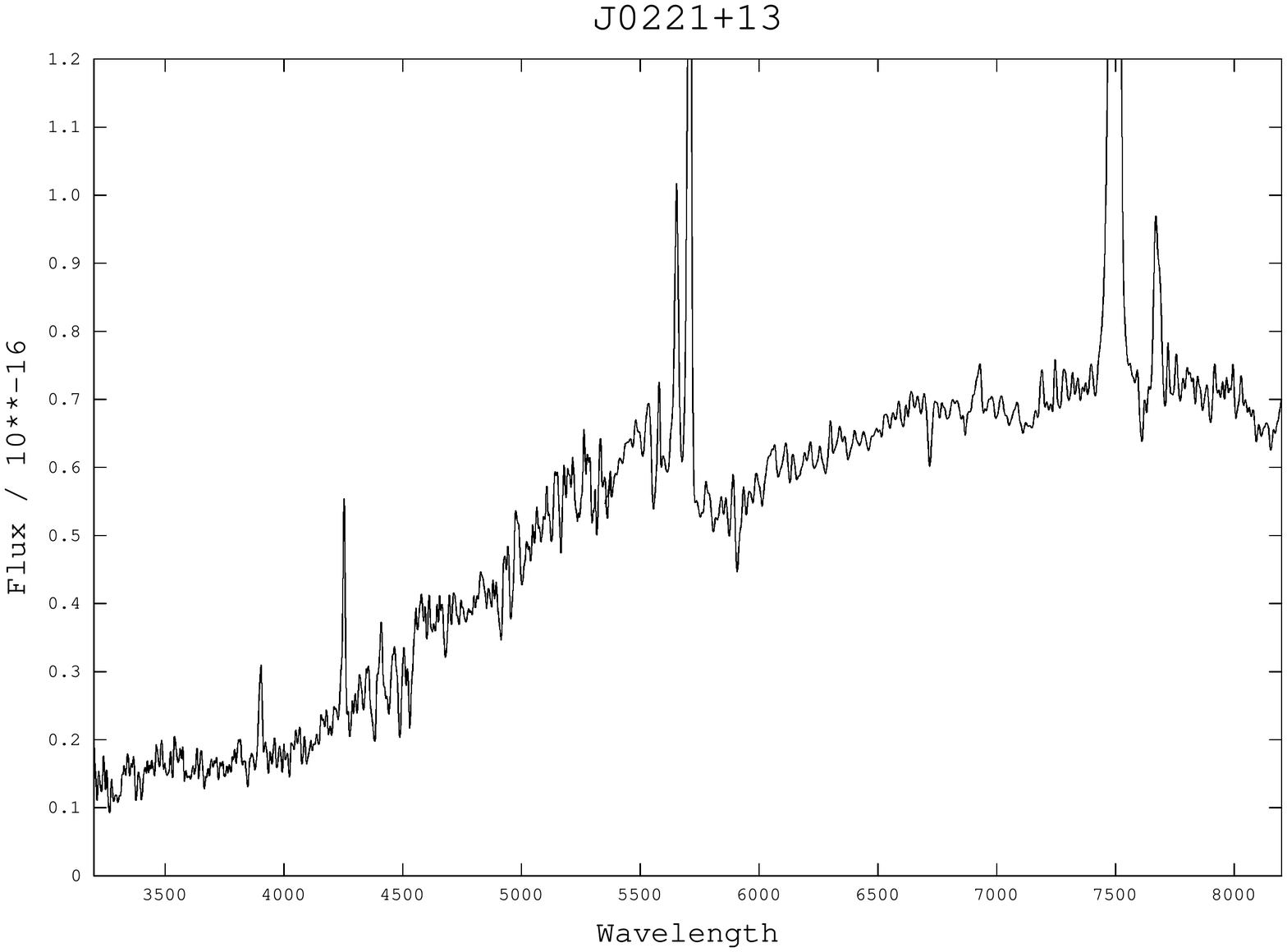}
\includegraphics[scale=0.25]{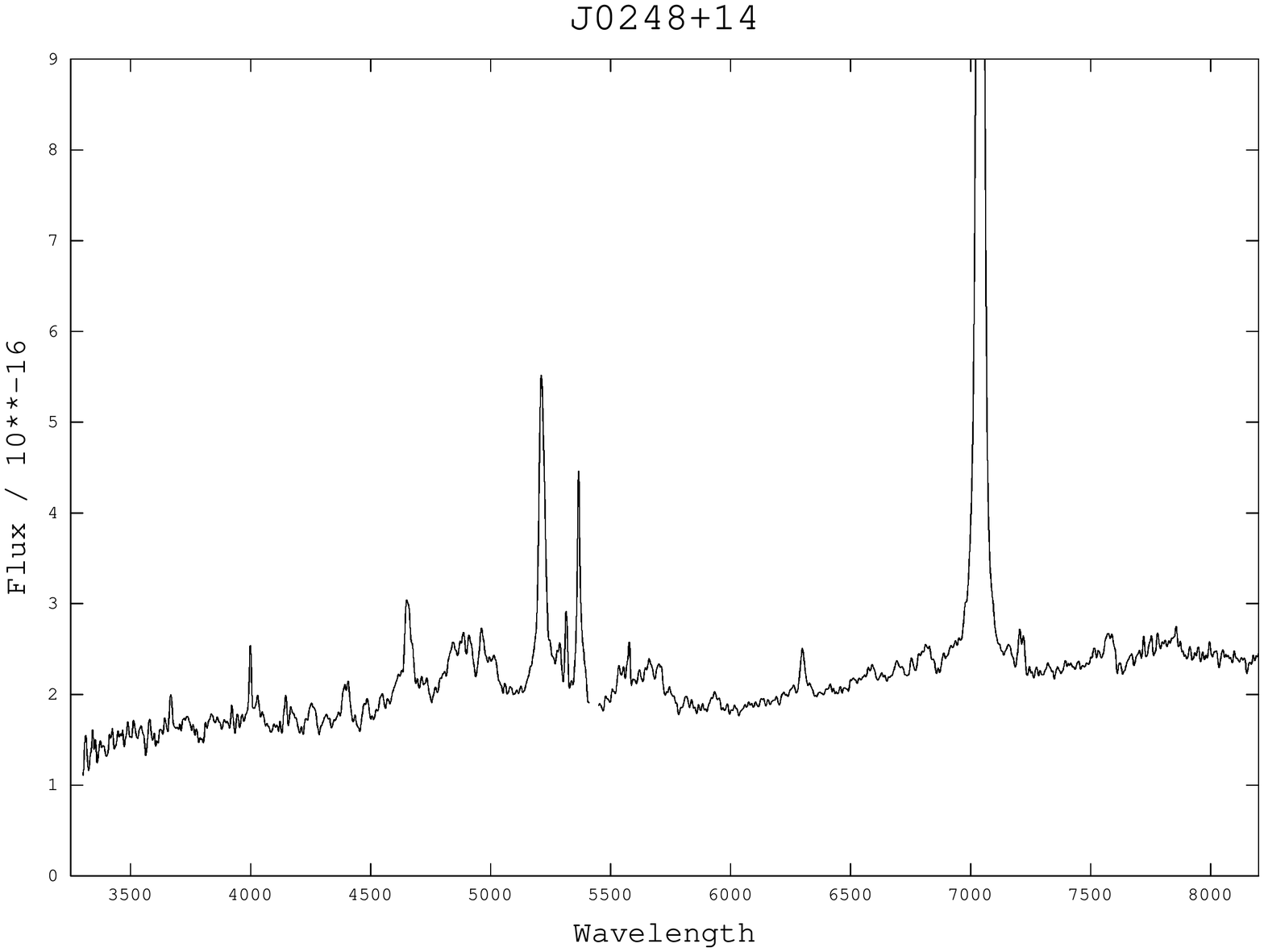}
\includegraphics[scale=0.25]{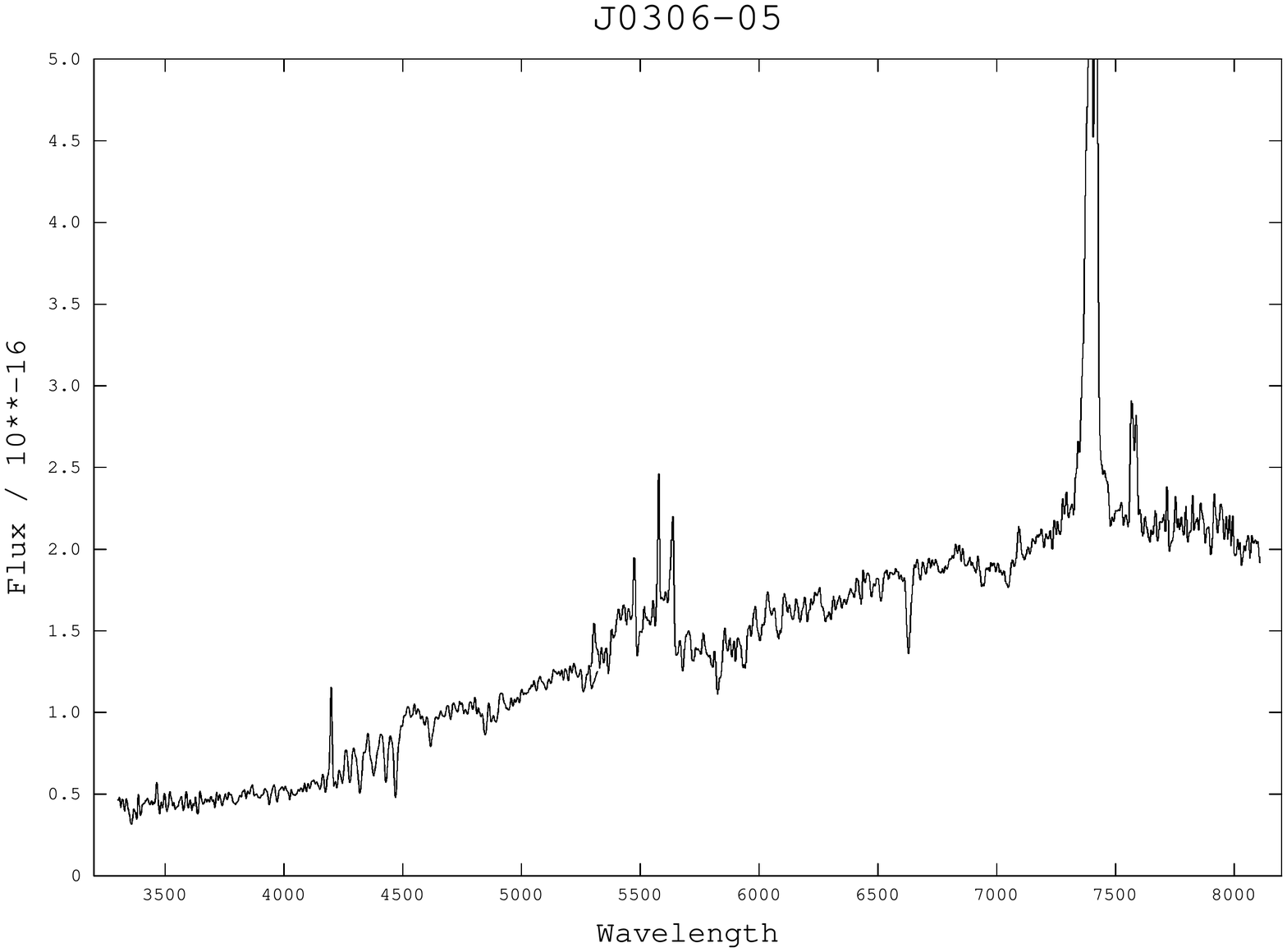}
\includegraphics[scale=0.25]{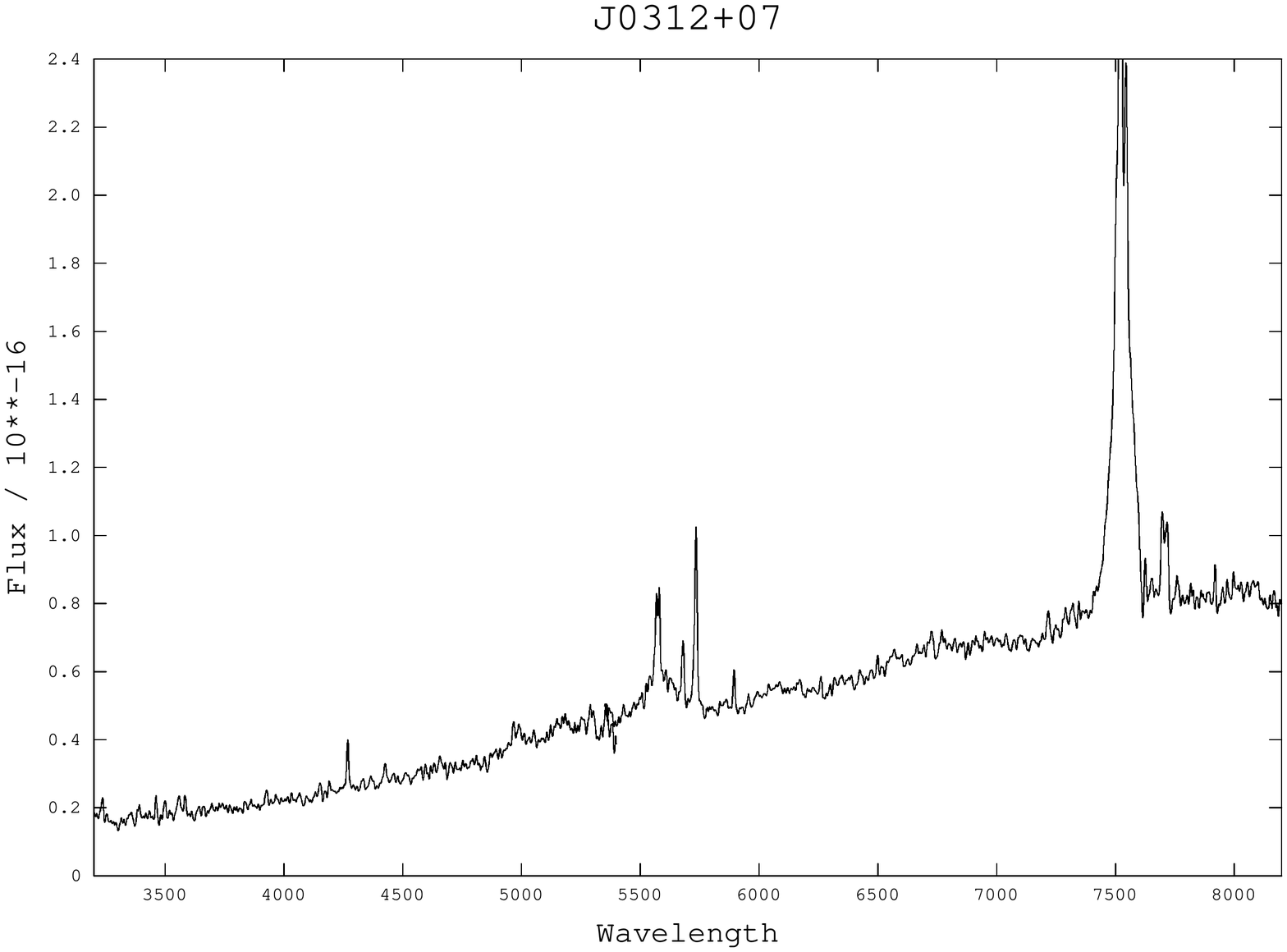}
\includegraphics[scale=0.25]{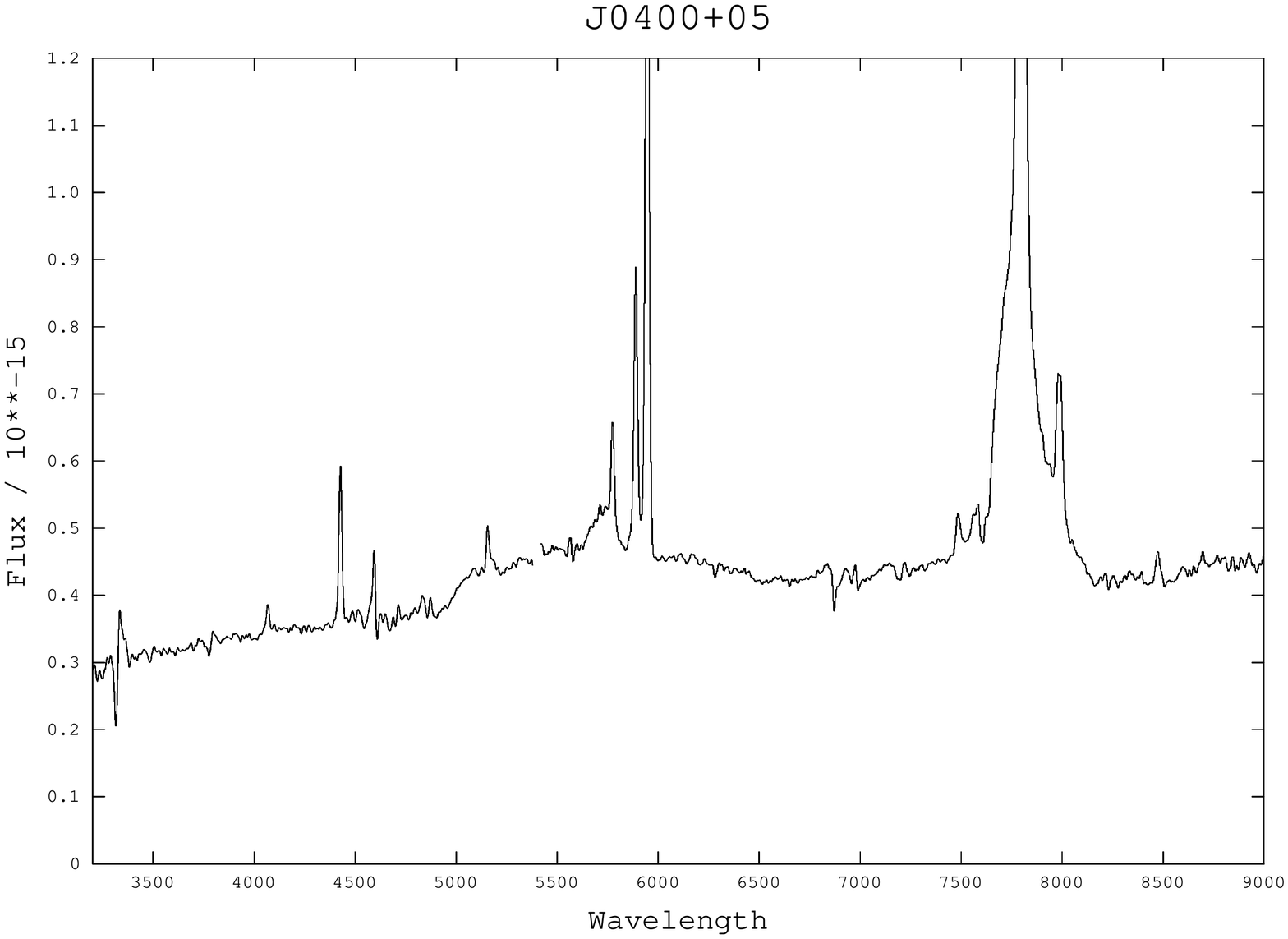}
\includegraphics[scale=0.25]{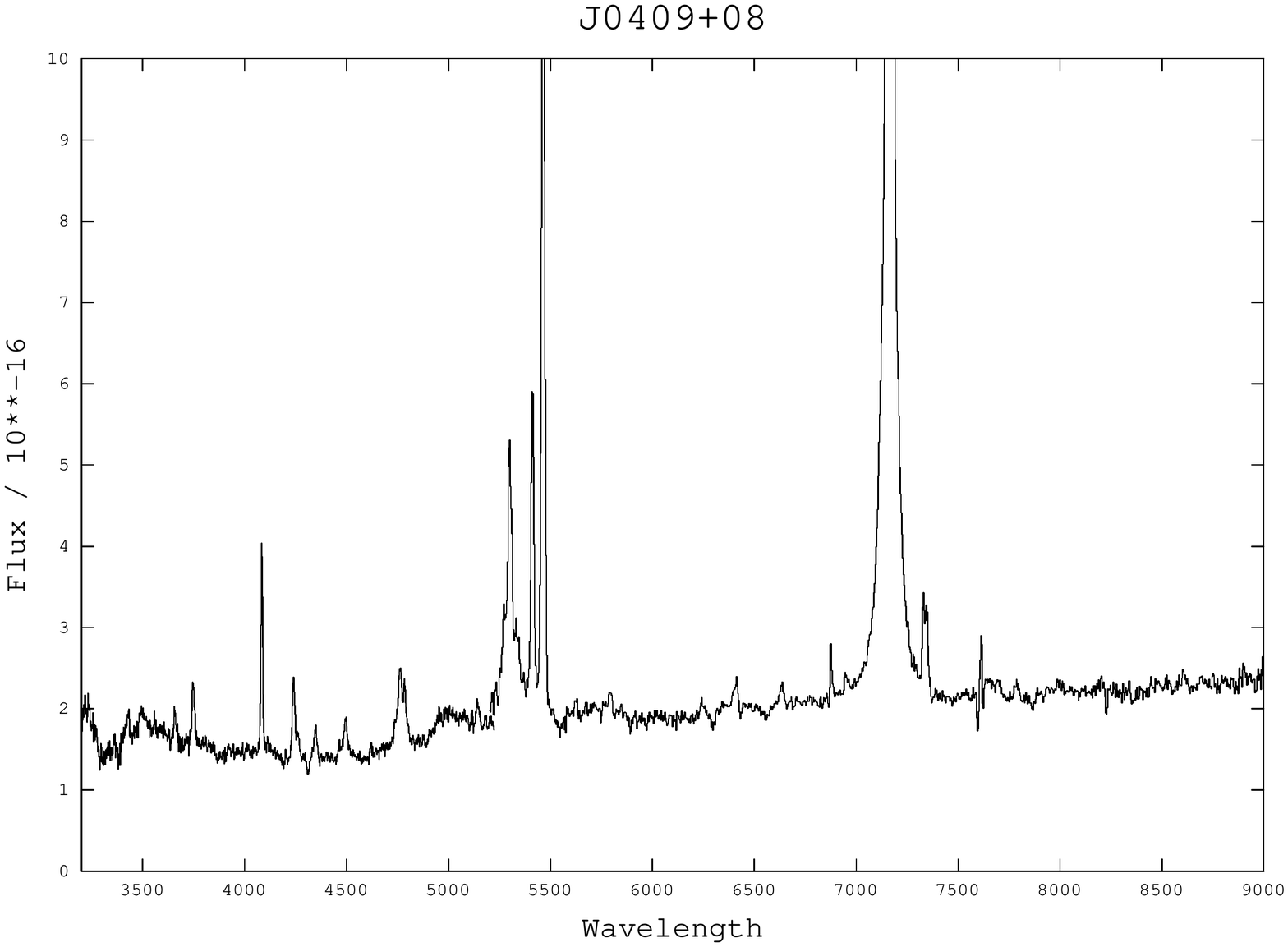}
\caption{The combined (blue and red arm) spectra of the 2MASS sample discussed in this paper. Note the remarkable variety in the spectra: the spectral types range from Seyfert 2 galaxies that lack the broad emission lines (e.g. J0504-19 \& J1057-13), through highly reddened broad line objects (e.g. J1040+59 \& J1127+24), to objects that show optical spectra similar to $\lq$normal' blue quasars (e.g. J1338-0438 \& J2124-17), peculiar objects with strong forbidden high ionization lines (FHILs) in their spectra (e.g. J1131+16; \citealt{rose}) and even LINER/HII region-like objects (e.g. J0411-01 \& J1001+41). J1637+25 and J2124-17 have not had the telluric absorption features removed, because no telluric standard stars were observed during the particular run in which they were measured. {\sc Continued on next page.}}
\end{figure*}

\begin{figure*}
\centering
\ContinuedFloat
\includegraphics[scale=0.25]{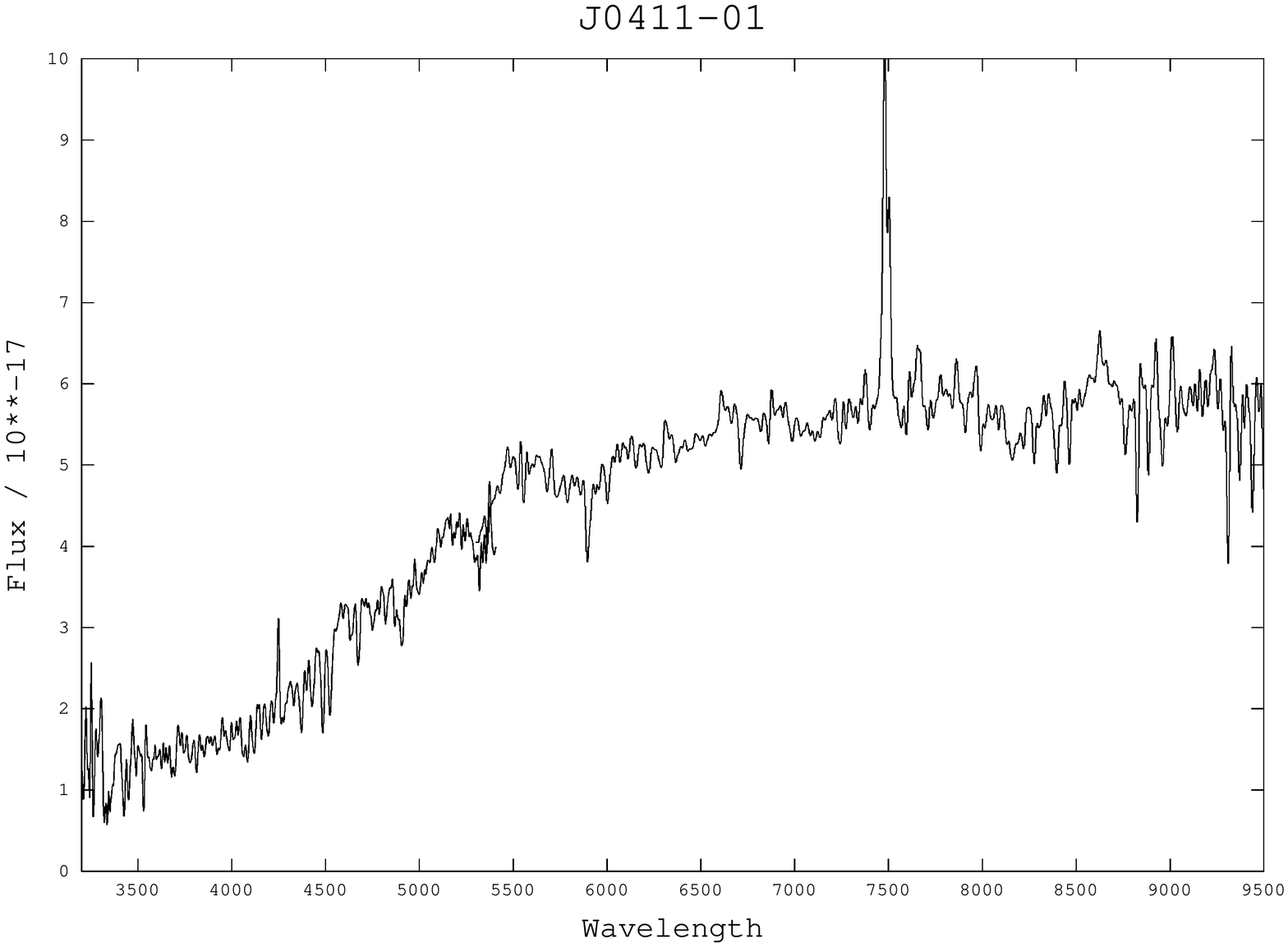}
\includegraphics[scale=0.25]{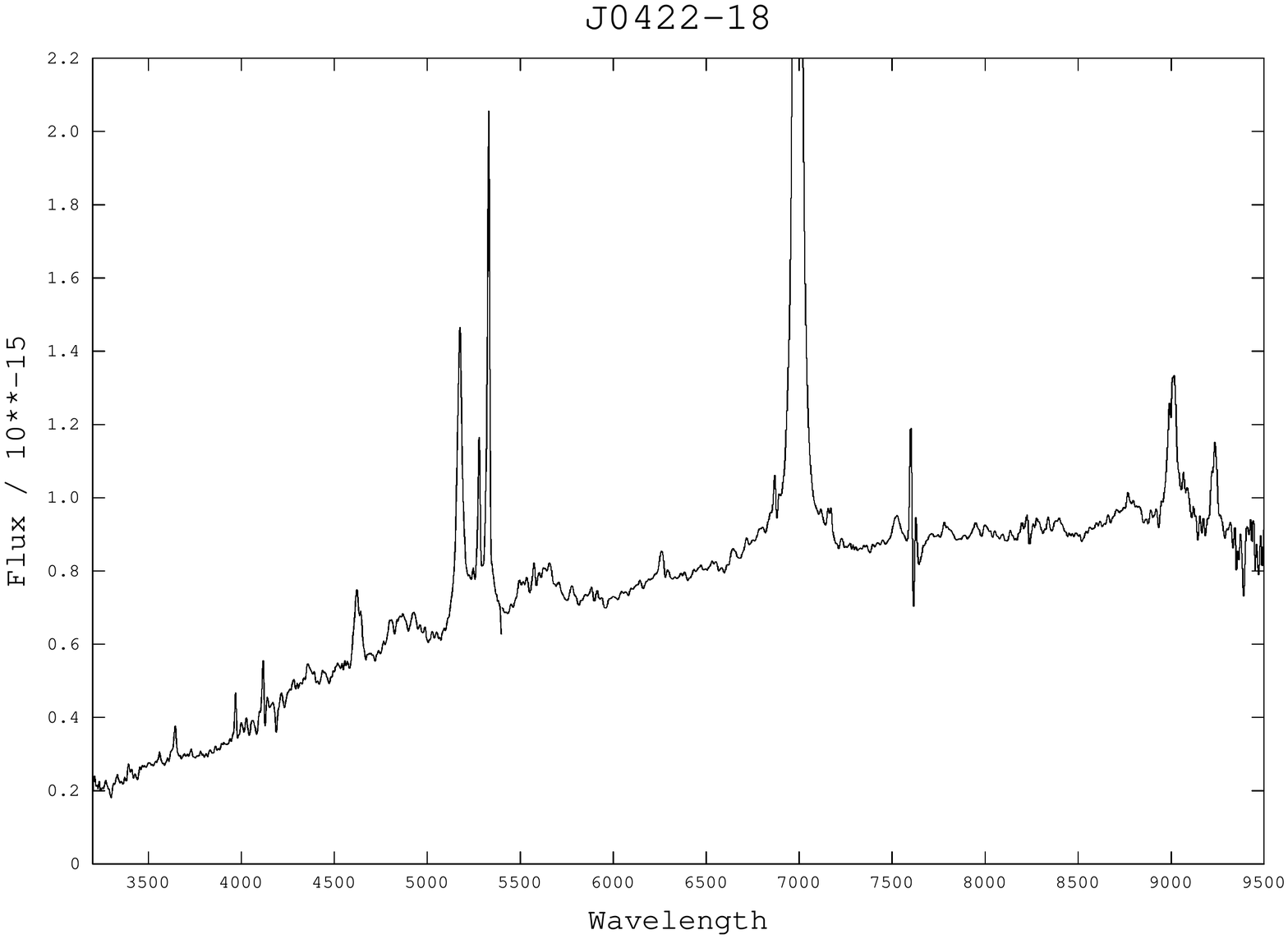}
\includegraphics[scale=0.25]{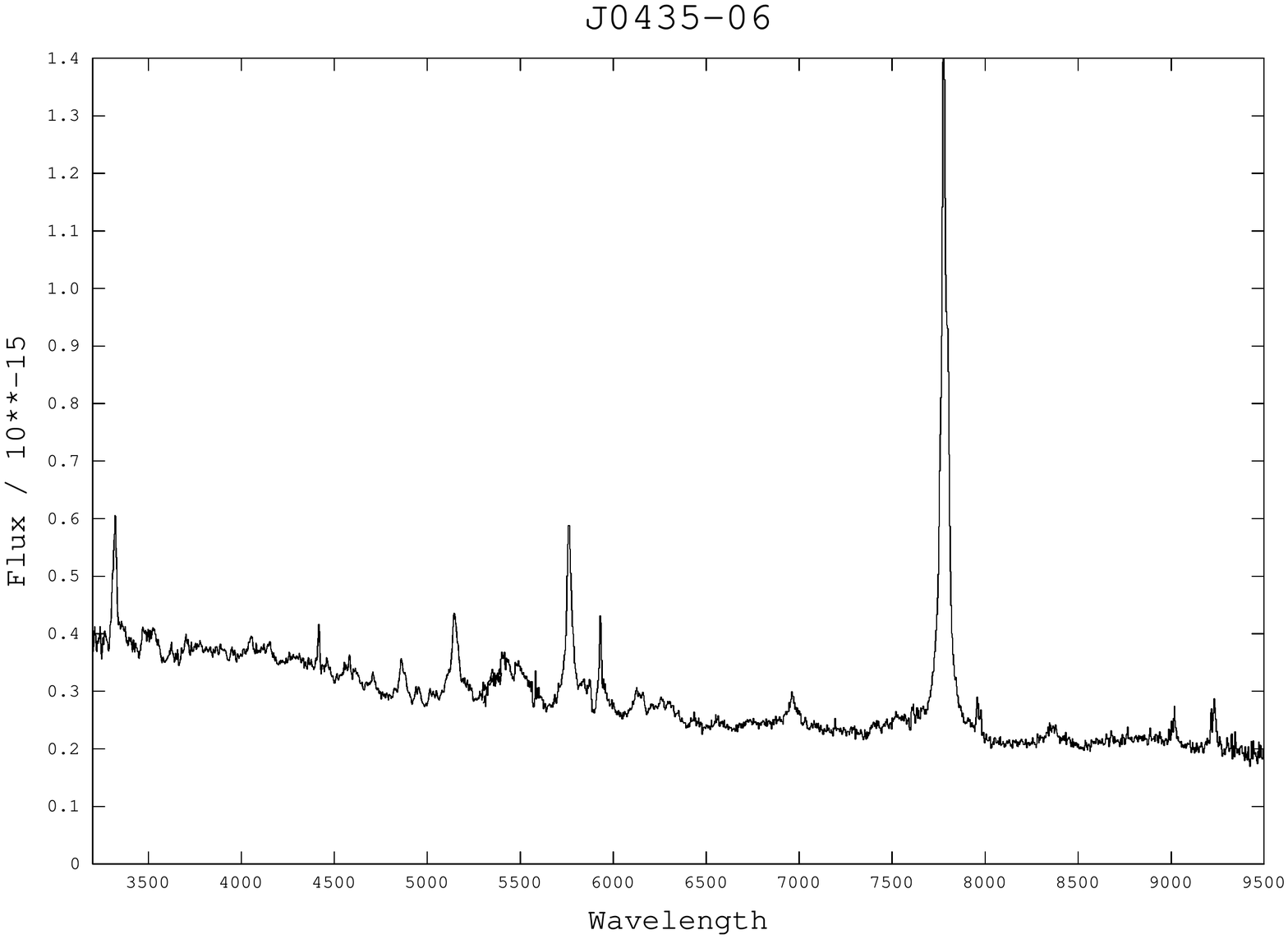}
\includegraphics[scale=0.25]{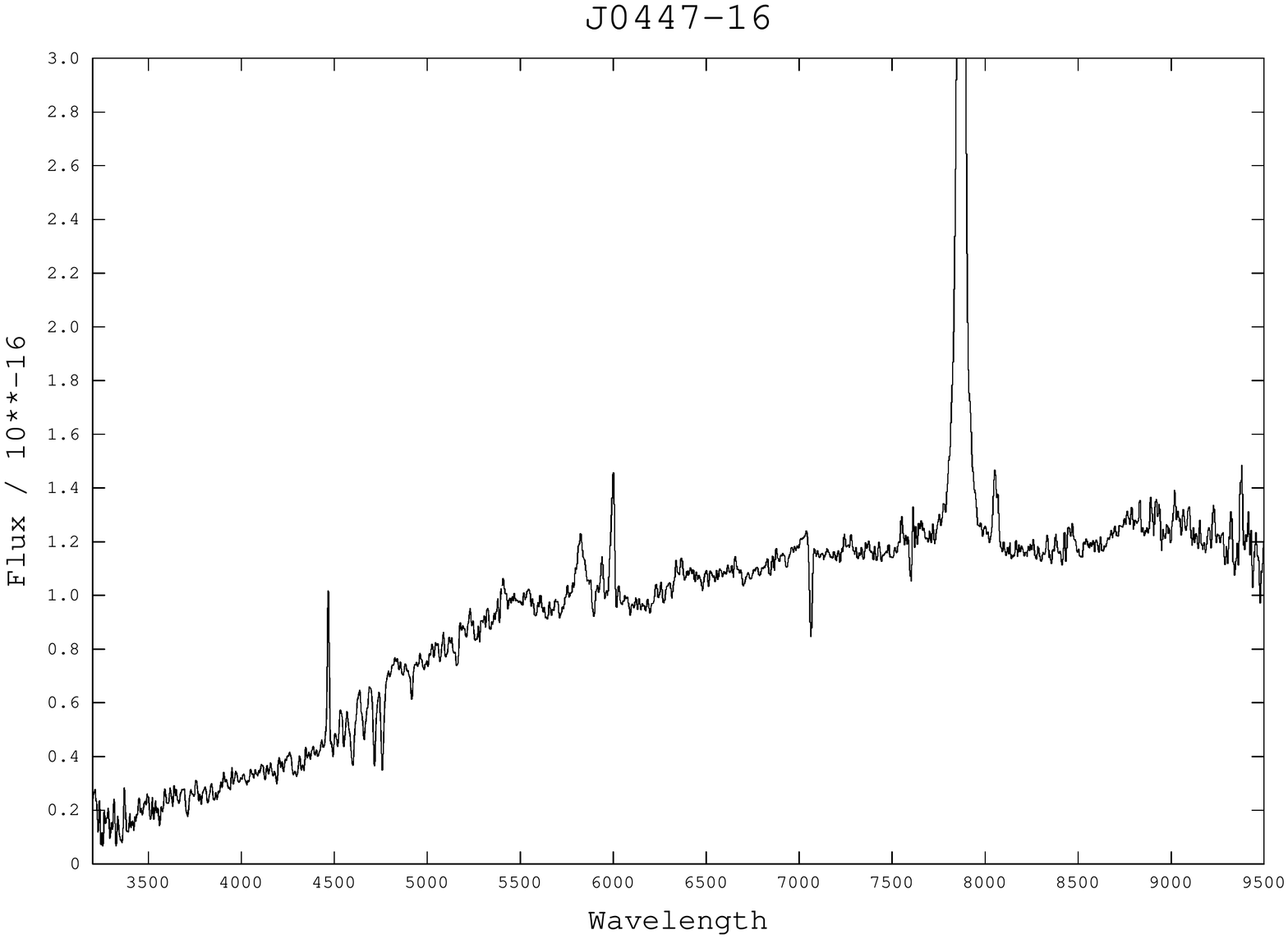}
\includegraphics[scale=0.25]{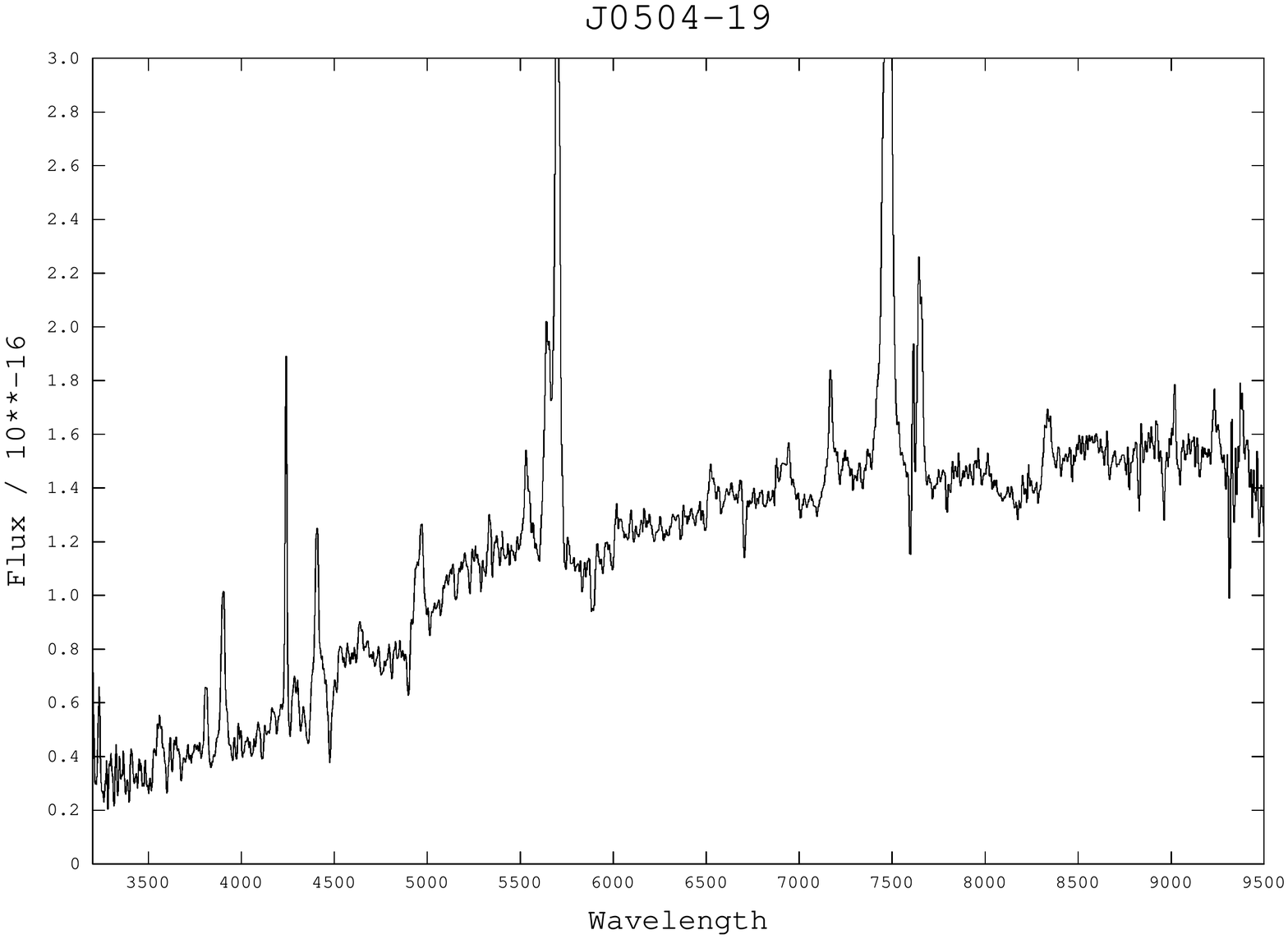}
\includegraphics[scale=0.25]{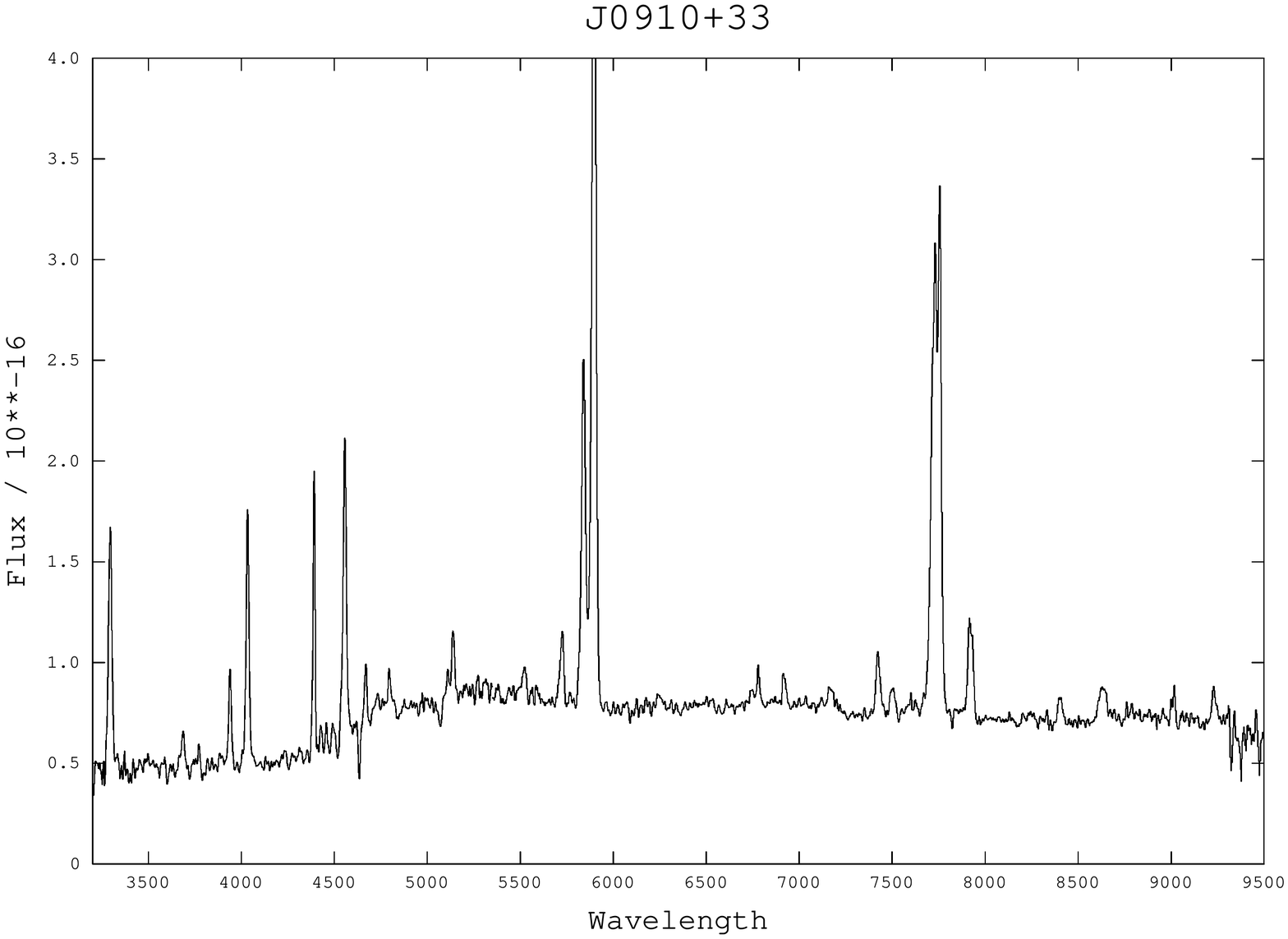}
\includegraphics[scale=0.25]{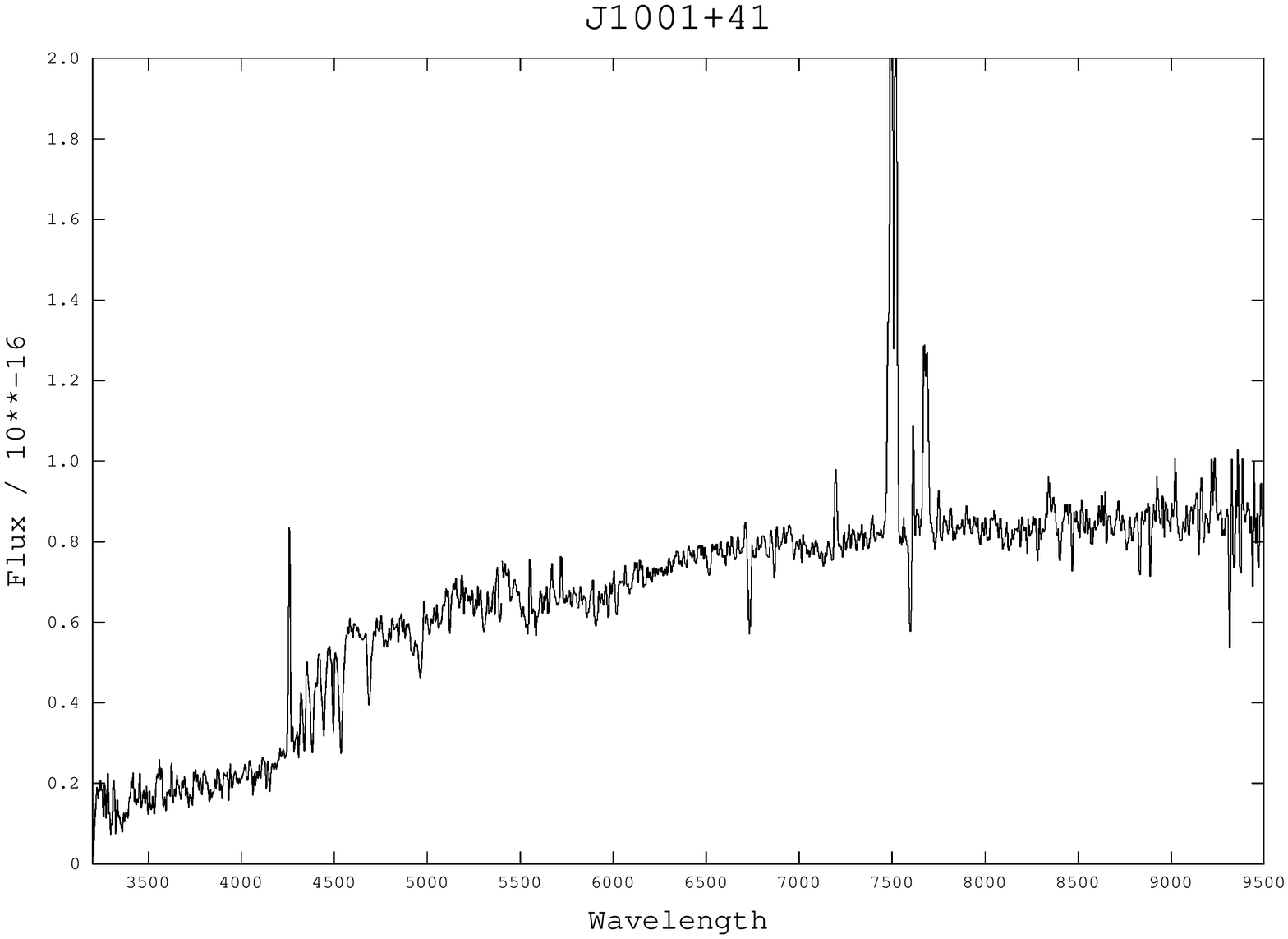}
\includegraphics[scale=0.25]{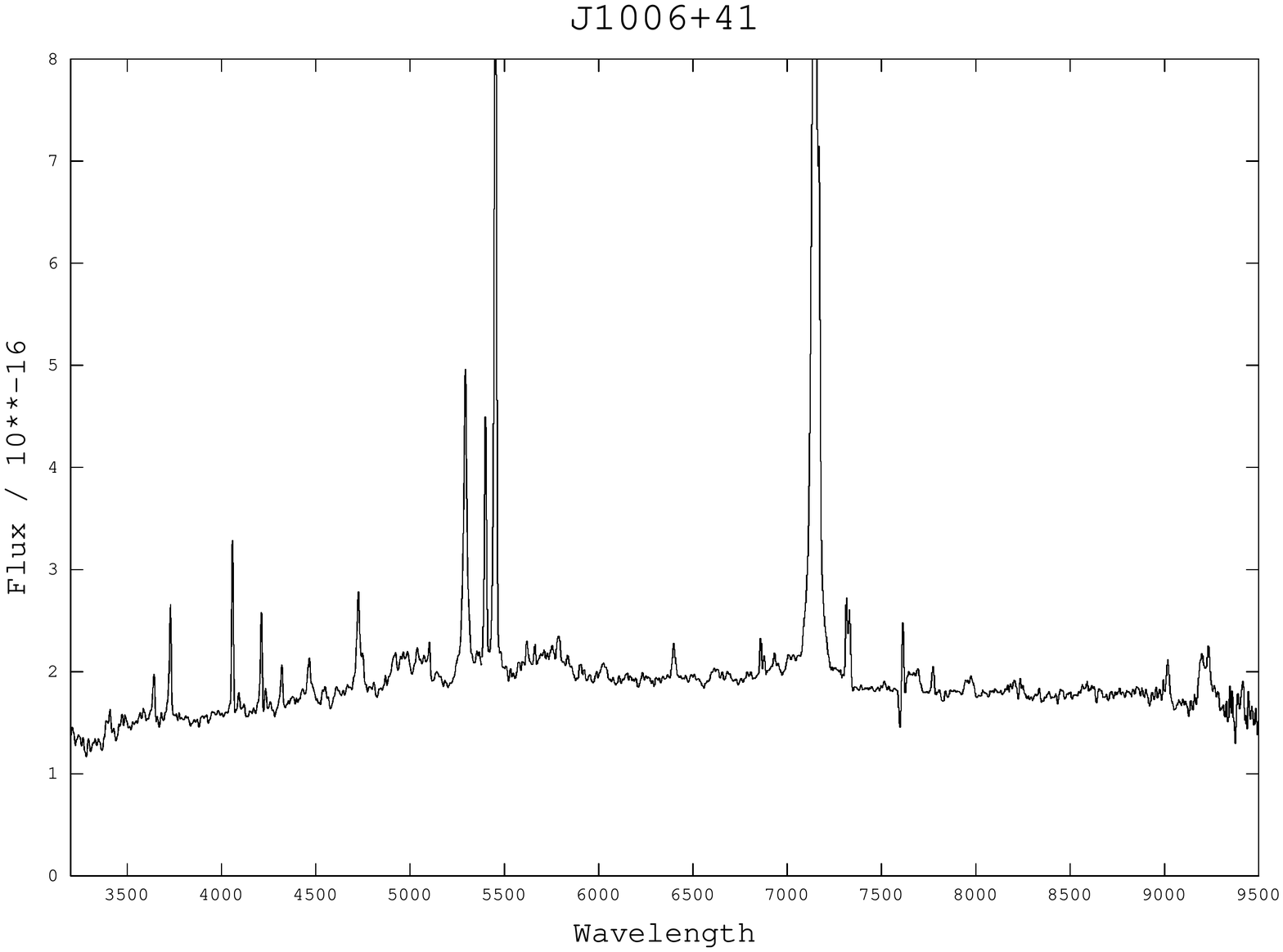}
\caption{}
\end{figure*}

\begin{figure*}
\centering
\ContinuedFloat
\includegraphics[scale=0.25]{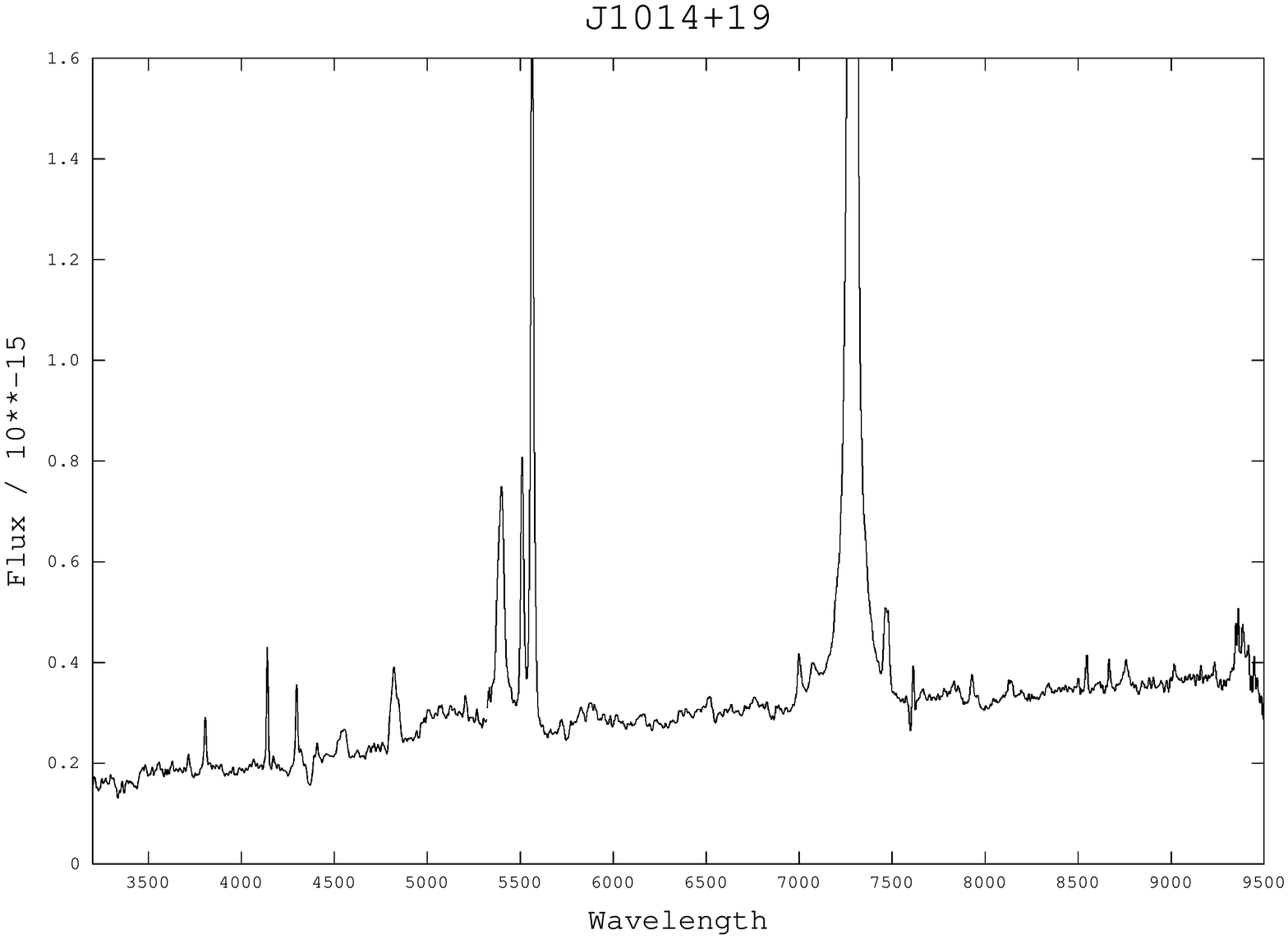}
\includegraphics[scale=0.25]{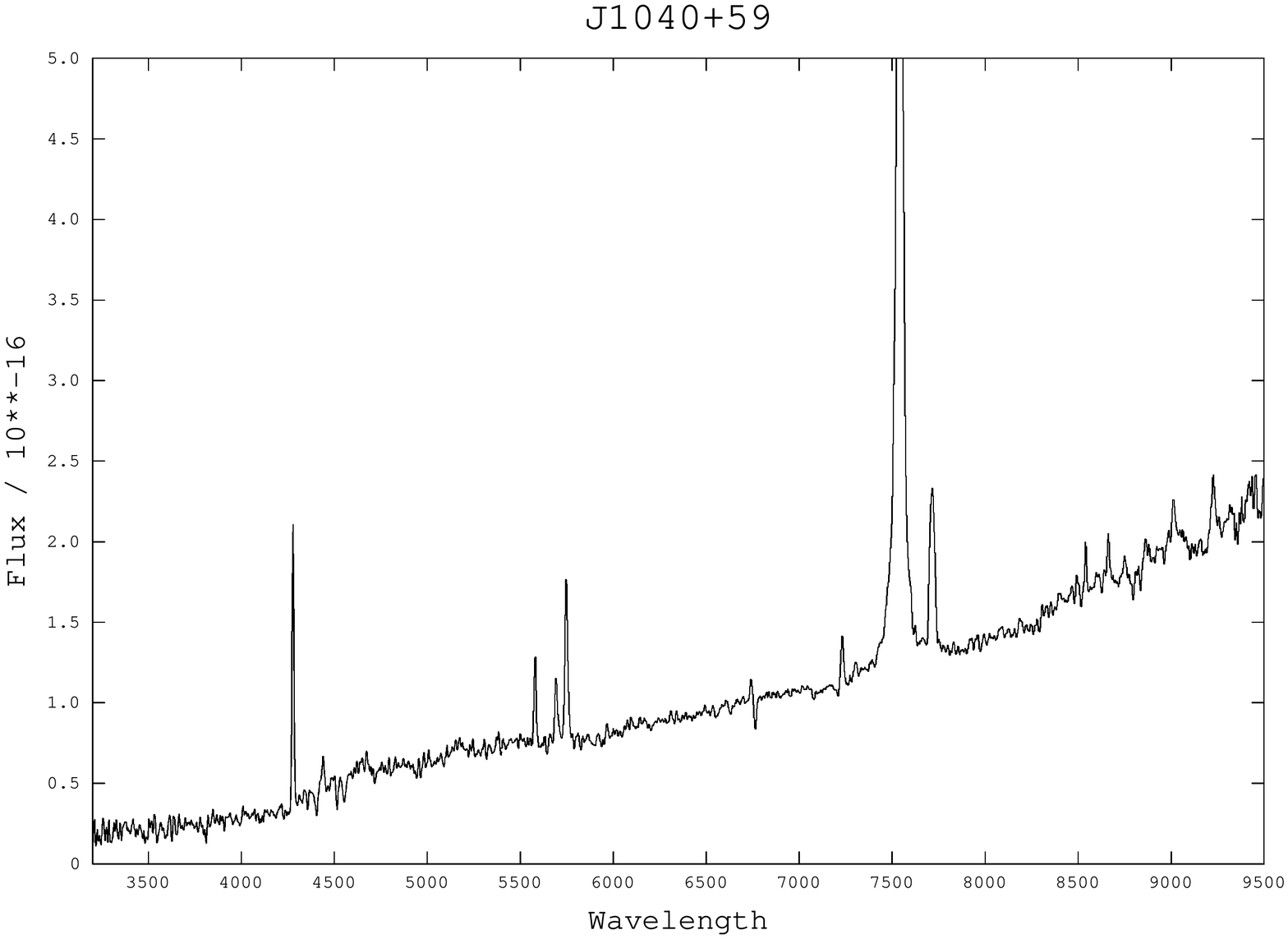}
\includegraphics[scale=0.25]{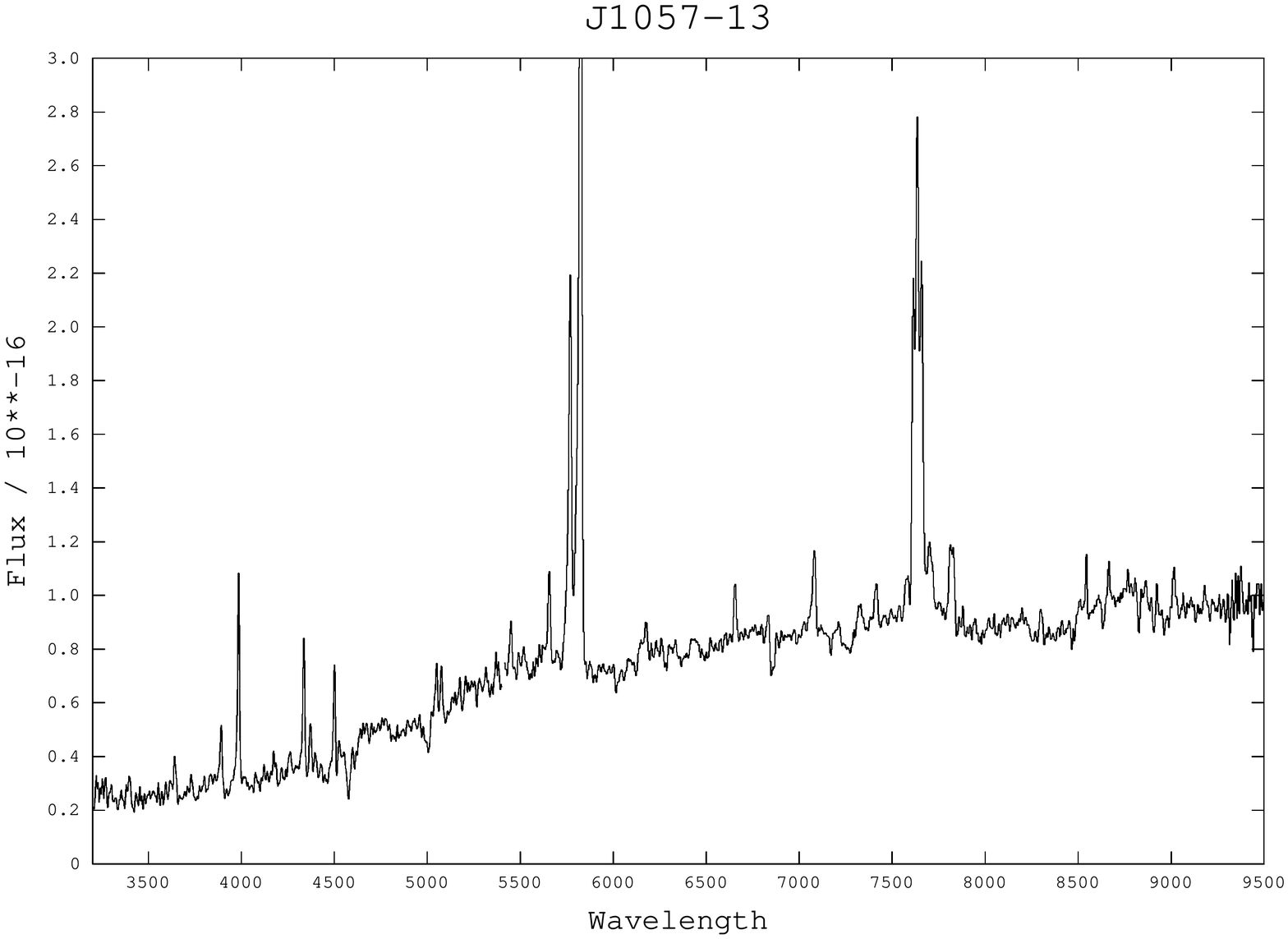}
\includegraphics[scale=0.25]{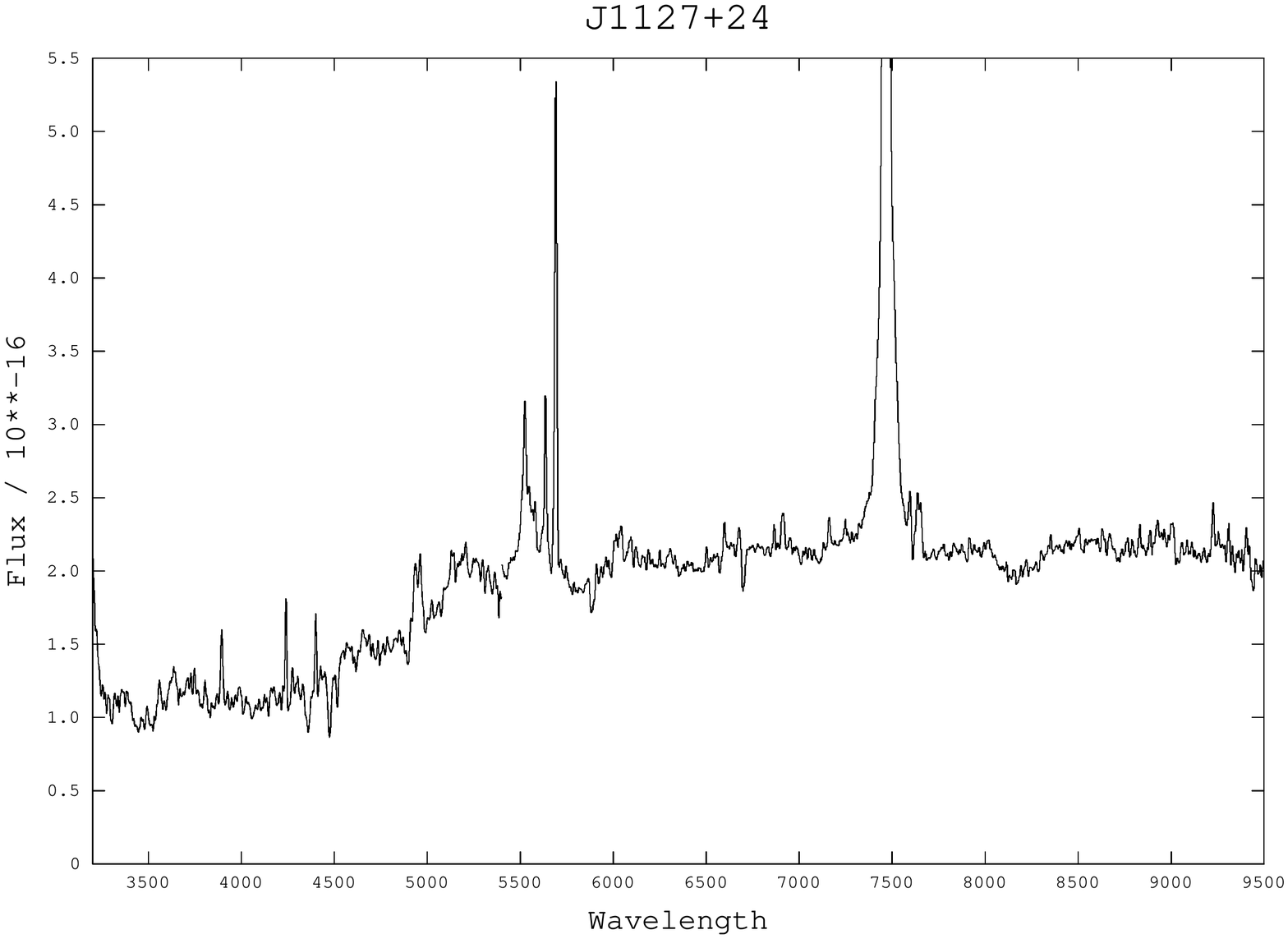}
\includegraphics[scale=0.25]{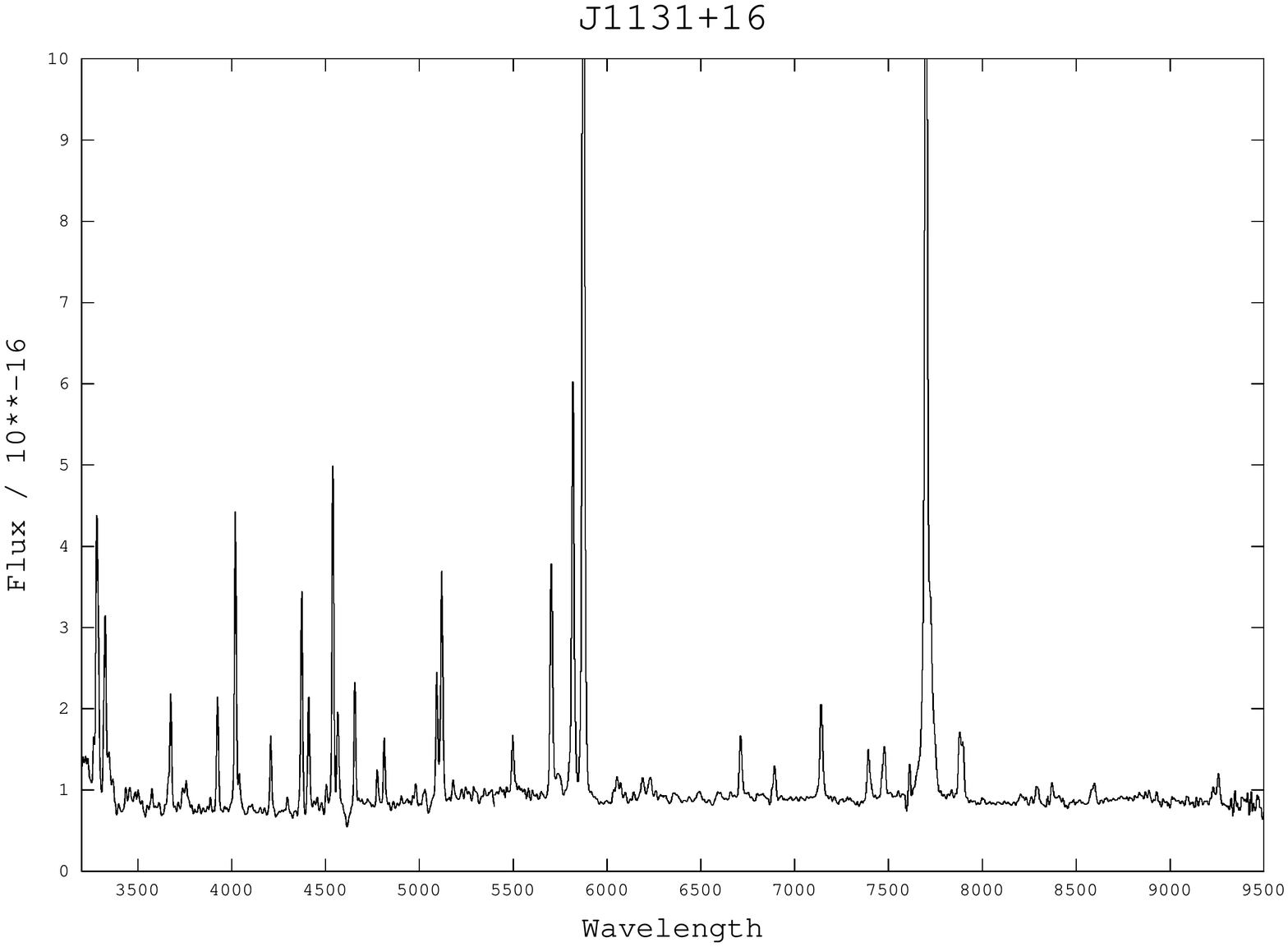}
\includegraphics[scale=0.25]{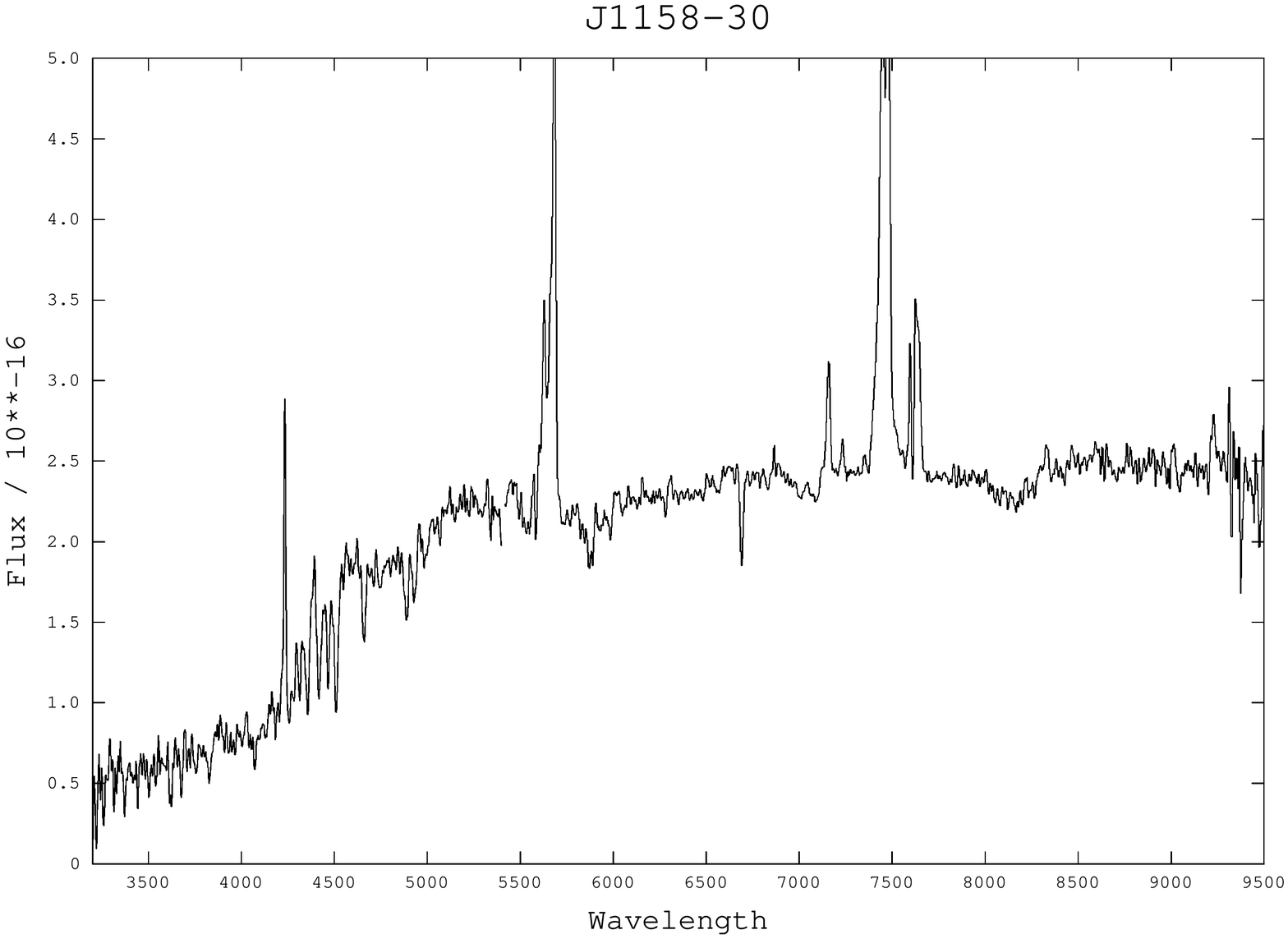}
\includegraphics[scale=0.25]{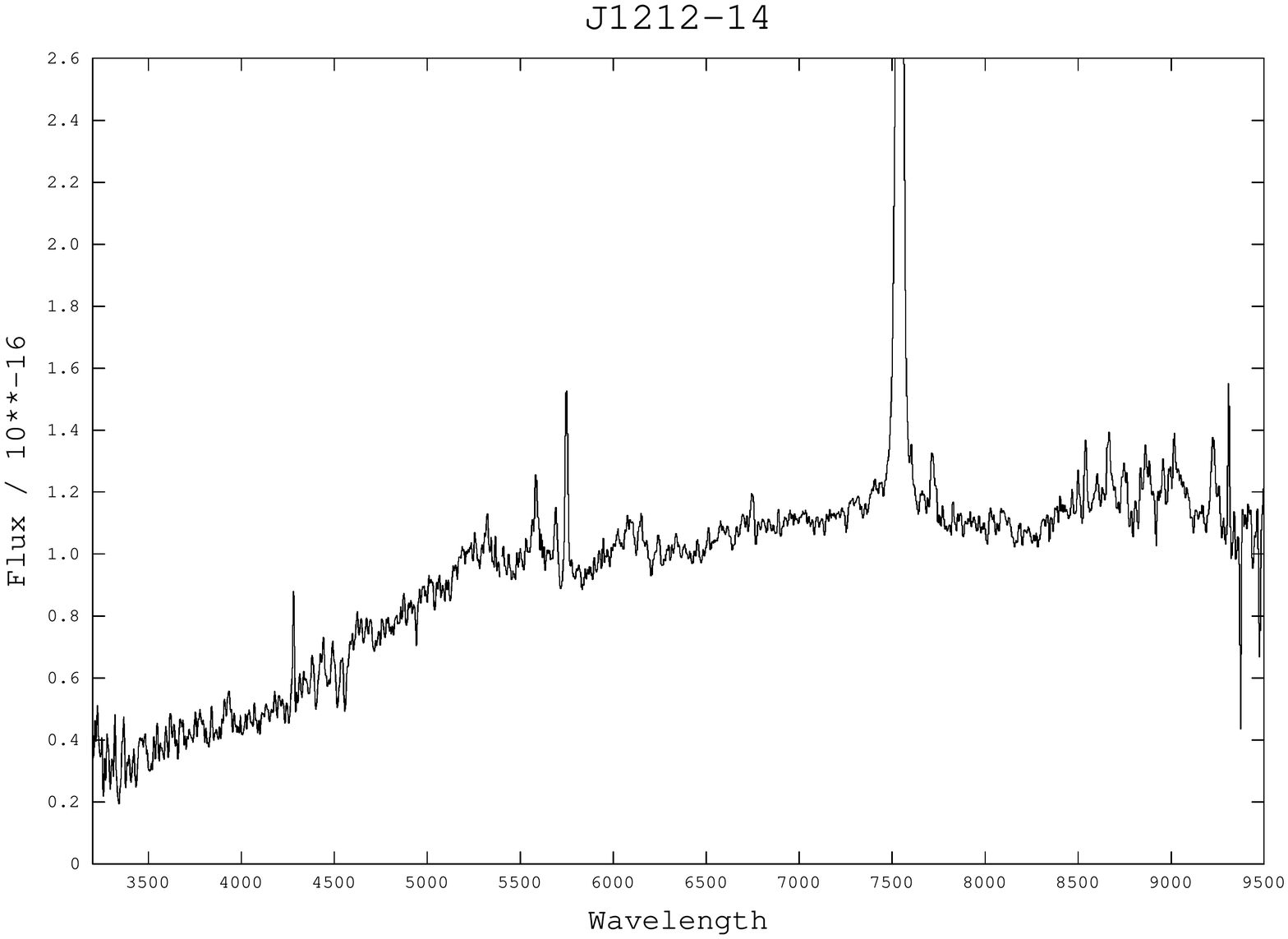}
\includegraphics[scale=0.25]{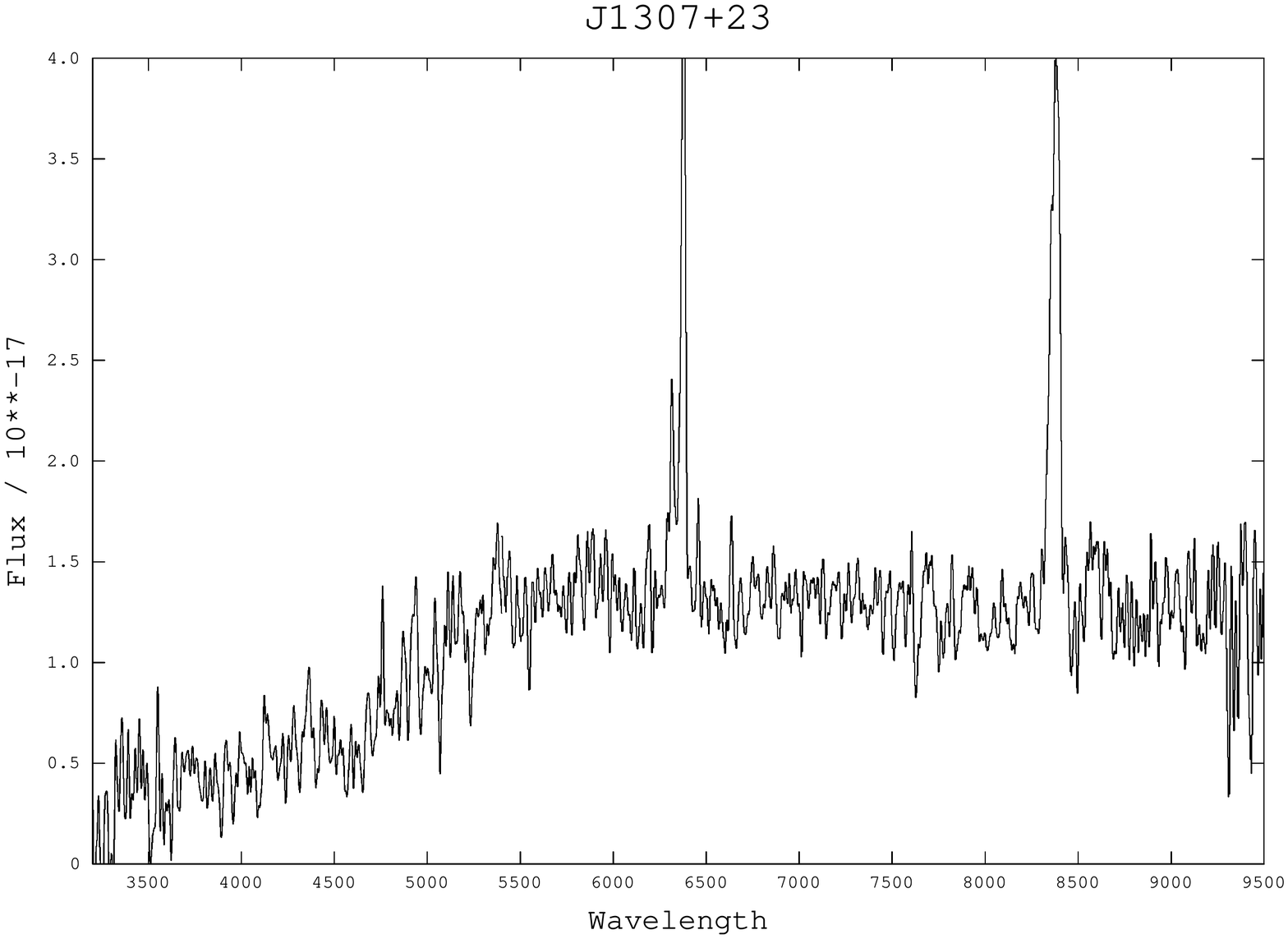}
\caption{}
\end{figure*}

\begin{figure*}
\centering
\ContinuedFloat
\includegraphics[scale=0.25]{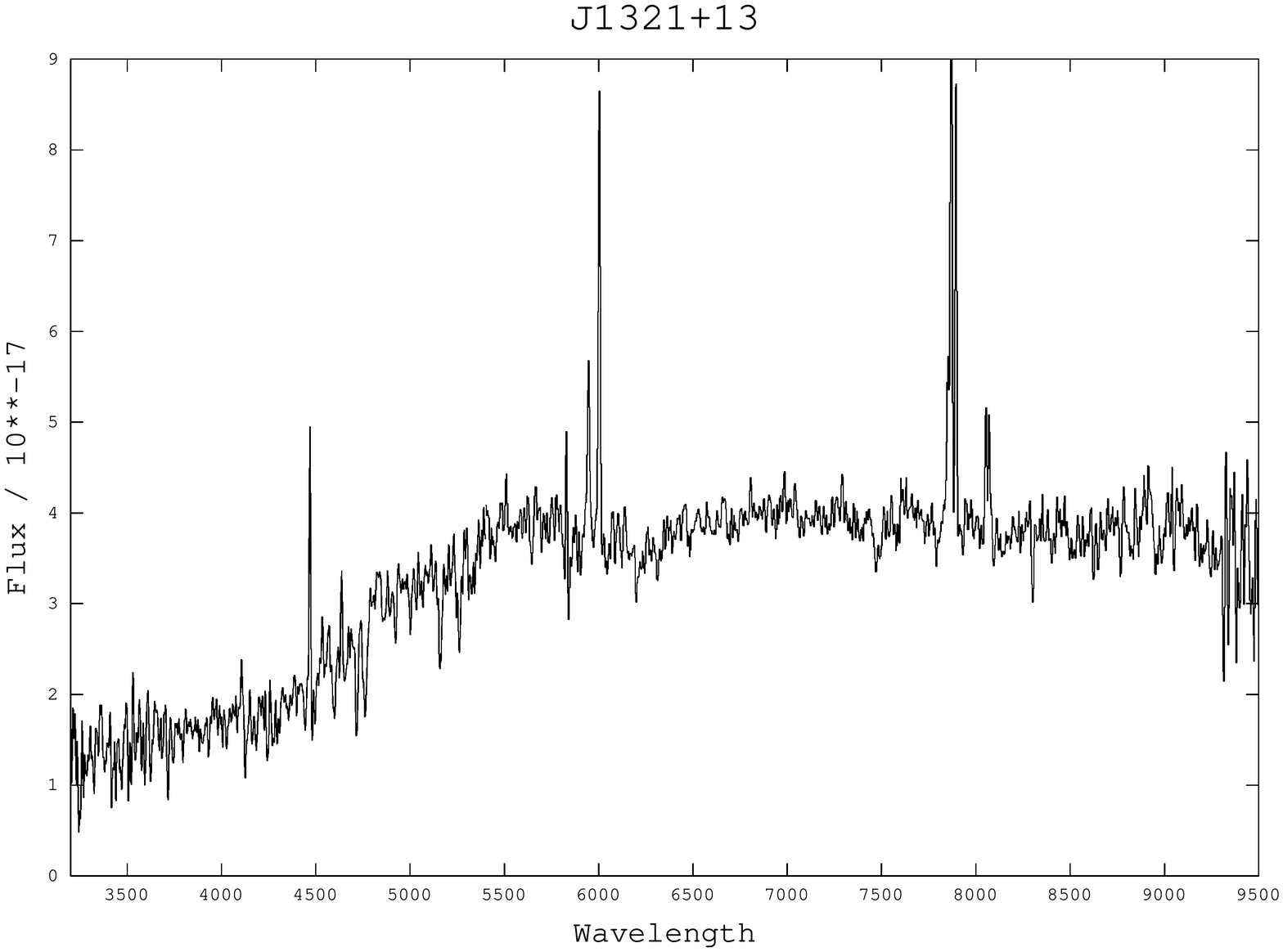}
\includegraphics[scale=0.25]{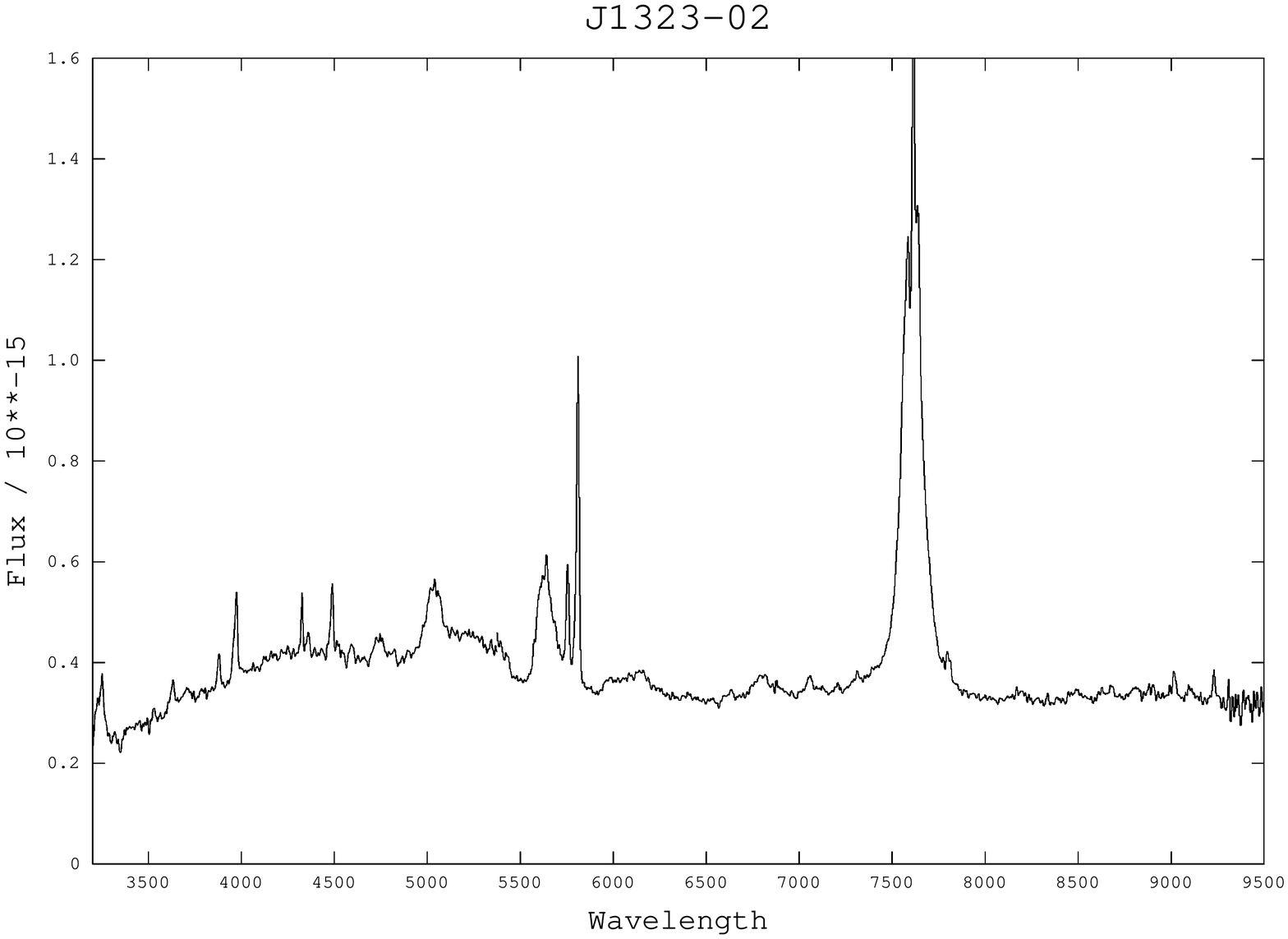}
\includegraphics[scale=0.25]{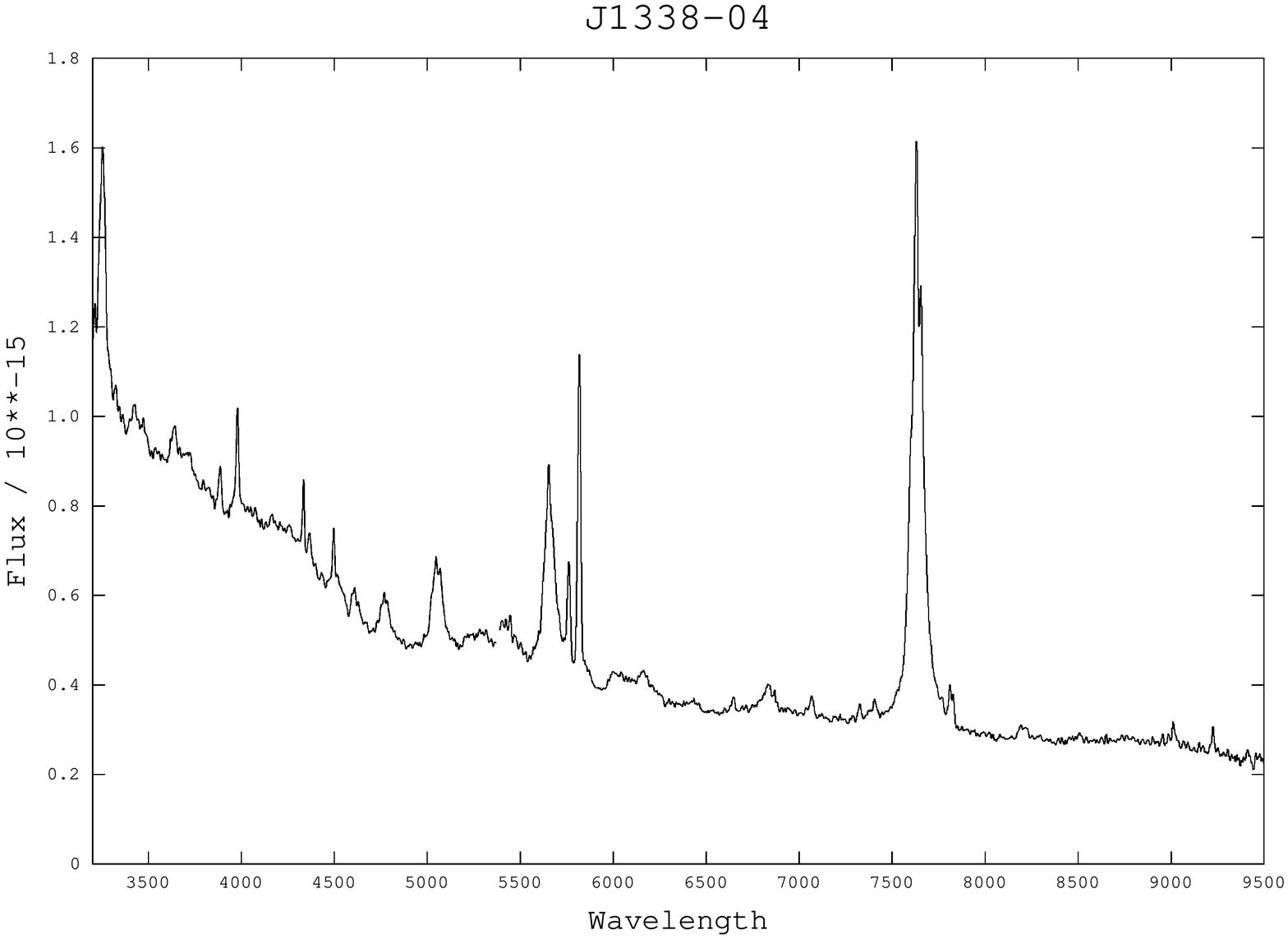}
\includegraphics[scale=0.25]{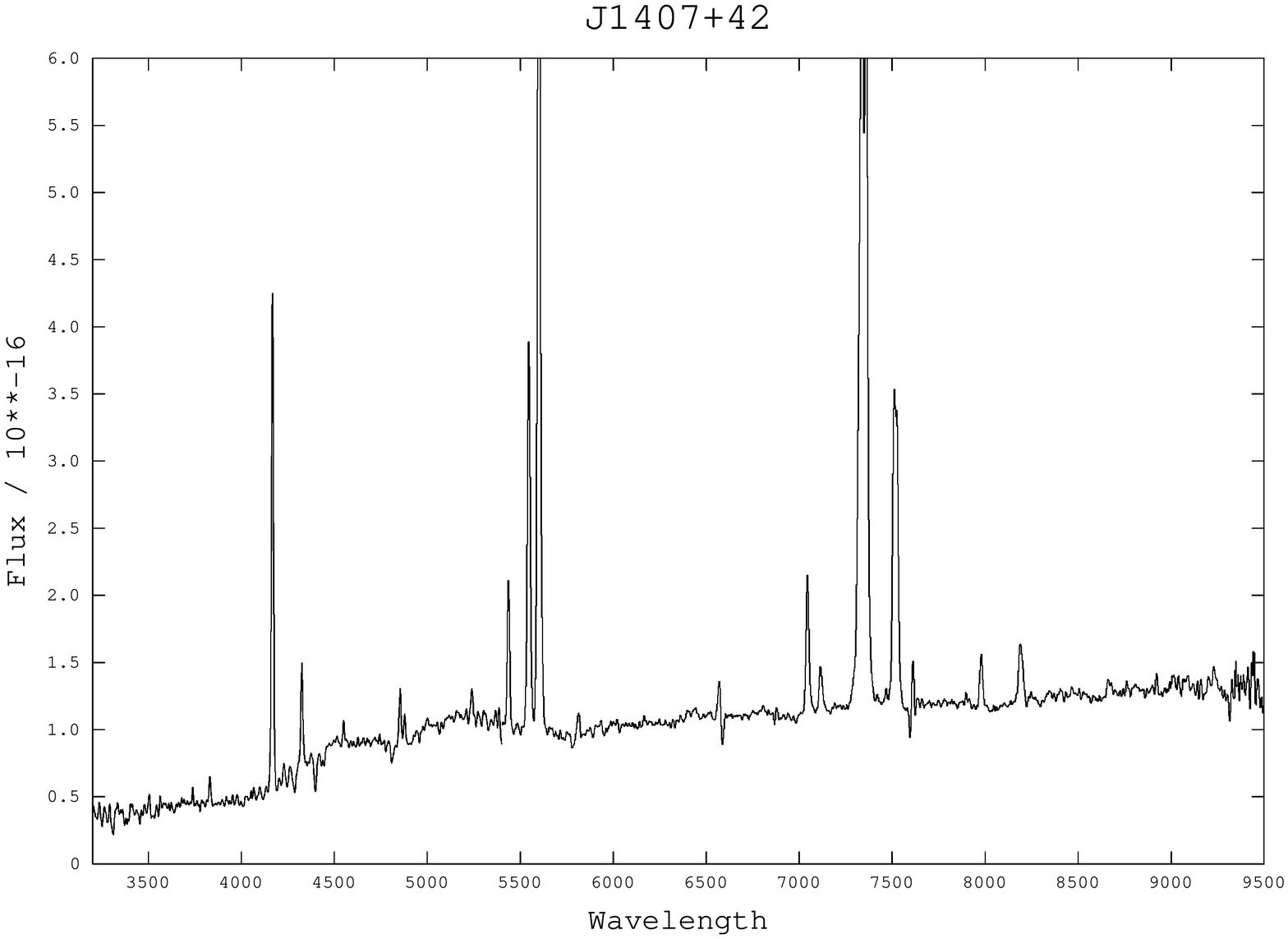}
\includegraphics[scale=0.25]{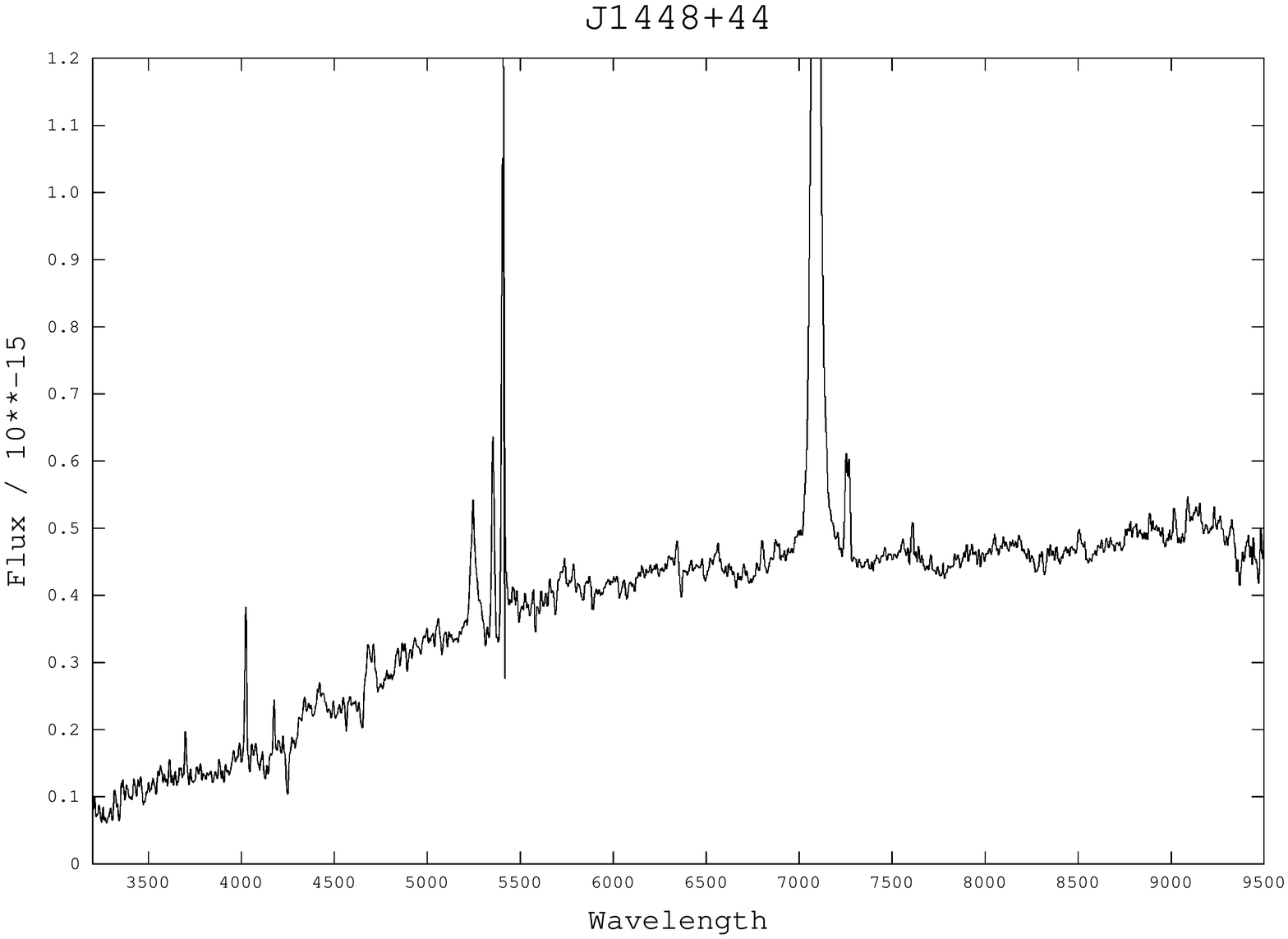}
\includegraphics[scale=0.25]{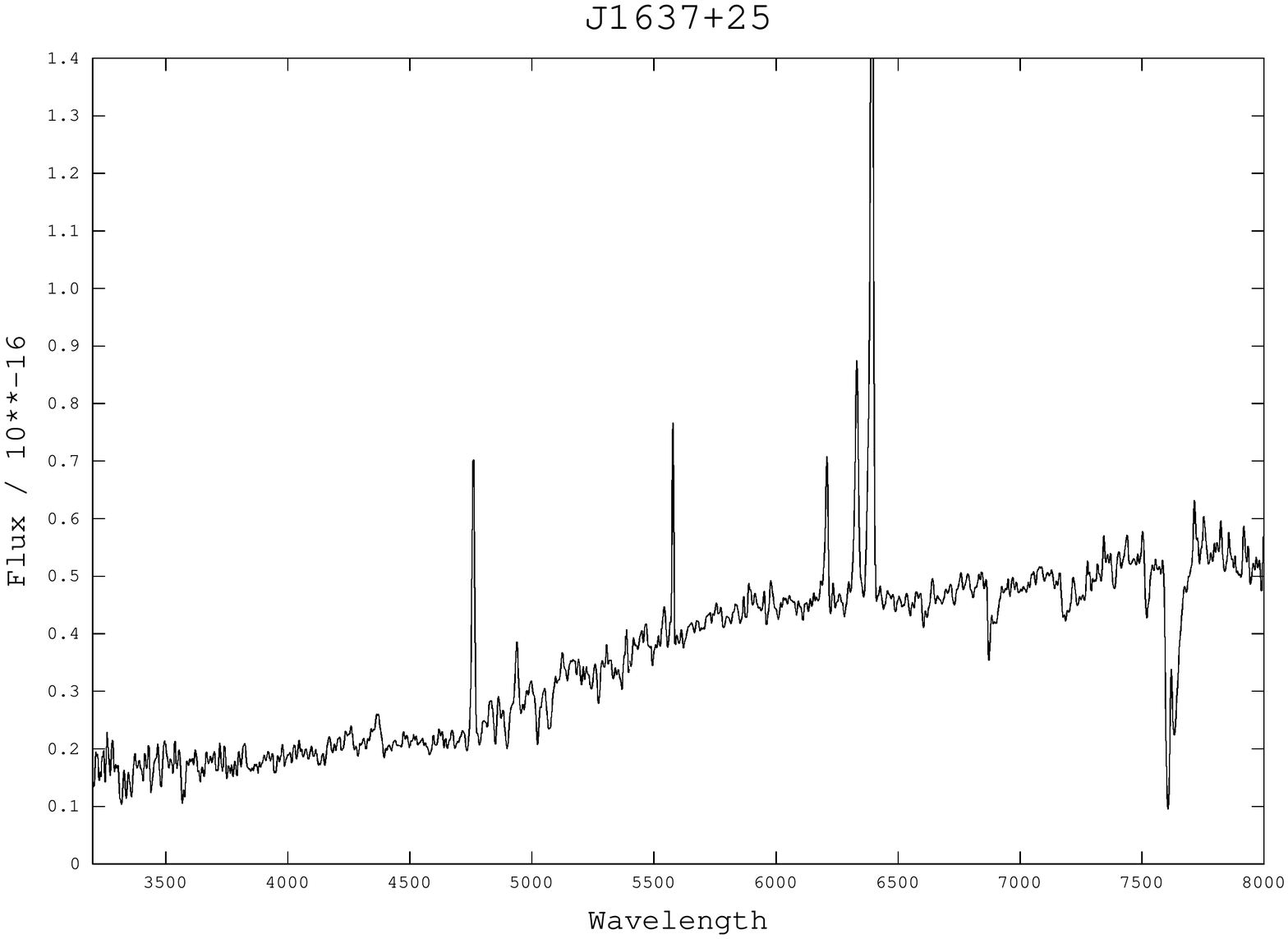}
\includegraphics[scale=0.25]{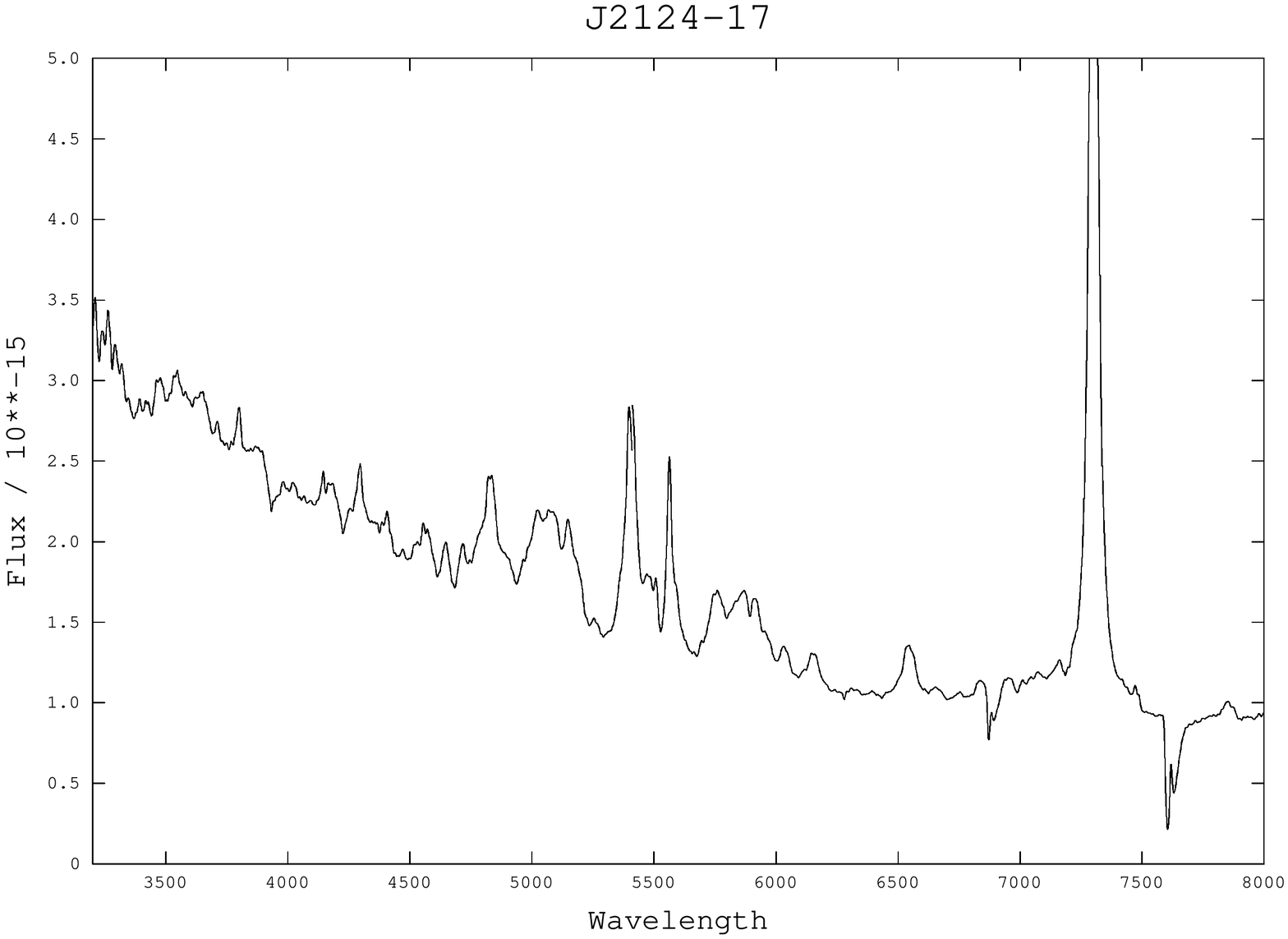}
\caption{}
\label{specs}
\end{figure*}

The full sample for this survey comprises a complete RA-limited (2 $<$ RA $<$ 15 hr) sub-sample of 27 objects with red near-IR colours (J-K$_S$ $\ga$ 2), K-band magnitudes 11.0 $<$ Ks $<$ 14.9 mag, and redshifts $z < 0.2$ selected from the list of \citet{hutchings03}, which
is itself representative of the population of red, 2MASS-selected quasars. We also observed a further 2 objects from \citet{hutchings03} that fall outside the original redshift and/or RA range: J1637+25 (z=0.277), and J2124-17 (z=0.111). Therefore, the total sample consists of 29 objects, however two of these objects have redshifts z$>$0.2. Although we present the data for the full 29 objects, we only consider the results for the 27 objects with redshifts z$<$0.2 in the analysis. Table \ref{prop} shows the basic properties of the sample.

Low resolution optical spectroscopic observations of the sample were taken with the ISIS dual-arm spectrograph on the 4.2-m William Herschel
Telescope (WHT) on La Palma in three runs: the 28th of July 2006, 8th and 9th of
February 2007, and the 26th and 27th of September 2011. The data were taken with the ISIS Dual arm spectrograph on the WHT. On the red arm, the R158R grating was used
with the REDPLUS CCD, and on the blue arm, the R300B grating was used with
the EEV12 CCD. A dichroic cutting at 5300\AA~was employed to obtain spectra simultaneously in the blue ($\sim$3250-5250\AA), and
in the red ($\sim$5200-9500\AA). To reduce the effects of
differential refraction, all exposures were taken when the objects were at
low airmass (sec $z$ $<$ 1.1) and/or with the slit aligned close to the
parallactic angle. The seeing for the nights of the observations varied over the range 0.8 $<$ FWHM $<$ 1.3 arcseconds.

Sets of three exposures were taken on both arms simultaneously for each object using a 1.5 arcsecond slit. An additional wider slit (5 arcseconds) exposure was taken on both arms for 23 of the objects\footnote{Wide slit exposures were not taken for 6 of the objects in the sample: J0221+13, J0248+14, J0306-05, J0312+07, J1637+25 and J2124-17. These objects were observed on separate runs to the rest of the sample and no wide slit exposures were taken for them.}. This was to assess the effect of possible slit losses on the emission line flux. To eliminate contamination from second order emission, a GG495 blocking filter was introduced into the ISIS red arm. The exposure times for each object are given in Table \ref{prop}.

The data were reduced in the standard way (bias subtraction, flat
fielding, cosmic ray removal, wavelength calibration, flux
calibration) using packages in {\sc iraf}\footnote{IRAF is distributed by the National Optical Astronomy Observatory, which is
operated by the Association of Universities for the Research in Astronomy,
Inc., under cooperative agreement with the National Science Foundation (
http://iraf.noao.edu/).}. The
two-dimensional spectra were also corrected for spatial distortions. To reduce wavelength calibration errors due to flexure of the
telescope and instrument, arc spectra were taken at the position of the
object on the sky. Based on measurements of the night sky lines, the wavelength calibration uncertainties are 0.1\AA\ on both the blue and red arms, and the narrow slit spectral resolution is estimated to be $\sim$5.6\AA\, and $\sim$10.5\AA\ on the blue and red arms respectively for 23/29 of the objects, and $\sim$6.6\AA\ and $\sim$6.1\AA\ for the blue and red arms for the rest of the sample (see Table \ref{prop}). Atmospheric absorption features were removed by dividing the red spectrum of each object by that of a telluric standard, taken close in time and airmass to the observations of 27/29 of the objects. No telluric standards were taken for J1637+25 and J2124-17, and therefore the atmospheric features were not removed for those objects. The spatial pixel scales of the 2D spectra are 0.4 arcseconds in the blue and 0.44 arcseconds in the red, and the relative flux calibration uncertainty -- based on 11 observations of 8 flux standard stars
taken throughout the runs -- is estimated to be $\pm$5\%. 

The spectra were extracted and analysed using the {\sc starlink}
packages {\sc figaro} and {\sc dipso}. All extracted spectra were corrected for Galactic extinction using the Galactic extinction maps of \citet{schlegel} and the extinction laws of of \citet{seaton} for the UV, and \citet{howarth} for the optical/IR, prior to the analysis. In order to investigate the nuclear spectra, we extracted apertures centred on the nuclei of the targets in the 2D spectra, with aperture sizes chosen such that the full flux from the nucleus of each object was extracted. The aperture sizes are given in Table \ref{prop} and the extracted spectra are shown in Figure \ref{specs}. 

\begin{landscape}
\begin{table}
\begin{center}
\caption{Basic properties of the 2MASS sample. The second column gives the abbreviated name used for each object in this paper. 'Type' gives the spectral type of the AGN. We only classify an object as Type 1 if it shows significant (i.e. well above the noise) broad wings that cannot be accounted for by the [OIII] model fitted
to the H$\alpha$+[NII] lines. Redshifts and [OIII] emission line luminosities (L$_{[OIII]}$) were determined using the [OIII]$\lambda$5007 emission line, and J-K$_S$ values were taken from the 2MASS point source catalogue. The key for the \lq Run' column is: a -- observed on 8-9th of February 2007, b -- 28th of July 2006 and c -- 26-27th of September 2011.}
\begin{tabular}{lcccllclclll}
\\
\hline
&  & &				&	&			&		&	Exp	&	B/N  &		&			&	\\								
Full Object Name	& Name & Type &	z			&	J-K$_S$	&	L$_{[OIII]}$			&	Run	&	Time	&	slit  &	Blue	&	Red		& Aperture	\\					
& &	&			&		& (10$^{34}$ W)			&	&	(s)	& ratio  &	(\AA)&	(\AA) &  (arcsec$^2$)	\\												
																											
\hline																											
J02215058+1327409	&	J0221+13	&	Type 2	&	0.1402	$\pm$	0.00021	&	2.37	&	0.76	$\pm$	0.09	&	c	&	600	&	-			&	6.80	&	6.14 & 2.2x1.5	\\
J02480733+1459577	&	J0248+14	&	Type 1	&	0.0718	$\pm$	0.00024	&	2.13	&	0.21	$\pm$	0.02	&	c	&	300	&	-			&	6.87	&	6.35 & 2.2x1.5	\\
J03065242-0531569	&	J0306-05	&	Type 2	&	0.1261	$\pm$	0.00089	&	2.24	&	0.95	$\pm$	0.83	&	c	&	600	&	-			&	6.21	&	6.41 & 2.2x1.5	\\
J03123105+0706547	&	J0312+07	&	Type 1	&	0.1455	$\pm$	0.00033	&	1.98	&	0.7	$\pm$	0.03	&	c	&	600	&	-			&	6.57	&	6.51 & 1.76x1.5	\\
J04001974+0502149	&	J0400+05	&	Type 1	&	0.1876	$\pm$	0.00014	&	2.00	&	22.39	$\pm$	0.85	&	a	&	600	&	1.03	$\pm$	0.13	&	5.02	&	10.50 & 2.2x1.5	\\
J04092488+0758560	&	J0409+07	&	Type 1	&	0.0914	$\pm$	0.00019	&	2.17	&	3.67	$\pm$	0.19	&	a	&	600	&	1.14	$\pm$	0.37	&	5.26	&	10.13 & 1.76x1.5	\\
J04112647-0118056	&	J0411-01	&	LINER/HII	&	0.1395	$\pm$	0.0035	&	2.22	&	0.06	$\pm$	0.02	&	a	&	600	&	1.13	$\pm$	0.31	&	5.23	&	10.40 & 2.2x1.5	\\
J04225654-1854424	&	J0422-18	&	Type 1	&	0.0646	$\pm$	0.00018	&	2.21	&	2.35	$\pm$	0.03	&	a	&	300	&	1.08	$\pm$	0.06	&	5.59	&	10.56 & 2.2x1.5	\\
J04352254-0635256	&	J0435-06	&	Type 1	&	0.1846	$\pm$	0.00074	&	1.98	&	1.74	$\pm$	0.07	&	a	&	600	&	1.87	$\pm$	0.59	&	5.56	&	10.40 & 2.2x1.5	\\
J04474760-1649344	&	J0447-16	&	Type 1	&	0.1985	$\pm$	0.00092	&	2.10	&	1.04	$\pm$	0.11	&	a	&	600	&	0.85	$\pm$	0.33	&	5.55	&	10.83 & 2.2x1.5	\\
J05042569-1909258	&	J0504-19	&	Type 2	&	0.1376	$\pm$	0.00092	&	2.04	&	3.4	$\pm$	0.04	&	a	&	450	&	1.00	$\pm$	0.22	&	5.48	&	10.23 & 2.2x1.5	\\
J091000.74+334809	&	J0910+33	&	Type 2	&	0.1779	$\pm$	0.00021	&	2.33	&	9.01	$\pm$	0.28	&	a	&	600	&	1.15	$\pm$	0.12	&	5.31	&	10.23 & 1.76x1.5	\\
J100139.51+410423	&	J1001+41	&	LINER/HII	&	0.1427	$\pm$	0.0021	&	2.00	&	0.11	$\pm$	0.02	&	a	&	450	&	2.04	$\pm$	0.36	&	5.71	&	10.94 & 1.76x1.5	\\
J10065783+4104064	&	J1006+41	&	Type 1	&	0.0890	$\pm$	0.00034	&	1.92	&	1.99	$\pm$	0.1	&	a	&	600	&	1.12	$\pm$	0.48	&	5.17	&	10.54 & 1.76x1.5	\\
J101400.47+194614	&	J1014+19	&	Type 1	&	0.1112	$\pm$	0.00018	&	2.04	&	8.7	$\pm$	0.23	&	a	&	300	&	1.12	$\pm$	0.11	&	5.56	&	10.80 & 2.2x1.5	\\
J10404364+5934092	&	J1040+59	&	Type 1	&	0.1476	$\pm$	0.00076	&	3.02	&	0.98	$\pm$	0.09	&	a	&	600	&	1.03	$\pm$	0.36	&	5.33	&	10.24 & 2.2x1.5	\\
J10572861-1353597	&	J1057-13	&	Type 2	&	0.1633	$\pm$	0.0002	&	2.50	&	5.17	$\pm$	0.18	&	a	&	600	&	1.22	$\pm$	0.21	&	5.73	&	10.49 & 2.2x1.5	\\
J11275114+2432081	&	J1127+24	&	Type 1	&	0.1366	$\pm$	0.00021	&	1.97	&	0.81	$\pm$	0.01	&	a	&	500	&	1.15	$\pm$	0.03	&	5.52	&	10.06 & 2.64x1.5	\\
J113111.05+162739	&	J1131+16	&	Type 2	&	0.1732	$\pm$	0.0001	&	2.15	&	10.02	$\pm$	0.29	&	a	&	600	&	1.05	$\pm$	0.12	&	6.11	&	11.37 & 1.76x1.5	\\
J11582462-3003350	&	J1158-30	&	Type 2	&	0.1352	$\pm$	0.00057	&	2.02	&	6.31	$\pm$	0.21	&	a	&	450	&	1.30	$\pm$	0.17	&	5.52	&	9.83 & 2.64x1.5	\\
J12121449-1422161	&	J1212-14	&	Type 1	&	0.1481	$\pm$	0.0006	&	2.06	&	0.19	$\pm$	0.01	&	a	&	600	&	1.33	$\pm$	0.07	&	5.34	&	10.22 & 2.2x2	\\
J13070062+2338052	&	J1307+23	&	Type 2	&	0.2741	$\pm$	0.0012	&	3.31	&	0.41	$\pm$	0.06	&	a	&	600	&	4.18	$\pm$	2.43	&	5.54	&	10.63 & 2.2x1.5	\\
J13213908+1342304	&	J1321+13	&	Type 2	&	0.199	$\pm$	0.00032	&	2.15	&	0.59	$\pm$	0.02	&	a	&	600	&	1.15	$\pm$	0.04	&	5.25	&	10.64 & 1.76x1.5	\\
J13231468-0219013	&	J1323-02	&	Type 1	&	0.1606	$\pm$	0.0003	&	2.25	&	7.82	$\pm$	0.27	&	a	&	600	&	0.97	$\pm$	0.15	&	5.33	&	10.66 & 2.2x1.5	\\
J13384530-0438530	&	J1338-04	&	Type 1	&	0.1625	$\pm$	0.00095	&	1.99	&	12.6	$\pm$	0.53	&	a	&	450	&	1.43	$\pm$	0.49	&	5.57	&	10.98 & 2.2x1.5	\\
J14073748+4256162	&	J1407+42	&	Type 2	&	0.1183	$\pm$	9.9E-05	&	2.16	&	13.48	$\pm$	0.59	&	a	&	600	&	1.24	$\pm$	0.14	&	5.34	&	10.03 & 2.2x1.5	\\
J14481932+4432324	&	J1448+44	&	Type 1	&	0.0795	$\pm$	0.00031	&	2.03	&	4.48	$\pm$	0.08	&	a	&	300	&	1.12	$\pm$	0.11	&	5.59	&	9.92 & 1.76x1.5	\\
J16373652+2543028	&	J1637+25	&	Type 2	&	0.2769	$\pm$	0.0001	&	2.33	&	0.16	$\pm$	0.02	&	b	&	900	&	-			&	6.50	&	5.20 & 1.6x1.5	\\
J21244163-1744458	&	J2124-17	&	Type 1	&	0.1110	$\pm$	0.00055	&	2.37	&	53.37	$\pm$	1.58	&	b	&	300	&	-			&	6.50	&	6.00 & 2x1.5	\\
\hline
\label{prop}
\end{tabular}
\end{center}
\end{table}
\end{landscape}

\section{Samples and emission line fitting}

\subsection{2MASS sample}

The basic properties of the 2MASS sample are given in Table \ref{prop}. To give an indication of the intrinsic powers of the AGN, the [OIII] emission line luminosities (L$_{[OIII]}$) were calculated using fluxes measured from the wide slit data for most objects; the narrow slit data were used where wide slit data was not available. For the purpose of calculating L$_{[OIII]}$ the [OIII] fluxes were corrected for Galactic reddening but not intrinsic reddening caused by dust in the host galaxies. L$_{[OIII]}$ was chosen because the [OIII]$\lambda$5007 emission line is regarded as a good indicator of AGN bolometric power (\citealt{reyes08}; \citealt{dicken}; \citealt{lamassa10}), and can be used to distinguish between quasars and Seyfert galaxies \citep{zakamska}. The redshifts of the objects were determined by fitting single Gaussian profiles to the [OIII]$\lambda$5007 emission lines. 

\begin{figure}
\centering
\includegraphics[scale=0.38, trim=0cm 4cm 0cm 0cm]{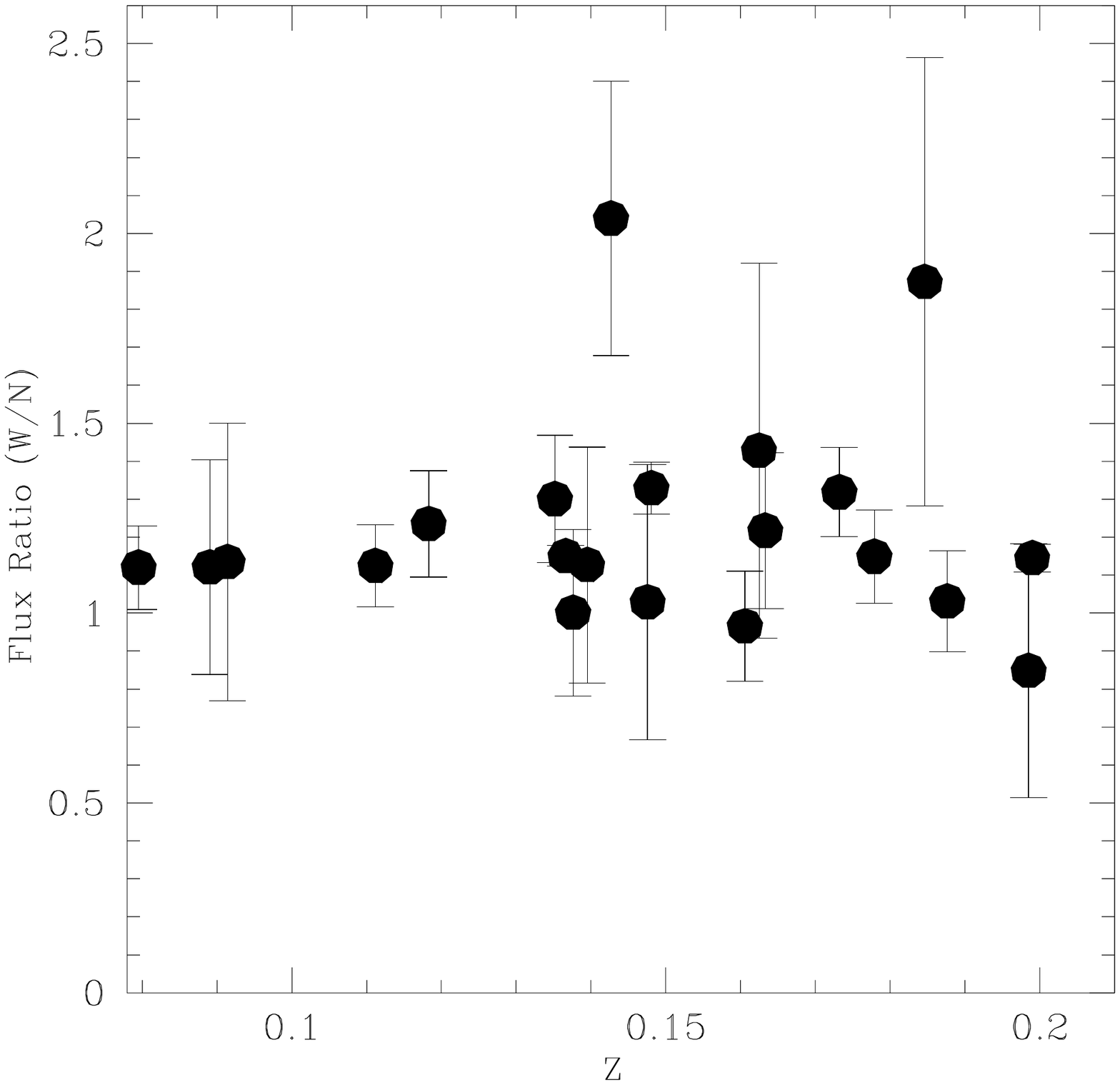}
\caption{Wide to narrow [OIII] flux ratio plotted against redshift for the 2MASS sample.}
\label{w1}
\end{figure}

The wide-to-narrow slit [OIII] flux ratios are shown in Figure \ref{w1}. The majority of objects have wide and narrow slit fluxes that agree within $\pm$40\%, suggesting that the narrower slits captured the majority of the flux from the emission regions: the median wide to narrow flux ratio is 1.14, showing that the wide slit captured slightly more flux than the narrow slit, as expected because the wide slit data probes a larger physical scale (4.9 to 22.7 kpc for the redshift range of this data) when compared to the the narrow slit data (1.5 to 6.8 kpc). There are two outliers: J0435-06 and J1001+41, which have ratios above 1.5. However, the wide to narrow flux ratios for these objects agree within $\sim$2$\sigma$ of the median ratio. In these cases the [OIII]$\lambda$5007 emission lines have a low equivalent width. Therefore the large wide to narrow flux ratios in these objects may reflect the large uncertainties in [OIII] fluxes, as measured from the noisy wide slit data, rather than uncertainties in the narrow slit fluxes. The similarity between the flux captured in the wide and narrow slit observations gives confidence that the spectra for the 5 objects without wide slit data (runs \lq b' and \lq c' in Table \ref{prop}) give an accurate indication (within $\sim$40\%) of L$_{[OIII]}$ for these objects.

\subsection{Comparison Samples}

In attempting to understand the nature of the red 2MASS AGN, some of the
key questions we are trying to address are: if one takes a sample of 
AGN selected on the basis of their red colours at near-IR wavelengths
(J-K$>$2.0), how do the optical spectroscopic properties of such a sample
differ from those of samples of typical type 1 AGN selected using their optical/UV colours, and can any differences between the samples be explained
in terms of reddening effects? Therefore, we have compared our results with two comparison samples of $\lq$typical' quasars. For any comparisons we make, the samples we use are limited to redshifts z$<$0.2, so they are consistent with our 2MASS AGN sample.

\subsubsection{PG Quasars}

One of the comparison samples used throughout this paper is the PG quasar sample of \citet{boroson}. The PG quasars comprise a complete sample of 87 UV-selected objects from the Palomar bright quasar survey \citep{sg}, selected to have redshifts z $<$ 0.5, and UV/optical colours U-B $<$ -0.44 to ensure an excess of flux in the UV. This sample is also selected to have quasar-like luminosities, with a total brightness of M$_B$ $<$ -23 for the AGN and host galaxy together, and broad permitted emission lines clearly present in the optical spectra \citep{boroson}. The PG quasars provide a useful comparison with the 2MASS sample because they represent \lq typical' blue selected quasars at their redshifts (z $<$ 0.5). By limiting the redshift range of the sample, there are 61 PG quasars suitable for comparison with the red 2MASS sample. Spectra of the \citet{boroson} sample were kindly supplied by T. Boroson, and remeasured by MR. The data cover a spectral range of $\lambda\lambda$4300--5700 and have spectral resolutions in the range 6.5--7.0 \AA. Detailed properties of the sample and the FeII subtraction procedure are outlined in \citet{boroson}.

\begin{figure}
\centering
\includegraphics[scale=0.38, trim=0cm 4cm 0cm 0cm]{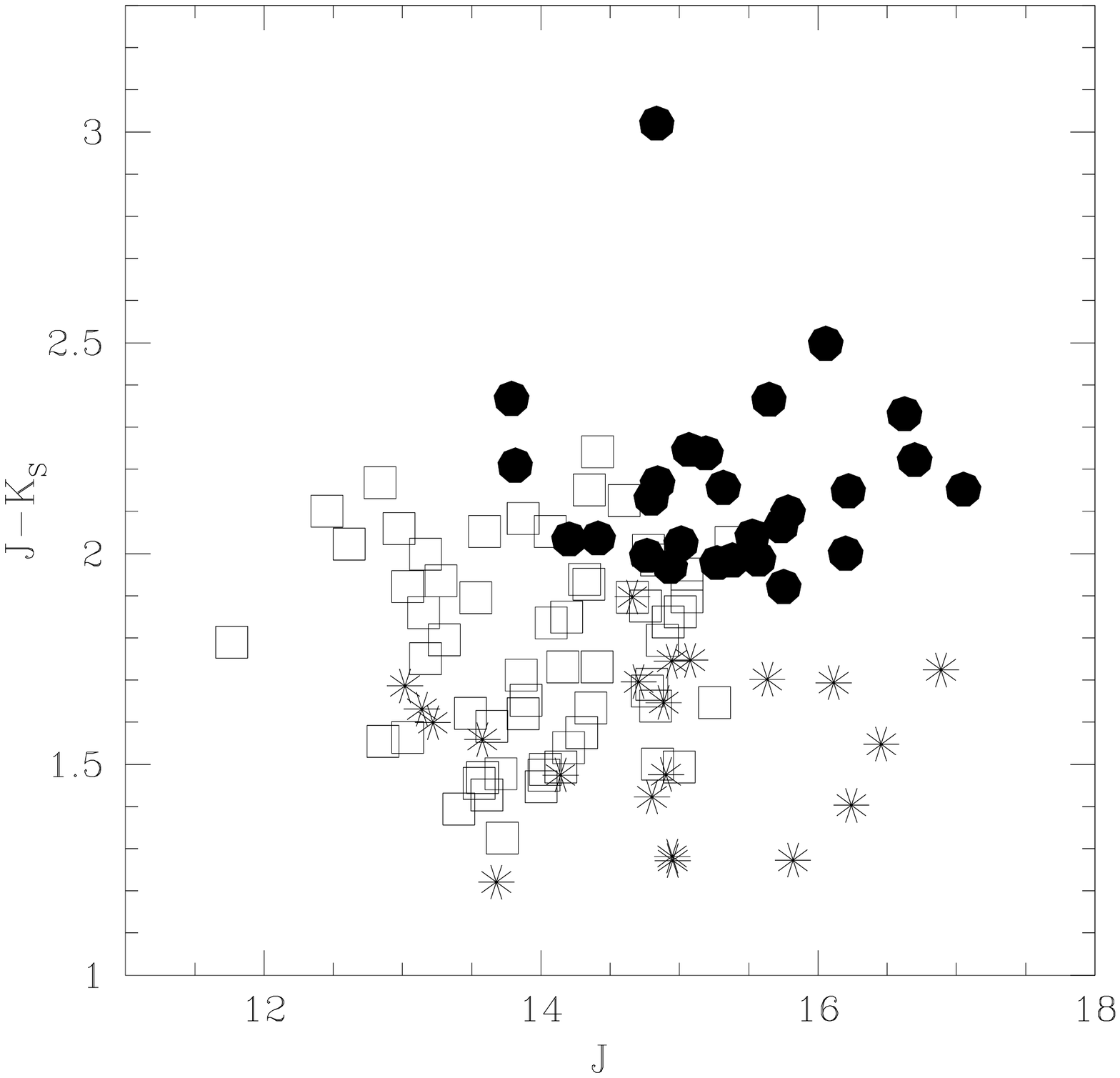}
\caption{J-K$_S$ versus J for the 2MASS and comparison samples. The 2MASS population measurements are indicated by the filled circles, the PG quasar population \citep{boroson} are indicated by the unfilled squares and the unobscured type 1 AGN \citep{jinall} are indicated by asterisks. Higher values of J-K$_S$ indicate redder colours.}
\label{colour}
\end{figure}

The spectra of the PG quasars were fitted in a similar manner to the 2MASS objects (see $\oint$ 3.3). However, in contrast to the 2MASS sample, only H$\beta$ and [OIII]$\lambda\lambda$5007,4959 were fitted for the PG quasars, due to the limited spectral range of their spectra \citep{boroson}.

A sub-sample of these quasars (17) have high quality Sloan Digital Sky Survey (SDSS; \citealt{york00}) spectra available, which cover a wider wavelength range than the spectra from the \citet{boroson} study. We will refer to this as the PG/SDSS sample. Where possible, these spectra have been used as part of the comparison with the 2MASS sample.

As shown in Figure \ref{colour}, which plots the J magnitude against the J-K$_S$ colours, the PG quasars are comparatively $\lq$bluer' than the 2MASS sample at near-IR wavelengths (PG median: J-K$_S$=1.79$\pm$0.03; 2MASS median: J-K$_S$=2.15$\pm$0.08), with minimal overlap in J-K$_S$ colours between the samples: only 20$\%$ of the PG sample have J-K$_S$ $>$ 2.0. In addition, the 2MASS objects are fainter in the J band. Given that the redshifts of the 2MASS objects in our sample are lower on average than those of the PG quasars, this suggests that the 2MASS objects have lower luminosities and therefore we may not be able to consider them as true quasars (see $\oint$ 4.1.1 for a full discussion).

\subsubsection{Unobscured Type 1 AGN}

The second comparison sample comprises a sample of 51 relatively nearby (z $<$ 0.38), unobscured type 1 AGN selected by \citet{jinall} on the basis of low optical reddening, and low gas columns as indicated by their X-ray spectra. The optical properties of this sample were obtained from high quality SDSS DR7 spectra \citep{jinall}. The SDSS DR7 spectra cover a wavelength range of 3800-9200\AA\ and have spectral resolution of $\sim$3\AA. The analysis techniques are outlined in the work of \citet{jinall}. There are 14 PG quasars in this sample, therefore we have removed these from any comparison which includes the PG quasar sample to avoid overlap. In addition, when considering the remaining objects with redshifts z$<$0.2, there are 21 unobscured type 1 AGN suitable for comparison. As these objects comprise unobscured AGN, we consider this a suitable sample to compare to the red 2MASS quasars.

The spectral fitting technique outlined in \citet{jina} is similar to the methods we have used to fit the emission lines of our 2MASS sample (see $\oint$ 3.3). Therefore we are confident that we can use the data presented in \citet{jina} for a direct comparison with our results. 

As shown in Figure \ref{colour}, like the PG quasars, the unobscured type 1 AGN are also relatively $\lq$blue' compared with the 2MASS sample at near-IR wavelengths (median J-K$_S$=1.60$\pm$0.04), with no overlap in J-K$_S$ colours between the samples. Unlike the PG quasars, this sample has comparable J-band magnitudes to the 2MASS sample. This suggests that, given the redshifts of the objects in the sample (median z$\sim$0.16; for the entire sample), this sample has similar luminosities to the 2MASS objects. 

\subsection{Continuum Subtraction and Emission Line Modelling}

\begin{figure*}
\centering
\includegraphics[scale=0.85, trim=2cm 12cm 2cm 5cm,]{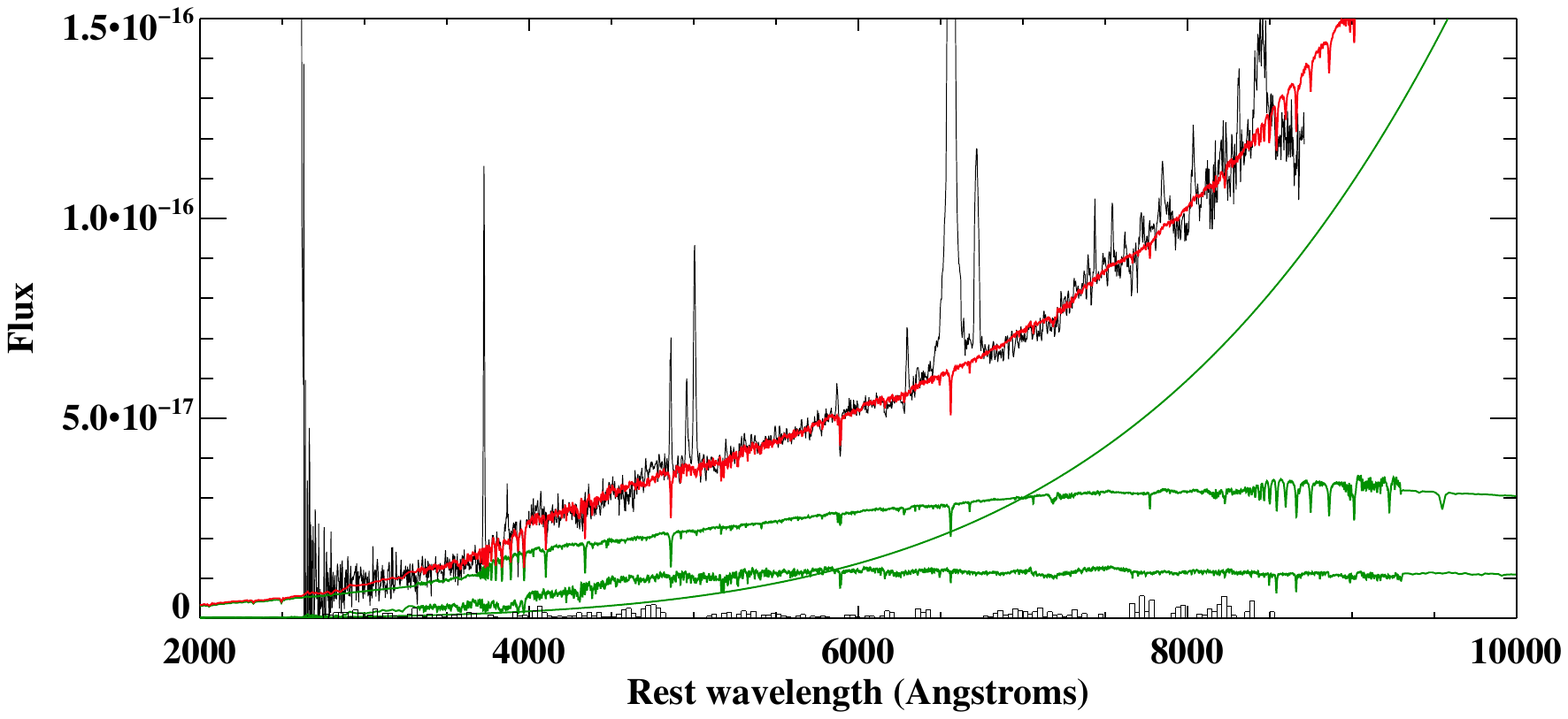}
\subfloat[Balmer features]{\includegraphics[scale=0.45, trim=0cm 12cm 0cm 9cm]{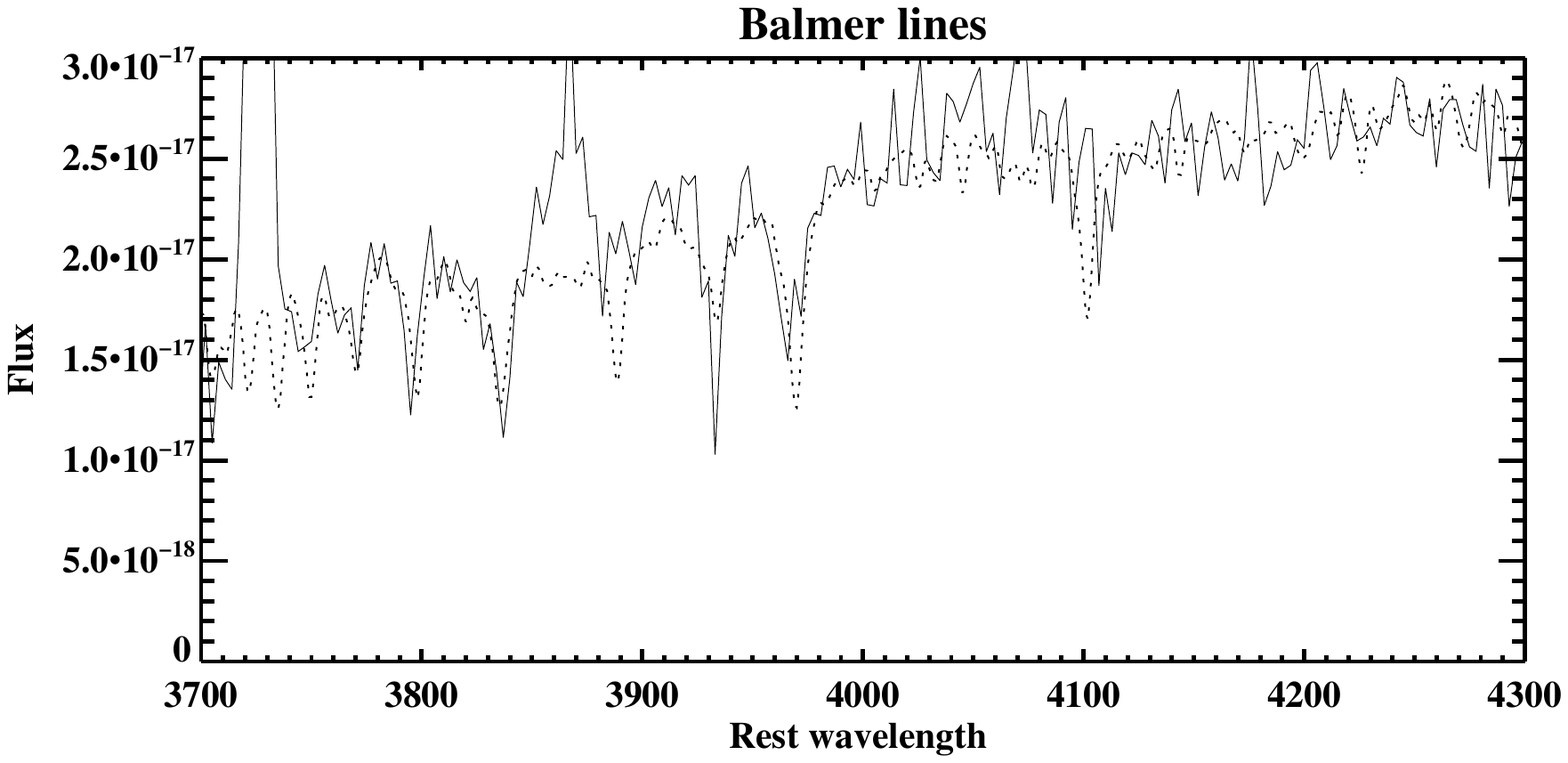}}
\subfloat[Na D feature]{\includegraphics[scale=0.45, trim=0cm 12cm 0cm 9cm]{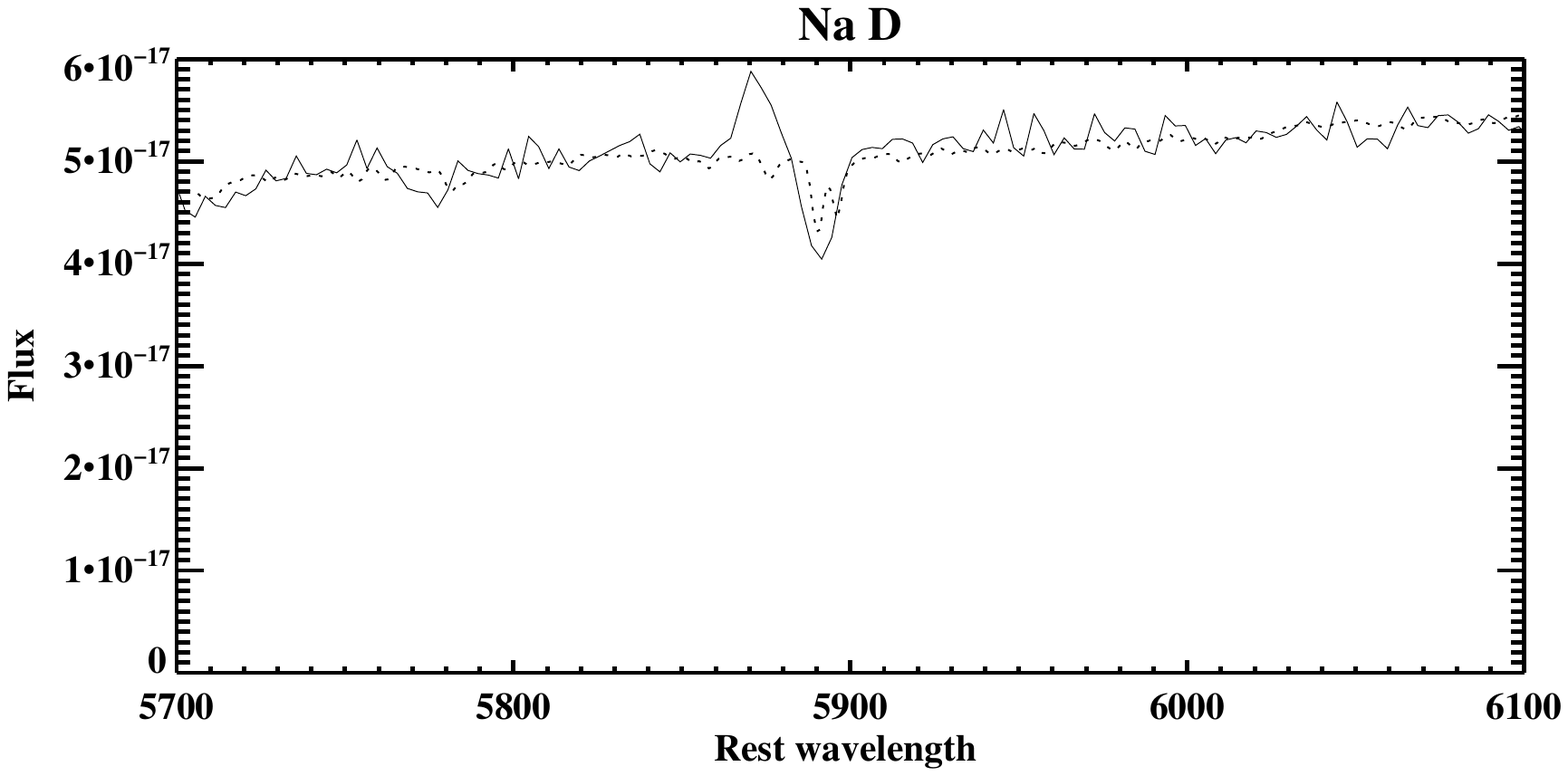}}\\
\subfloat[G band]{\includegraphics[scale=0.45, trim=0cm 12cm 0cm 8cm]{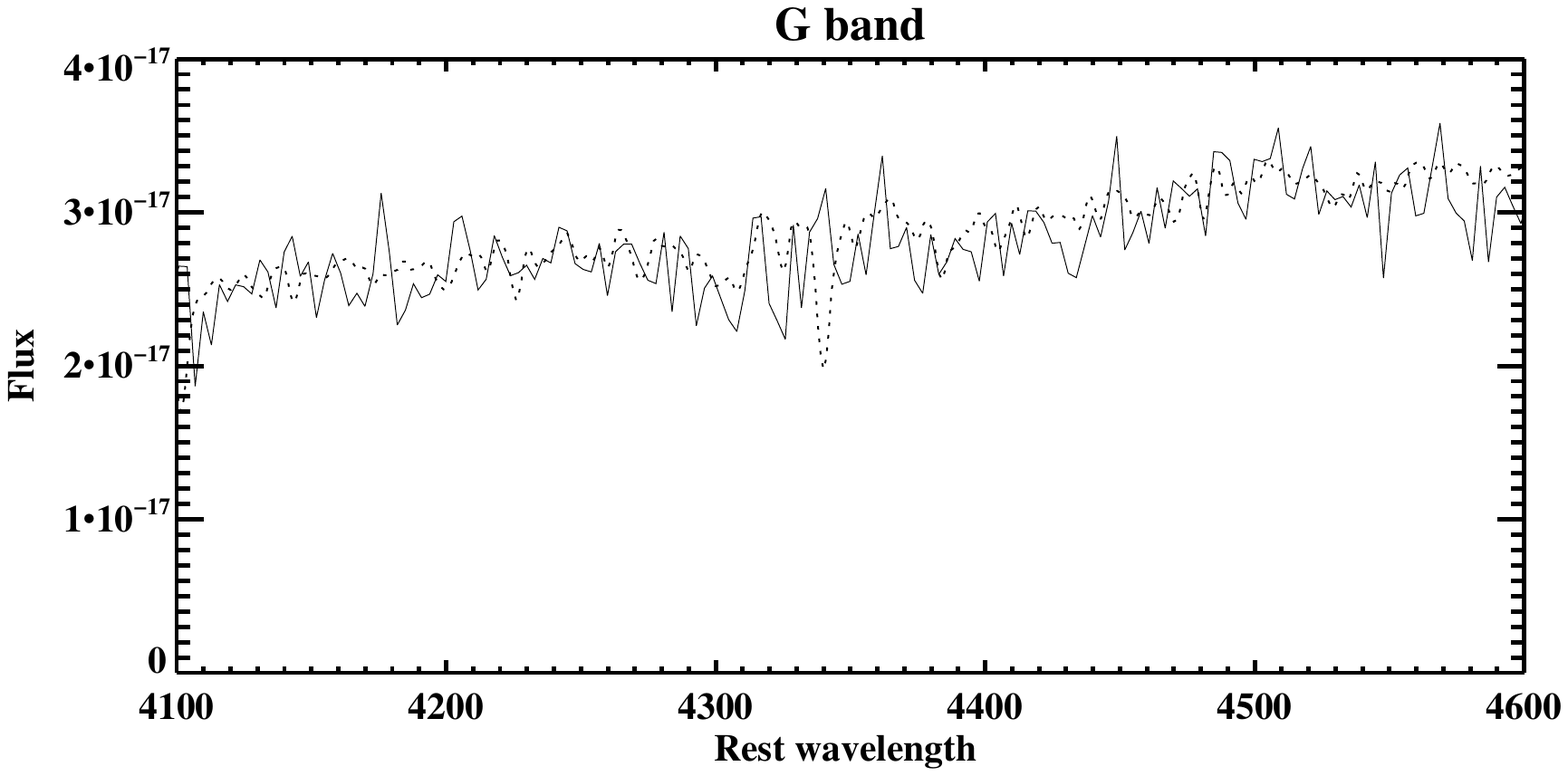}}
\subfloat[Mg Ib band]{\includegraphics[scale=0.45, trim=0cm 12cm 0cm 8cm]{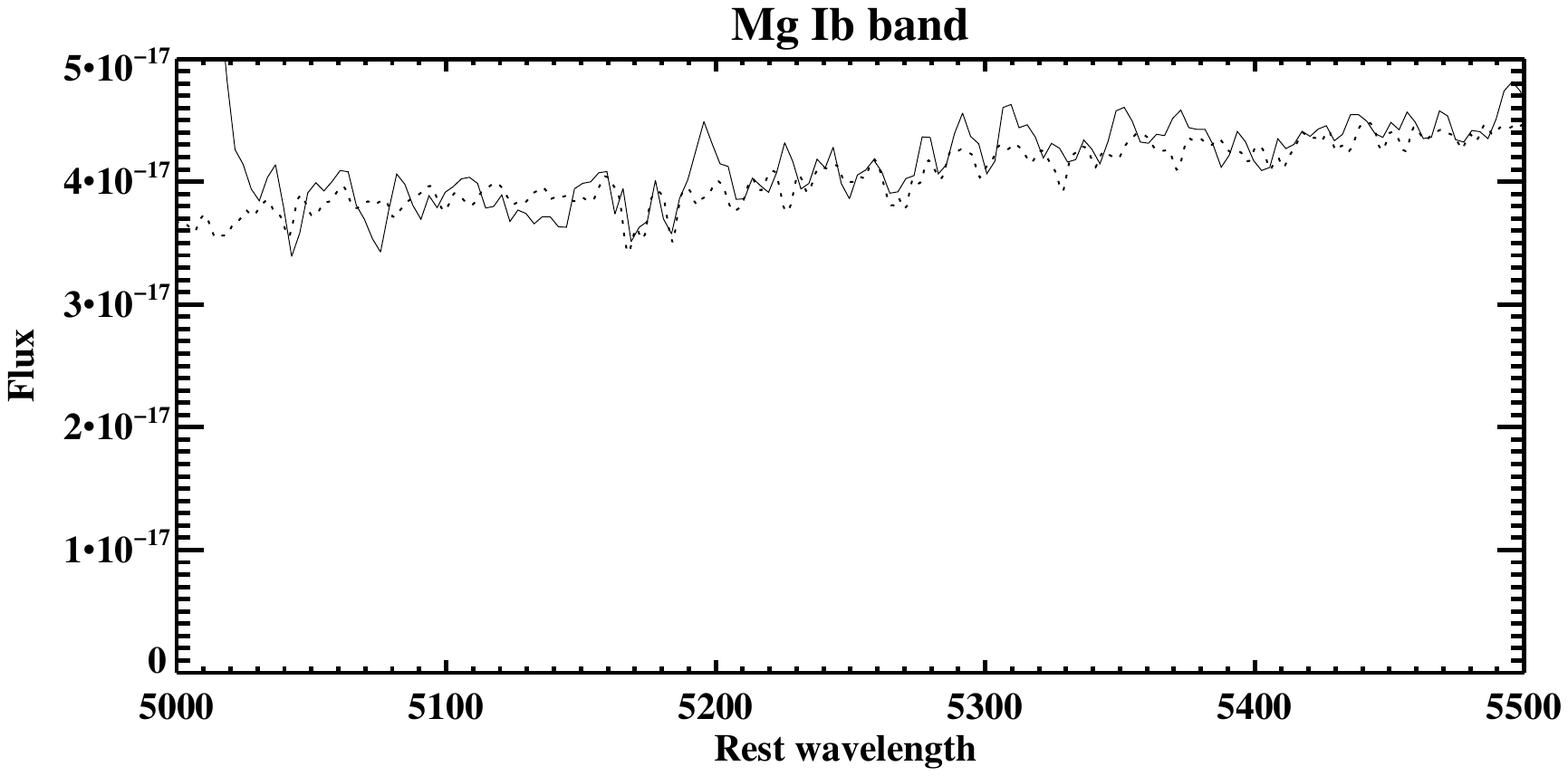}}
\caption{A best-fit {\sc confit} model for J10404364+5934092 ($\chi$$^2$=0.54), plots (a)-(d) show how the combined stellar populations fit the detailed features. The object requires an OSP, a YSP, and a power-law component for an adequate fit. The YSP has an age of 0.005 Gyr, an extinction of E(B-V)=1.0 and represents 58.7$\%$ of the flux in the normalising bin. The power-law component has $\alpha$=5.11 and represents 16.4$\%$ of the flux in the normalising bin. Finally, the OSP represents 24.8$\%$ of the flux in the normalising bin. The flux is in units of erg s$^{-1}$ cm$^{-2}$ \AA$^{-1}$, and wavelength is in units of \AA.}
\label{pow_comb}
\end{figure*}

Fitting emission features in AGN spectra is often complicated by the presence of the host galaxy continuum in the spectrum. Subtle features such as weak broad components can often be overlooked, and the presence of underlying Balmer absorption from the stellar continuum can lead to underestimation of the intensity of the Balmer and other emission line strengths (e.g. \citealt{sarzi06}).

For the purposes of studying the emission regions, the underlying continuum was modelled and subtracted in all objects in which the stellar continuum is significant --- mainly LINER/Seyfert 2 type objects (e.g. J0221+13) and objects with weak broad lines (e.g. J0422-18). These steps required the use of the {\sc starlink} package {\sc dipso} \citep{howarth04} and a customised {\sc idl} minimum $\chi$$^2$ fitting program called {\sc confit} (\citealt{robinson}; \citealt{tadhunter}; \citealt{holt}; \citealt{javier09}).

We started by experimenting with subtracting a nebular continuum component, which might be significant at near-UV wavelengths \citep{dickson95}. The contribution of the nebular continuum was estimated using the narrow H$\beta$ emission line flux. However, its strength is sensitive to the reddening of the NLR, which is not accurately known in most cases. Indeed, we found that in all cases we obtained better fits if we did not subtract a nebular continuum. Therefore the results below refer to models without nebular continuum subtraction. 

\begin{center}
\begin{table}
\caption{The [OIII] model parameters used when fitting the emission lines of the spectra. The second and third columns present the rest-frame widths of the narrow and broad kinematic components respectively. The final column gives the rest-frame velocity separation of the broad relative to the narrow kinematic components.}
\begin{tabular}{lcccccc}
\label{models} \\
\hline
Object	&	FWHM$_N$			&	FWHM$_B$	&	Separation		\\
	&	 (km s$^{-1}$)		&	 (km s$^{-1}$)	& (km s$^{-1}$)		 \\
\hline
J0221+13	&	Inst.			&	799	$\pm$	36	&	-217	$\pm$	27	\\
J0248+14	&	448	$\pm$	39	&	945	$\pm$	61	&	-168	$\pm$	44	\\
J0306-05	&	Inst.			&	826	$\pm$	148	&	-426	$\pm$	152	\\
J0312+07	&	Inst.			&	718	$\pm$	95	&	-227	$\pm$	72	\\
J0400+05	&	326	$\pm$	13	&	1197	$\pm$	51	&	-36	$\pm$	11	\\
J0409+07	&	620	$\pm$	22	&	1842	$\pm$	340	&	-44	$\pm$	68	\\
J0411-011$^a$	&	544	$\pm$	150	&	-			&	-			\\
J0422-18$^b$	&	Inst.			&	1700	$\pm$	56	&	-298	$\pm$	31	\\
J0435-06	&	256	$\pm$	19	&	376	$\pm$	84	&	-655	$\pm$	93	\\
J0447-16	&	194	$\pm$	7	&	1072	$\pm$	165	&	-569	$\pm$	187	\\
J0504-19$^b$	&	Inst.			&	1778	$\pm$	40	&	191	$\pm$	32	\\
J0910+33	&	715	$\pm$	21	&	1652	$\pm$	48	&	-276	$\pm$	33	\\
J1001+41$^a$	&	418	$\pm$	76	&	-			&	-			\\
J1006+41$^b$	&	Inst.			&	554	$\pm$	159	&	-353	$\pm$	152	\\
J1014+19	&	428	$\pm$	13	&	1637	$\pm$	63	&	127	$\pm$	23	\\
J1040+59	&	181	$\pm$	22	&	993	$\pm$	84	&	214	$\pm$	62	\\
J1057-13	&	707	$\pm$	14	&	1477	$\pm$	139	&	-855	$\pm$	116	\\
J1127+24	&	267	$\pm$	7	&	-			&	-			\\
J1131+16$^c$	&	Inst.			&	720	$\pm$	31	&	-92	$\pm$	5	\\
J1158-30	&	693	$\pm$	27	&	1177	$\pm$	139	&	-1195	$\pm$	66	\\
J1212-14$^a$	&	455	$\pm$	25	&	-			&	-			\\
J1321+13$^a$	&	390	$\pm$	14	&	-			&	-			\\
J1323-02	&	366	$\pm$	16	&	1512	$\pm$	108	&	-488	$\pm$	94	\\
J1338-04$^b$	&	Inst.			&	780	$\pm$	24	&	-224	$\pm$	34	\\
J1407+42	&	584	$\pm$	8	&	1316	$\pm$	48	&	190	$\pm$	28	\\
J1448+44$^a$	&	441	$\pm$	15	&	-			&	-			\\
J2124-17$^a$	&	753	$\pm$	22	&	-			&	-			\\
\hline
\end{tabular}
\begin{tablenotes}
       \item[a]$^a$ The model for this object requires just one component.
       \item[b]$^b$ The narrow component of the model can be modelled with the instrumental width.
       \item[c]$^c$ This object was studied in detail in \citet{rose}.
     \end{tablenotes}
\end{table}
\end{center}

Once the spectra had been prepared, the continua were modelled using the customised {\sc idl} procedure {\sc confit} (\citealt{robinson}; \citealt{holt}). To minimise degeneracy when fitting our spectra, we fit the smallest number of stellar and/or power-law components to the continuum as possible. For each object we ran 3 sets of models. In the first run we fit an unreddened old stellar population (OSP, 12.5 Gyr) and a young stellar population (YSP), with ages in the range 0.001--5 Gyr, and reddening: 0 $<$ E(B-V) $<$ 2.0\footnote{We used the \citet{calzetti} reddening law to redden the YSP spectra.}. The inclusion of a YSP is justified on the basis that many of the objects with significant stellar continua show higher order Balmer lines in absorption (e.g. J1001+41). The stellar population templates were taken from the instantaneous burst spectral synthesis results of \citet{bruzal} for a Salpeter IMF and solar metallicities. For the second run a power-law component, which represents either the direct or scattered AGN component, was also included in the model fitting. Finally, an OSP with and without a power-law was fitted to each spectrum, independent of a YSP.

\begin{center}
\begin{table*}
\caption{Best fit {\sc confit} model results for the 2MASS sample. The simple stellar population models were taken from \citet{bruzal}. The subscripts `U' and `L' refer to the upper and lower cases of the model parameters that provide a good fit to the continuum. For the cases where the best fits include power laws we use the form F$_{\nu}$ $\propto$ $\nu$$^{+\alpha}$, where $\alpha$ represents the power law index.}
\begin{tabular}{lcccccc}
\label{conftab} \\

\hline
Object & YSP$_{U}$ (Gyr) & YSP$_{L}$ (Gyr) & E(B-V)$_{U}$ & E(B-V)$_{L}$ &  $\alpha$$_{U}$ & $\alpha$$_{L}$  \\ \hline																	
																	
J0221+13	&					1.0	&	0.7	&	0.5	&	0.3	&	2.18	&	0.457\\	
J0306-05	&					2.0	&	0.8	&	0.5	&	0.3	&	3.95	&1.40\\		
J0411-01	&					0.006	&	0.005	&	1.1	&	0.6	&	-	&	-\\	
J0422-18	&					0.005	&	0.004	&	1.1	&	0.9	&	2.35	&	0.32	\\
J0447-16	&					1.2	&	0.8	&	0.3	&	0.2	&	1.31	&	1.28	\\
J0504-19	&					0.01	&	0.006	&	0.9	&	0.7	&	-	&	-\\	
J0910+33	&					0.9	&	0.3	&	0.3	&	0.1	&	-0.347	&	-0.818	\\
J1001+41	&					0.2	&	0.1	&	0.7	&	0.6	&	-	&	-\\	
J1040+59	&					0.005	&	0.004	&	1.1	&	0.8	&	5.97	&	5.11\\	
J1057-13	&					2.0	&	0.2	&	0.6	&	0.0	&	-0.30	&	-1.33\\	
J1158-30	&					0.4	&	0.2	&	0.5	&	0.2	&	2.86	&	1.74\\	
J1212-14	&					1.2	&	0.8	&	0.4	&	0.2	&	1.07	&	0.627\\	
J1321+13	&					0.05	&	0.005	&	0.7	&	0.4	&	-	&	-\\	
J1407+42	&					0.06	&	0.04	&	0.6	&	0.5	&	-	&	-\\	
\hline

\end{tabular}
\end{table*}
\end{center}

The spectra were fitted using fluxes measured in 30\AA\ bins, chosen to have as even a distribution in wavelength as possible, and avoiding the brighter atmospheric emission and stronger atmospheric absorption, and AGN emission, features in the spectra. The entire useful wavelength ranges of the spectra were fitted: observed wavelengths 3200--9000\AA. In addition, a normalising bin was chosen which is also free of any emission/absorption features ($\sim$5050--5200 \AA\ for all objects in the rest frame).

Fits to the overall continuum shape were chosen with low reduced $\chi$$^2$ $<$ 1 (see discussion in \citealt{tadhunter}). Then a visual inspection of fits to individual absorption lines, including CaII K (3934\AA), MgI (5175 \AA), Balmer and NaI D (5890-5896 \AA), was used to select the best overall fits. Note that not all objects were successfully fitted. For example, we could not produce an adequate fit to the type 2 object J1131+16 because there are relatively few genuine continuum bins suitable to model the spectrum of this object. It should be noted that this object has an unusual spectrum, which contains a large number of forbidden high ionisation lines (FHILs; see \citealt{rose}). Figure \ref{pow_comb} shows an example of a successful {\sc confit} fit to a 2MASS object. Details of the models chosen for the continuum subtraction in individual objects are given in Table \ref{conftab}.  

Once the continuum had been subtracted from the spectra (where applicable), the emission lines were fitted using the {\sc starlink dipso} package \citep{howarth04}.

The emission lines in the spectra of the 2MASS and PG comparison sample AGN were initially fitted with single Gaussian profiles, starting with [OIII]$\lambda$$\lambda$5007,4959 doublet. We used the [OIII]$\lambda$$\lambda$5007,4959 emission lines because they are often the strongest unblended lines emitted by the NLR. To produce the models, Gaussian profiles were fitted to the [OIII] doublet, with the separation and intensity ratio set by atomic physics for each kinematic component (as defined by the velocity shifts and line widths). While single Gaussian fits were sufficient in some objects, such fits did not adequately model the wings of the emission lines of most of the objects; in such objects double Gaussian model were fitted to the spectral lines. When a second Gaussian was added, the line ratios, wavelength separations and FWHM were fixed for the two doublet components as before, however the shift and FWHM were allowed to vary relative to the first component, and the widths were allowed to vary. In all cases, one or two Gaussian components proved sufficient, and in all cases where two Gaussian profiles were required, the width of one Gaussian was significantly broader than the other. In some cases, the widths of the narrow components for the [OIII] emission lines were found to be consistent with the instrumental width of their spectra. In such cases, the narrow components of the models that were fitted to the other lines were fixed to the instrumental width. However, in the other (resolved) cases, the measured FWHM was used to obtain an intrinsic velocity width for the narrow component by correcting its FWHM in quadrature using the instrumental width. The same technique was applied to the broader components of the model. Also, the broad component is often shifted with respect to the narrow component; this is accounted for in the models. 

Once the model parameters had been determined from the fits to [OIII], they were used to fit the other lines in the spectra. The [OIII] model parameters for each object are presented in Table \ref{models}. Known blends (e.g. H$\alpha$ $\&$ [NII]$\lambda\lambda$6548,6583) and doublets (e.g. [SII]$\lambda\lambda$6716,6731 and [OII]$\lambda\lambda$3726,3729) were modelled with the [OIII] model using constraints provided by atomic physics (i.e. fixed FWHM, line separation and, where appropriate, the intensity ratios set by atomic physics). In most cases, the [OIII] model successfully fit all the emission lines/blends in the spectra. However, in a few cases where emission lines have a low equivalent width (e.g. weaker lines such as [OI] or [FeX]), one of the components in the 2 component [OIII] model provides a better fit to the emission line. This happened in some objects with strong broad Balmer emission lines.

\begin{figure}
\centering
\includegraphics[scale=0.28]{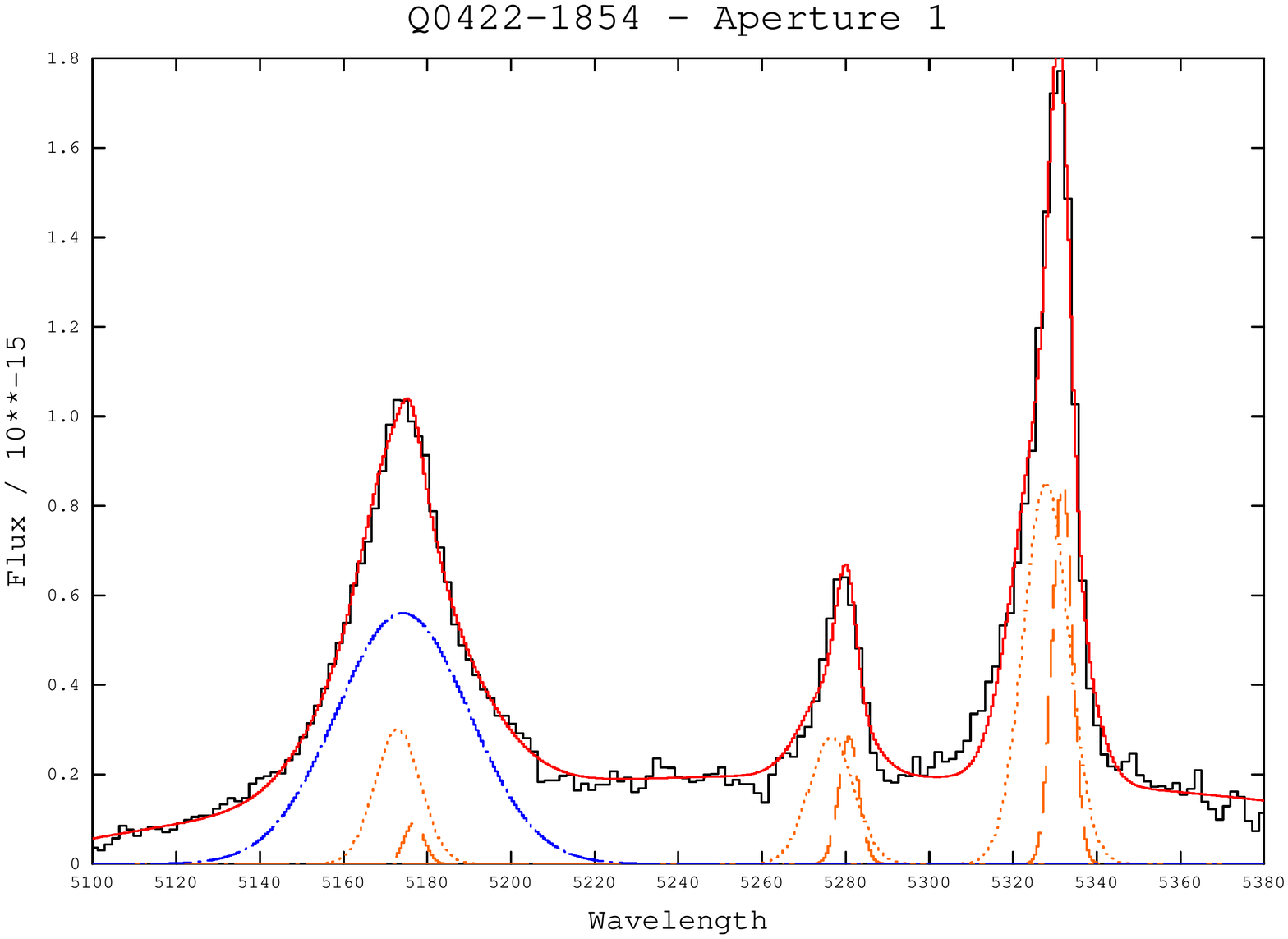}
\caption{Gaussian fits to the H$\beta$ and [OIII]$\lambda\lambda$5007,4959 emission lines of J0422-18. The overall fit is represented by a red line.  Both the NLR [OIII] model (orange lines; the narrow component is represented by a dashed line and the broad component is presented by a dotted line) and BLR components (represented by a blue dashed and dotted line) are displayed. The NLR model fit to the H$\beta$ blend highlights the degeneracy issues with blended emission from the BLR and the NLR (this is described in more detail in $\oint$ 3.3).}
\label{blfit}
\end{figure}

The broad permitted lines emitted by many of the objects (mainly H$\alpha$ and H$\beta$) were initially fitted with a single Gaussian component, where the intrinsic velocity widths (FWHM) were kept the same for the all BLR emission lines in each object. The NLR emission components of the Balmer emission were still fitted with the [OIII] model. However, for 11 objects in the 2MASS sample, the broad and narrow line components of the Balmer lines could not be confidently separated because, at the resolution of our spectra, they merge seamlessly together, leading to degeneracies in fits that attempt to model the lines as single narrow and broad Gaussians (see Figure \ref{blfit}). In addition, for the PG/SDSS sample, we could only confidently separate the broad and narrow emission in 6 objects. Figure \ref{blfit} highlights the problem of degeneracy in the blending of the NLR with BLR in the Balmer lines, using the example of J0422-18. Although the overall fit to the H$\beta$ feature is adequate, the narrow line Gaussian fitted to the H$\beta$ emission line may be over- or underestimating the flux of the NLR H$\beta$ emission. In such cases, we only consider the total fluxes of the combined broad and narrow emission components of the Balmer lines.

The H$\beta$ and [OIII]$\lambda$$\lambda$5007,4959 may also be strongly blended with broad permitted FeII emission. In cases where broad permitted FeII emission is prominent in the spectra (e.g. J0248+14), the FeII emission was fitted separately from the affected emission lines as follows: the FeII emission lines in the same multiplets (F, S or G) were fitted with the same intrinsic velocity widths as the broad H$\beta$ emission line, and their expected intensity ratios were used to constrain the fits, as outlined in \citet{kova}. This approach provided a good fit to the FeII emission blends in all objects where they are significant.

\begin{figure}
\centering
\includegraphics[scale=0.28]{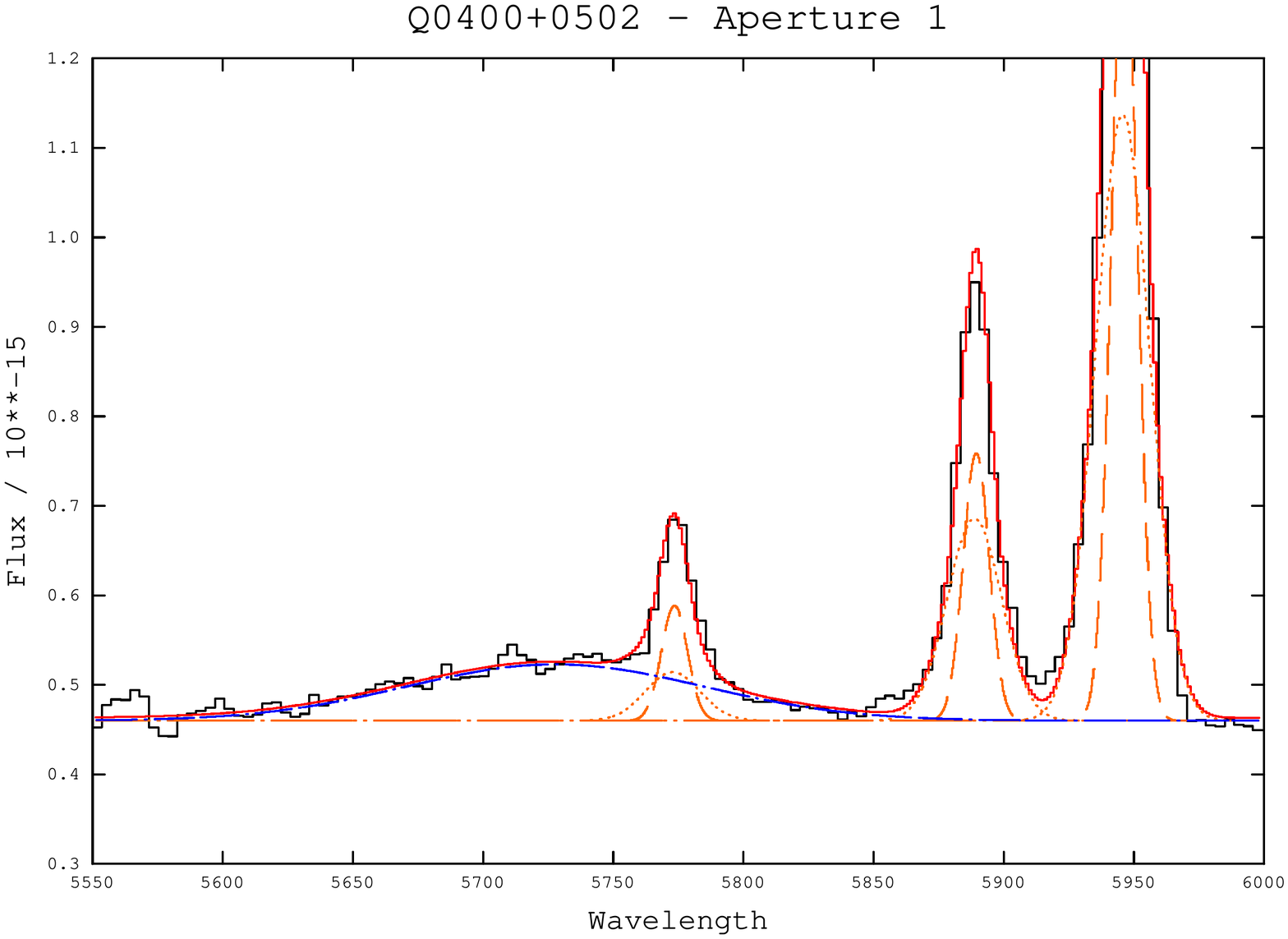}
\caption{Gaussian fits to the H$\beta$ and [OIII]$\lambda\lambda$5007,4959 emission lines of J0400+05. Both the NLR [OIII] model and BLR components are displayed. In this case the BLR and NLR components of H$\beta$ could be clearly separated. Interestingly, the BLR is substantially blueshifted relative to the NLR in this object (by -2870$\pm$160 km s$^{-1}$). The properties of the individual objects in the 2MASS sample with particularly interesting spectra, are discussed in the Appendix.}
\label{modfit}
\end{figure}

\begin{figure}
\centering
\includegraphics[scale=0.28]{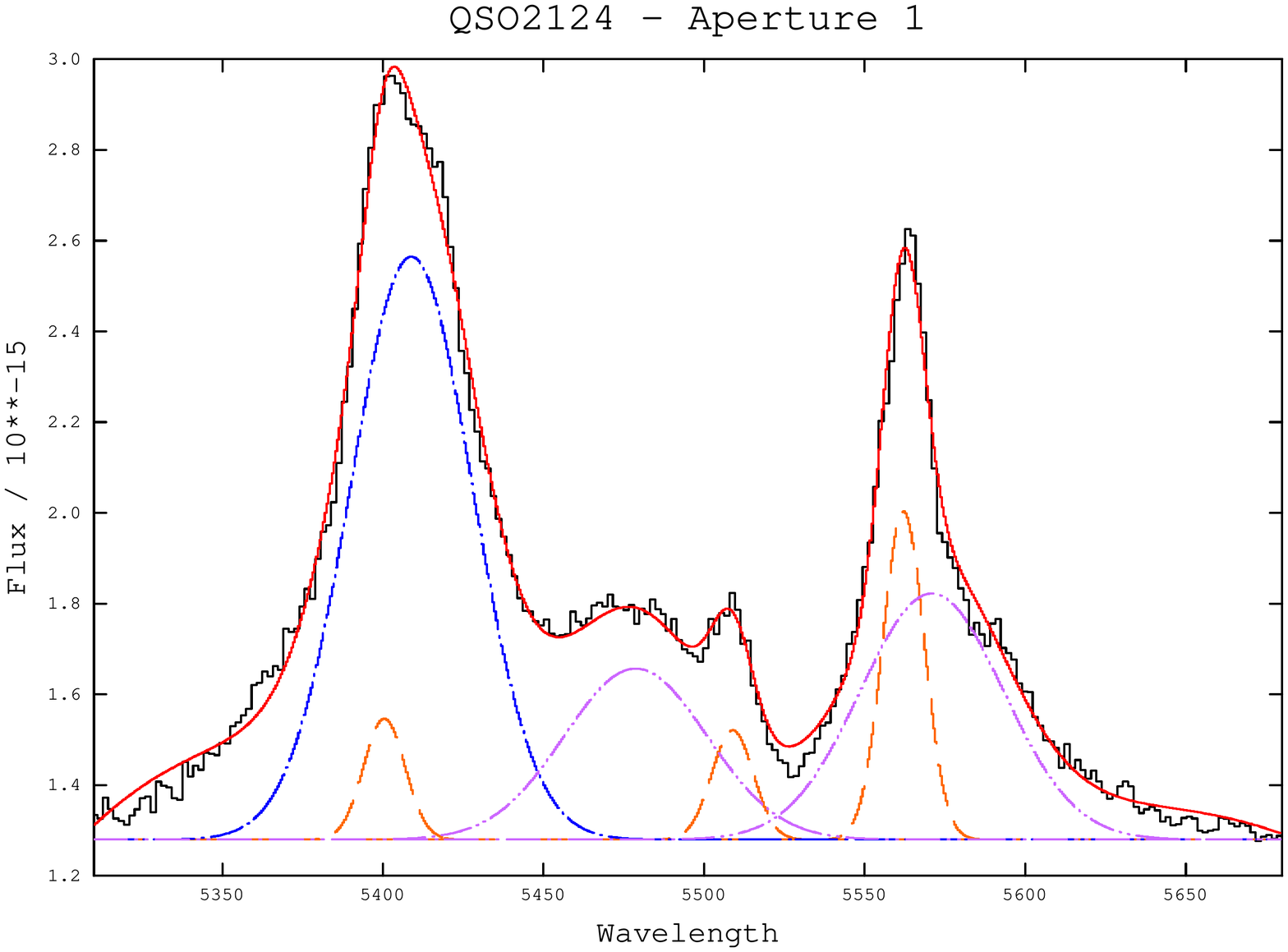}
\caption{Gaussian fits to the H$\beta$ and [OIII]$\lambda\lambda$5007,4959 emission lines of J2121-17. The NLR [OIII] model, BLR component and FeII emission (represented by a violet dashed and dotted line) are displayed. The overall fit is far better than when the emission lines are modelled without the FeII emission.}
\label{feiifit}
\end{figure}

Examples of model fits are given in Figures \ref{modfit} and \ref{feiifit}. Figure \ref{modfit} shows an example of a multi-component Gaussian fit to the H$\beta$ and [OIII]$\lambda$$\lambda$5007,4959, where the broad and narrow components of H$\beta$ were easy to separate. In contrast, Figure \ref{feiifit} shows a case where the broad and narrow emission could not be easily separated, and where FeII emission is prominent. Interestingly, the BLR of this object is substantially blueshifted relative to the NLR in this object (by -2870$\pm$160 km s$^{-1}$). The properties of the individual objects in the 2MASS sample with particularly interesting spectra are discussed in the Appendix.  

The results for the NLR [OIII] model fits to the other lines are presented in Table \ref{lines}; all the emission line ratios are presented relative to [OIII]$\lambda$5007, and the narrow and broad line region Balmer components are presented separately.

\begin{landscape}
\begin{center}
\begin{table}
\caption{Emission line fluxes of the key emission lines in the 2MASS sample, given as a ratio to the flux of the [OIII]$\lambda$5007 emission line (in units of erg s$^{-1}$ cm$^{-2}$, presented in column 2). The broad and narrow line region emission is indicated separately for the H$\alpha$ and H$\beta$ emission lines. In general, the uncertainties associated with the emission line fluxes do not exceed $\pm$10\%.} 
\begin{tabular}{l c c c c c c c c c c c c c c c c}
\hline
Object & F$_{[OIII]}$ & [NeV] & [OII] & [NeIII] & H$\delta$ & H$\gamma$ & [OIII] & H$\beta$$_{NLR}$ & H$\beta$$_{BLR}$ & [FeVII] & [OI] & [FeX] & H$\alpha$$_{NLR}$ & H$\alpha$$_{BLR}$ & [NII] & [SII] \\
 &  & 3425 & 3727 & 3868 &  &  & 4363 &  &  & 6086 & 6300 & 6374 &  &  & 6583 & 6716\&6731 \\
 \hline

0221+13	&	1.8E-15	&	0.226	&	0.264	&	3.971	&	-	&	-	&	-	&	0.203	& - &	0.128	&	0.030	&	-	&	0.610	& - &	1.139	&	0.420	\\
0248+14	&	2.0E-15	&	0.208	&	0.446	&	0.323	&	0.495	&	0.789	&	0.179	&	2.190	& 1.052 &	-	&	0.106	&	-	&	17.841	& 5.535 &	4.395	&	0.476	\\
0306-05	&	2.1E-15	&	-	&	0.285	&	-	&	-	&	-	&	-	&	0.761	& - &	-	&	0.163	&	-	&	6.751	& - &	3.578	&	0.950	\\
0312+07	&	7.6E-16	&	0.023	&	0.065	&	0.024	&	-	&	0.111	&	-	&	0.818	& 2.721 &	0.038	&	0.082	&	-	&	3.586	& 14.832 &	1.442	&	1.041	\\
0400+05	&	3.0E-14	&	0.027	&	0.151	&	0.059	&	0.013	&	0.046	&	-	&	0.095	& 0.391 &	-	&	0.025	&	0.061	&	0.668	& 3.157 &	0.420	&	0.259	\\
0409+08	&	2.1E-14	&	0.052	&	0.125	&	0.070	&	0.009	&	0.026	&	0.015	&	0.185	& 0.675 &	0.024	&	0.031	&	0.017	&	0.926	& 1.124 &	0.173	&	0.138	\\
0411-01	&	1.5E-16	&	-	&	1.044	&	-	&	-	&	0.487	&	-	&	1.639	& - &	-	&	0.139	&	-	&	7.031	& - &	3.224	&	1.578	\\
0422-18	&	2.8E-14	&	0.049	&	0.056	&	0.075	&	0.070	&	0.181	&	0.108	&	0.854	& 0.522 &	-	&	0.010	&	-	&	0.464	& 5.357 &	0.916	&	0.069	\\
0435-06	&	2.3E-15	&	-	&	0.338	&	-	&	0.362	&	1.059	&	-	&	2.420	& 3.752 &	-	&	0.035	&	-	&	12.498	& 8.809 &	0.970	&	0.496	\\
0447-16	&	1.2E-15	&	-	&	0.504	&	-	&	-	&	-	&	-	&	0.169	& 1.224 &	0.072	&	0.150	&	-	&	0.508	& 11.667 &	3.362	&	0.544	\\
0504-19	&	8.4E-15	&	0.166	&	0.191	&	0.219	&	-	&	0.088	&	0.105	&	0.179	& - &	-	&	0.038	&	-	&	0.686	& - &	0.736	&	0.275	\\
0910+33	&	1.3E-14	&	0.132	&	0.112	&	0.173	&	0.022	&	0.030	&	0.036	&	0.098	& - &	0.022	&	0.040	&	0.016	&	0.211	& - &	0.367	&	0.121	\\
1001+41	&	2.4E-16	&	-	&	2.532	&	4.897	&	-	&	0.812	&	0.128	&	2.102	& - &	-	&	1.204	&	-	&	16.816	& - &	9.636	&	5.907	\\
1006+41$^a$	&	1.2E-14	&	0.114	&	0.158	&	0.266	&	0.045	&	0.161	&	0.055	&	0.520	& - &	-	&	0.037	&	0.014	&	1.707	& 1.393 &	0.315	&	0.171	\\
1014+19	&	3.6E-14	&	0.045	&	0.076	&	-	&	0.011	&	0.020	&	0.047	&	0.385	& 0.517 &	0.008	&	0.035	&	0.046	&	2.141	& 2.881 &	0.303	&	0.128	\\
1040+59$^a$	&	2.1E-15	&	-	&	1.073	&	0.157	&	-	&	0.083	&	-	&	0.516	& - &	-	&	0.248	&	-	&	1.118	& 5.143 &	0.434	&	1.573	\\
1057-13	&	8.9E-15	&	0.123	&	0.078	&	0.071	&	0.021	&	0.035	&	0.014	&	0.052	& - &	0.081	&	0.055	&	-	&	0.368	& - &	0.105	&	0.105	\\
1127+24	&	5.1E-15	&	0.084	&	0.147	&	0.132	&	0.057	&	0.094	&	-	&	0.264	& 0.909 &	0.079	&	0.077	&	-	&	0.857	& 4.886 &	0.376	&	0.203	\\
1131+16	&	2.3E-14	&	0.192	&	0.138	&	0.209	&	0.046	&	0.083	&	0.171	&	0.190	& - &	0.094	&	0.037	&	0.044	&	0.948	& - &	0.149	&	0.103	\\
1158-30	&	9.2E-15	&	-	&	0.345	&	0.118	&	-	&	0.106	&	-	&	0.121	& - &	-	&	0.162	&	-	&	1.328	& - &	0.658	&	0.365	\\
1212-14	&	4.5E-15	&	0.104	&	0.415	&	-	&	-	&	-	&	-	&	0.325	& 1.182 &	-	&	0.005	&	-	&	1.646	& 5.565 &	1.593	&	0.507	\\
1307+23	&	9.4E-16	&	-	&	0.045	&	-	&	-	&	0.121	&	0.019	&	0.084	& - &	-	&	0.011	&	-	&	0.254	& - &	0.283	&	0.048	\\
1321+13	&	6.8E-16	&	0.089	&	0.417	&	0.156	&	-	&	-	&	-	&	0.254	& - &	-	&	0.105	&	-	&	1.444	& - &	0.900	&	0.491	\\
1323-02	&	1.4E-14	&	0.173	&	0.082	&	0.130	&	-	&	0.011	&	0.021	&	0.049	& 1.827 &	0.074	&	0.025	&	-	&	1.264	& 7.657 &	1.980	&	0.132	\\
1338-04	&	1.4E-14	&	0.233	&	0.140	&	0.124	&	0.011	&	0.015	&	0.042	&	0.092	& 1.461 &	0.080	&	0.074	&	0.045	&	0.426	& 5.397 &	0.222	&	0.108	\\
1407+42	&	1.6E-14	&	0.014	&	0.319	&	0.069	&	0.032	&	0.063	&	0.010	&	0.139	& - &	0.029	&	0.124	&	-	&	0.947	& - &	0.685	&	0.481	\\ 
1448+44	&	1.4E-14	&	0.046	&	0.203	&	0.077	&	-	&	-	&	-	&	0.074	& 0.722 &	-	&	0.046	&	-	&	0.473	& 3.422 &	0.461	&	0.260	\\
1637+25	&	2.3E-15	&	0.042	&	0.371	&	0.087	&	0.030	&	0.054	&	0.125	&	0.264	& - &	-	&	-	&	-	&	-	& - &	-	&	-	\\
2121-17	&	2.7E-14	&	0.113	&	0.112	&	-	&	-	&	0.434	&	0.913	&	0.498	& 2.685 &	-	&	0.089	&	-	&	5.152	& 1.680 &	3.141	&	0.083	\\
\hline
\label{lines}
\end{tabular}
\begin{tablenotes}
       \item[a]$^a$ Only H$\alpha$ was detected for the BLR of this object.
     \end{tablenotes}
\end{table}
\end{center}
\end{landscape}

\section{Results}

The optical spectra of the 2MASS sample, shown in Figure \ref{specs}, reveal a remarkable variety, ranging from Seyfert 2 galaxies that lack broad emission lines (e.g. J0504-19 \& J1057-13), through highly reddened broad line objects (e.g. J1040+59 \& J1127+24), to objects that show optical spectra similar to $\lq$normal' blue quasars (e.g. J1338-04 \& J2124-17) and even LINER/HII region-like objects (e.g. J0411-01 \& J1001+41). We now describe the results obtained from the detailed analysis of the spectra.

\subsection{Quasars, AGN or starbursts?}

\subsubsection{Quasars?}

Previous studies of red 2MASS AGN have shown that most red 2MASS AGN in the local Universe are not quasars, but have luminosities more consistent with those of Seyfert galaxies (\citealt{smith02}; \citealt{marble03}; \citealt{kuraab}).

\begin{figure}
\centering
\includegraphics[scale=0.38, trim=0cm 4cm 0cm 0cm]{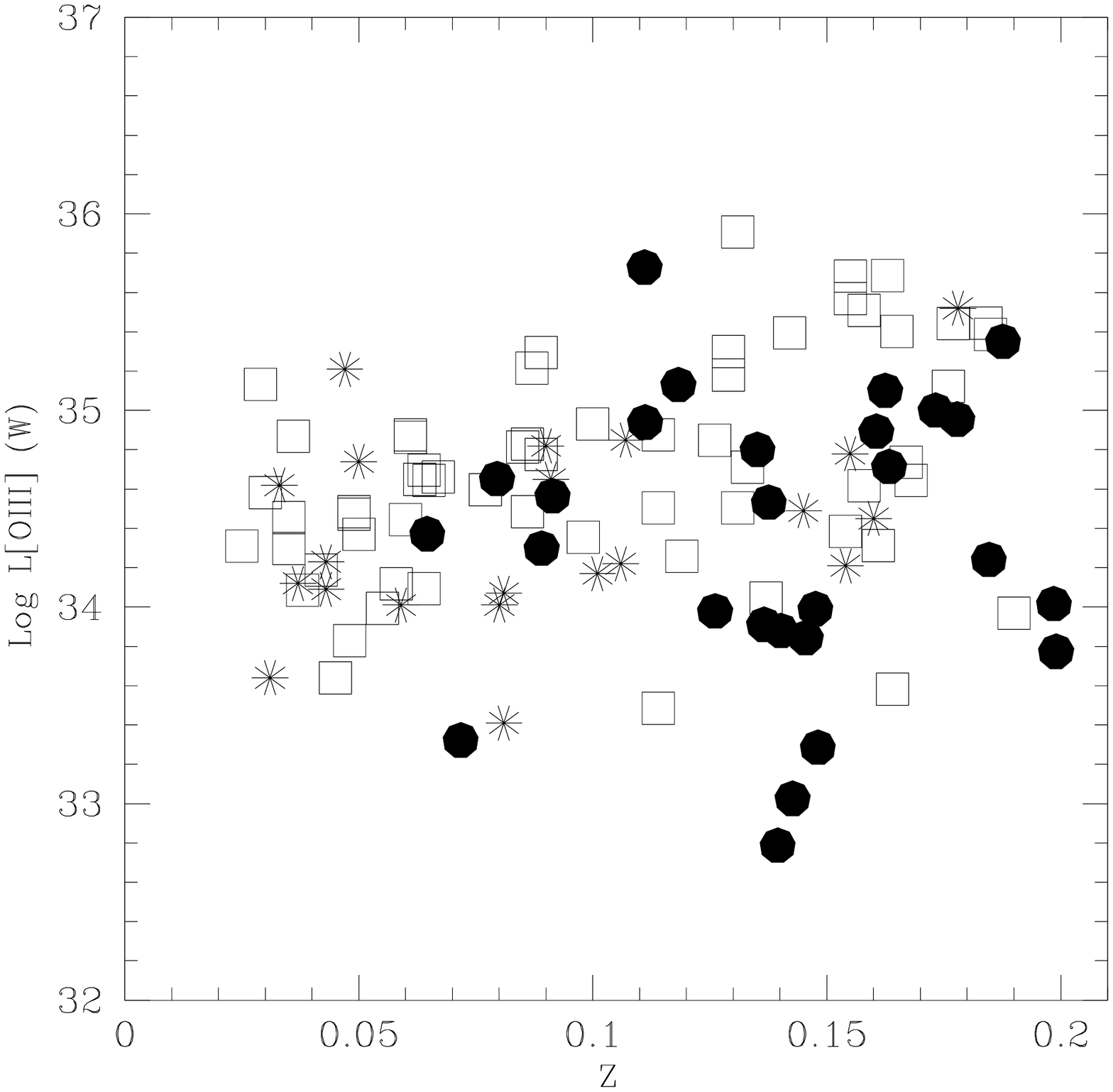}
\caption{[OIII] emission line luminosity (L$_{[OIII]}$) plotted against redshift for the 2MASS objects and the comparison samples. The 2MASS objects are represented by filled circles, the PG quasars are represented by unfilled squares, and the unobscured type 1 AGN are represented by asterisks. Objects with L$_{[OIII]}$ greater than 10$^{35}$ W are likely to have quasar-like bolometric luminosities.}
\label{loiii}
\end{figure}

To test the incidence of quasars in our low-z 2MASS sample, we used the [OIII]$\lambda$5007 emission line luminosity, which has been shown to be a good indicator of AGN bolometric power (see \citealt{heckman04}; \citealt{bian}; \citealt{dicken}; \citealt{lamassa10}). Other luminosity indicators, such as the L$_{5100}$ optical continuum luminosity, cannot be accurately calculated in most of the 2MASS objects because of a combination of reddening of the AGN continuum and host galaxy contamination. Following \citet{zakamska} we assume that an object is a quasar if it has an [OIII]$\lambda$5007 luminosity L$_{[OIII]}$ $\ga$ 10$^{35}$ W, corresponding to a nuclear continuum luminosity of M$_B$ $\la$ -23 (\citealt{sg}; \citealt{zakamska}). 

Figure \ref{loiii} shows L$_{[OIII]}$ plotted against redshift for the 2MASS objects and the comparison samples. Note that the luminosities have been determined using the observed frame [OIII]$\lambda$5007 fluxes and the luminosity distances appropriate for the assumed cosmology, but have not been corrected for intrinsic reddening. Where possible we use the wide slit (5\arcsec) data to ensure that the AGN [OIII] luminosity measurements are not affected by slit losses; for the 5 objects for which there are no wide slit data, we use the available 1.5 arcsecond data. We find that 5 objects in our sample (19\%) qualify as quasars according to the [OIII] emission line luminosity criterion: J0400+05, J1131+16, J1338-04, J1407+42 and J2124-17. Note that the latter object was previously identified as a ULIRG by \citet{k&s}. 

The rate of quasars for the 2MASS sample (filled circles) is comparable to that of the rate of quasars in the comparison samples: 28\% of the PG quasars (unfilled squares) and 10\% of the unobscured type 1 AGN (asterisks) have quasar-like luminosities based on the L$_{[OIII]}$ luminosity. Therefore the rate of occurrence of quasars in the PG and unobscured type I samples is not significantly different from that of the 2MASS sample. However, we note that 4 objects in our 2MASS sample have lower [OIII] emission line luminosities than any of the objects in the PG quasar and unobscured type 1 AGN samples.  

We can only be confident that 5 of the objects in the sample are genuine quasars. The remaining sample has [OIII] luminosities in the range 32.5 $<$ L$_{[OIII]}$ $<$ 35 W --- more typical of Seyfert galaxies. {If the 4 objects with low [OIII] luminosities are ignored, the 2MASS AGN have a similar range of luminosities as the PG quasars and the unobscured type 1 AGN}. Note that we have not corrected the [OIII] emission line luminosities for intrinsic dust extinction. If we correct the emission line luminosities using the NLR reddening estimates available for 16 objects ($\oint$ 4.2.1), we find that the rate of occurrence of quasars in the 2MASS sample increases to 27\% (8 objects). However, we emphasise that we can only accurately correct the narrow line fluxes for reddening in 55\% of the objects. 

\subsubsection{AGN or Starbursts?}

Both starbursts and AGN activity can potentially ionise the narrow emission regions in the nuclei of galaxies (\citealt{bpt}). It is therefore important to establish whether the NLR in the 2MASS objects are indeed photoionized by AGN, or whether recent starburst activity provides the photons needed to ionize the emission line regions. Therefore we have categorised the spectra based on the emission line ratios using Baldwin, Phillips \& Terlevich (BPT) diagnostic diagrams (e.g. see \citealt{bpt}; \citealt{kewley}). 

\begin{figure}
\centering
\includegraphics[scale=0.38, trim=0cm 4cm 0cm 0cm]{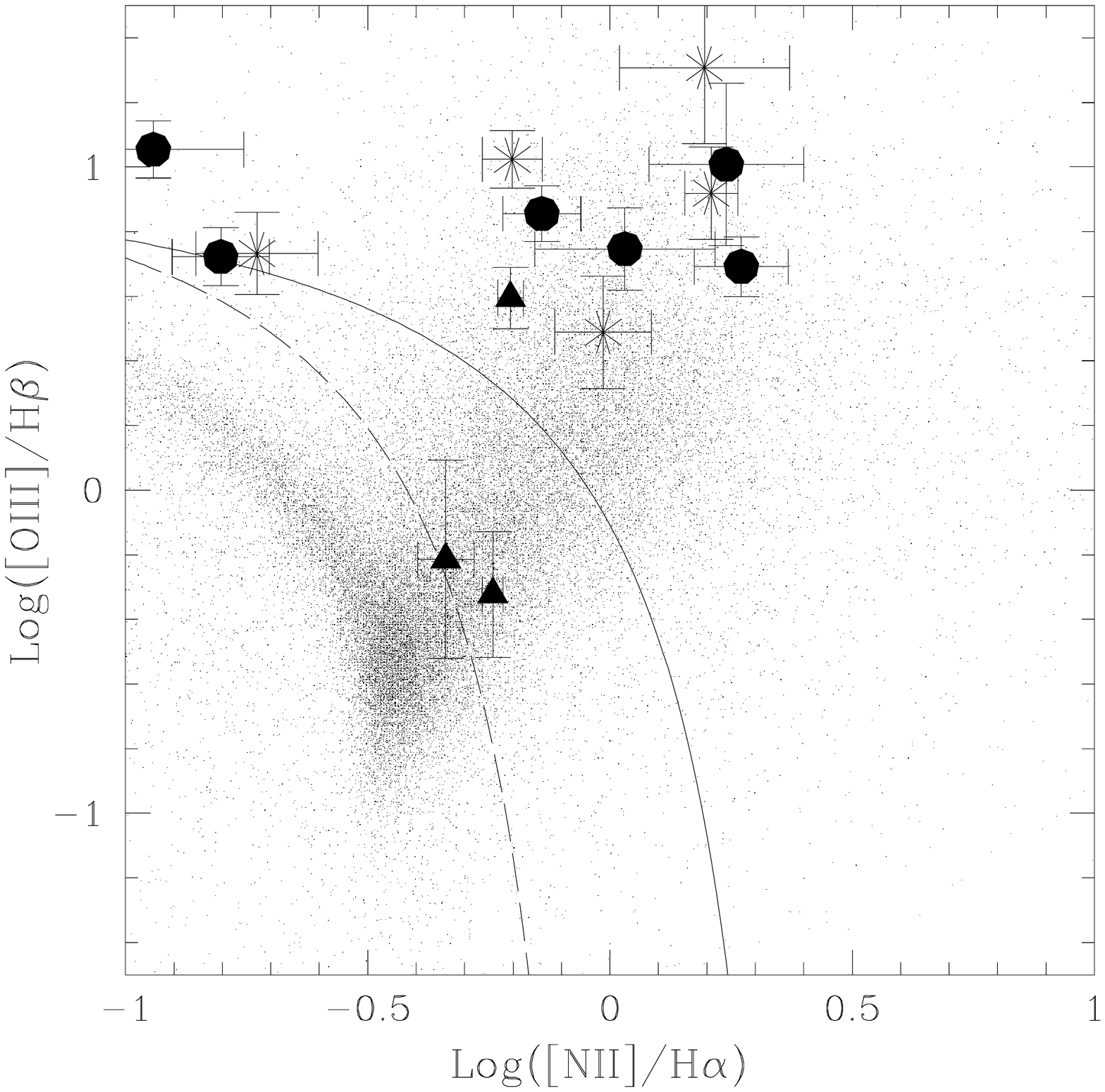}
\caption{Diagnostic plot of Log$_{10}$([OIII]/H$\beta$) vs Log$_{10}$([NII]/H$\alpha$). AGN are defined to lie above the solid line, HII-region like galaxies below the dashed line, and composite galaxies between these boundaries. The filled circles indicate the emission line measurements for type 2 objects in the 2MASS sample, the stars represent narrow-line emission from the type 1--1.9 objects, the triangles represent the objects in the 2MASS sample, which have fainter emission than the rest of the sample, and the small points are from SDSS-DR 8 data.}
\label{diag1}
\end{figure}

\begin{figure}
\centering
\includegraphics[scale=0.38, trim=0cm 4cm 0cm 0cm]{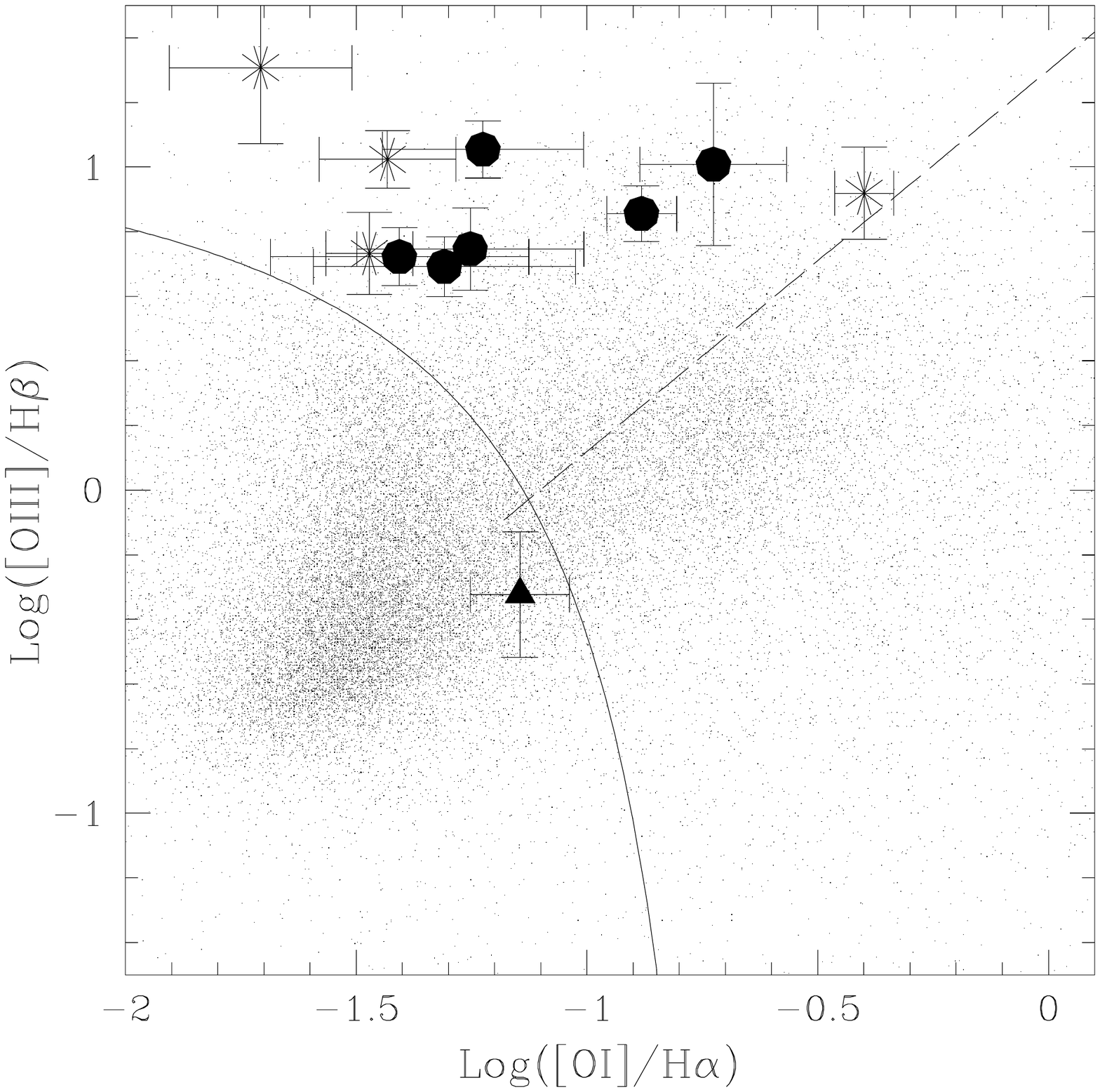}
\caption{Diagnostic plot of Log$_{10}$([OIII]/H$\beta$) vs Log$_{10}$([OI]/H$\alpha$). AGN are defined to lie above the solid line, with the dashed line separating Seyfert-type objects (above the line), and LINER-type objects (below). HII-region like galaxies fall below the solid line. The symbols are the same as Figure \ref{diag1}.}
\label{diag2}
\end{figure}

\begin{figure}
\centering
\includegraphics[scale=0.38, trim=0cm 4cm 0cm 0cm]{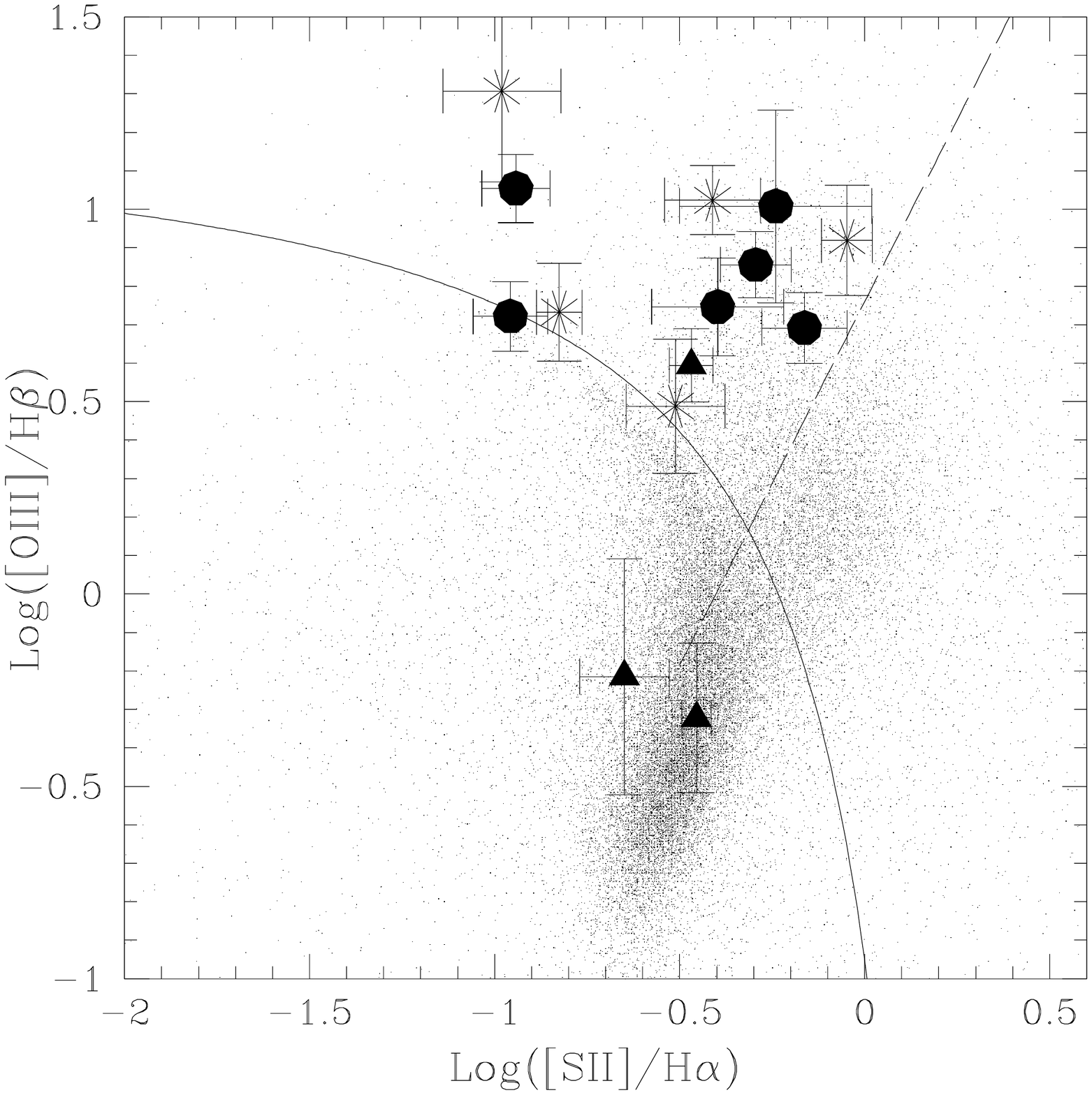}
\caption{Diagnostic plot of Log$_{10}$([OIII]/H$\beta$) vs Log$_{10}$([SII]/H$\alpha$). AGN are defined to lie above the solid line, with the dashed line separating Seyfert-type objects (above the line), and LINER-type objects (below). HII-region like galaxies fall below the solid line. The symbols are the same as Figure \ref{diag1}.}
\label{diag3}
\end{figure}

Figures \ref{diag1} - \ref{diag3} show the results for the 2MASS sample. The divisions between the different classes of objects in these figures are taken from \citet{kewley}. Note that only $\sim$50$\%$ of the 2MASS sample could be plotted on the BPT diagrams because of the difficulty of measuring accurate narrow H$\beta$ emission line fluxes in many of the objects, related to the de-blending of the emission of the BLR and the NLR (see discussion in $\oint$ 3.3).

Based on the diagnostic diagrams, all of the objects that could be plotted fall in the AGN part of the diagrams with the exceptions of J0411-01 and J1001+41, which fall in the composite/HII region part of the diagrams, suggesting a major contribution from stellar photoionization. We therefore exclude the latter two objects from any subsequent analysis involving AGN properties.

\begin{figure}
\centering
\includegraphics[scale=0.38, trim=0cm 4cm 0cm 0cm]{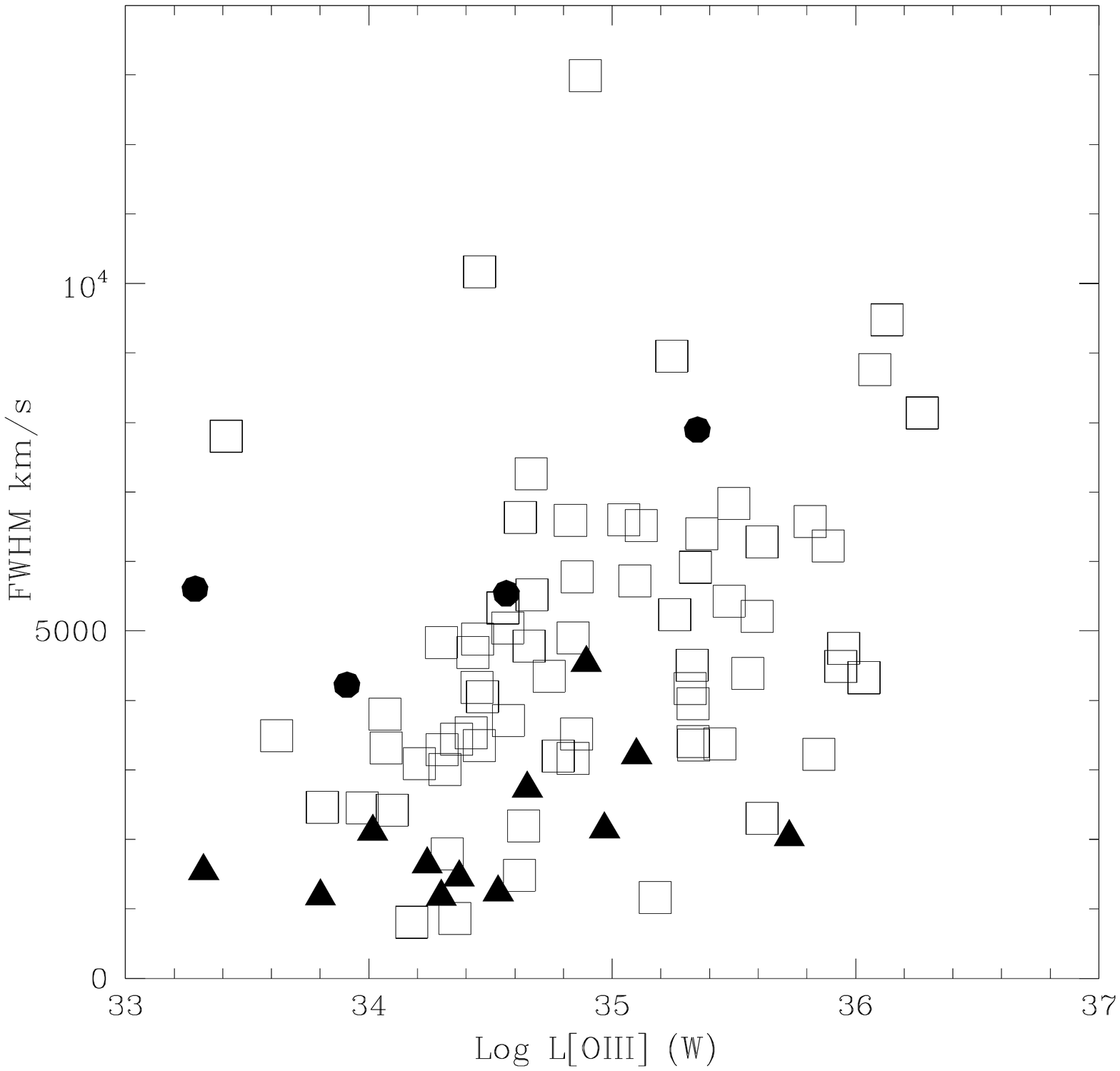}
\caption{Velocity width (FWHM) of the BLR components plotted against L$_{[OIII]}$. The 2MASS objects for which broad emission can be confidently separated from the narrow line emission are indicated by the filled circles, whereas the 2MASS objects for which the broad and narrow emission could not be confidently separated are indicated by the filled triangles. The PG quasar population are indicated by the unfilled squares. All the line measurements are based on single Gaussian fits.}
\label{blrwidths}
\end{figure}

In addition to emission line diagnostic ratios plotted in Figures \ref{diag1} - \ref{diag3}, the presence of FHILs in the spectra can indicate unambiguously the presence of an AGN. These lines include [NeV]$\lambda\lambda$3345,3426, [FeVII]$\lambda$6086 and [FeX]$\lambda$6374 which, due to their high ionisation potentials ($>$97 eV), are not generally associated with LINER nuclei or starbursts. We find that 23 of the objects in our sample (85\%) show evidence for at least one FHIL, suggesting significant AGN activity. 

Another indicator of the presence of an AGN is the detection of broad line emission from the BLR. Table \ref{truth} indicates whether broad emission lines are detected for each of the 2MASS objects. We assume that the BLR is detected if the Balmer recombination lines show significant (i.e. well above the noise) broad wings that cannot be accounted for by the [OIII] model, and have FWHM $>$ 1000 km s$^{-1}$. 16 out of 27 objects in our sample (59\%) show the presence of definite BLR emission. In addition, tentative evidence for weak H$\alpha$ broad line emission in J1131+16 was presented in \citet{rose}. The broad line component in J1131+16 is consistent with the presence of a scattered \citep{am}, or a directly observed, but weak, broad line region (BLR) component. The rest of the type 2 objects had no such emission. Figure \ref{blrwidths} presents the velocity widths (FWHM) of the BLR components as measured from single Gaussian fits to the broad Balmer lines\footnote{Note that the FWHM of the Balmer lines are based on fits to the H$\beta$ in all cases except of J1006+41 and J1040+59, which do not have any detectable broad emission in the H$\beta$ emission line, but clear BLR H$\alpha$ components.}, plotted against L$_{[OIII]}$ for both the 2MASS sample and PG sample. Where possible, the broad and narrow-line emission were fitted separately in order to obtain the BLR FWHM; such cases are indicated by the filled circles. Where the broad and narrow components could not be confidently separated we use the velocity widths of single Gaussian fits to the whole profiles; these cases are indicated by the black triangles. The median BLR velocity width (FWHM) for the 2MASS sample is 2100$\pm$400 km s$^{-1}$, which is lower than that of the PG sample (4500$\pm$300 km s$^{-1}$). However, the 2MASS median may be affected by the inclusion of NLR emission in the objects for which we could not confidently separate the emission from the BLR and the NLR. The median velocity width for the 2MASS objects in which the broad and narrow components could be confidently separated is 4400$\pm$1100 km s$^{-1}$ --- comparable with the PG quasars.

All the results from this section are summarised in the truth table of Table \ref{truth}. Overall, 25/27 ($\sim$93$\%$) objects show some sign of AGN activity; and only 2 objects --- J0411-01 and J1001+41 --- show no clear evidence of AGN activity. Given that it is not clear whether the latter two objects are powered by AGN activity, we do not include them in the analysis in the rest of the paper. In terms of a more detailed classification of the spectra of the 2MASS objects for which we detect significant AGN activity, 59\% qualify as Seyfert 1 galaxies, 30\% as Seyfert 2 galaxies, and 4\% have weak broad lines placing them between the Seyfert 1 and Seyfert 2 classification.

\begin{table}
\caption{Truth table presenting the evidence for AGN activity in the 2MASS sample objects. A true result for BLR indicates the detection of a broad component for either/both H$\alpha$ and H$\beta$. A true result for the FHIL indicates the positive detection of a high ionisation species ($>$54.4 eV). The [NII] column shows the results from Figure \ref{diag1}, the [OI] column shows Figure \ref{diag2} and the [SII] column shows Figure \ref{diag3}. T=true, F=false, P=tentative evidence for a BLR, A=AGN, S=Seyfert type, H=starburst galaxy, C=composite galaxy, D=unmeasured due to the degeneracy between the broad and narrow Balmer emission, and X=unmeasured because an appropriate emission line has not been detected.}
\begin{tabular}{lllllll}
\hline
Name 	&	quasar 	&	BLR	&	FHIL	&	[NII]	&	[OI]	&	[SII]	\\
\hline
0221+13	&	F	&	F	&	T	&	A	&	S	&	S	\\
0248+14	&	F	&	T	&	T	&	D	&	D	&	D	\\
0306-05	&	F	&	F	&	T	&	D	&	D	&	D	\\
0312+07	&	F	&	T	&	T	&	D	&	D	&	D	\\
0400+05	&	T	&	T	&	T	&	A	&	S	&	S	\\
0409+08	&	F	&	T	&	T	&	A	&	S	&	S	\\
0411-01	&	F	&	F	&	F	&	C	&	X	&	H	\\
0422-18	&	F	&	T	&	T	&	D	&	D	&	D	\\
0435-06	&	F	&	T	&	F	&	D	&	D	&	D	\\
0447-16 &   F   &   T   &   T   &   D   &   D   &   D   \\
0504-19	&	F	&	F	&	T	&	A	&	S	&	S	\\
0910+33	&	F	&	F	&	T	&	A	&	S	&	S	\\
1001+41	&	F	&	F	&	F	&	C	&	H	&	H	\\
1006+41	&	F	&	T	&	T	&	D	&	D	&	D	\\
1014+19	&	F	&	T	&	T	&	D	&	D	&	D	\\
1040+59	&	F	&	T	&	F	&	D	&	D	&	D	\\
1057-13	&	F	&	F	&	T	&	A	&	S	&	S	\\
1127+24	&	F	&	T	&	T	&	D	&	D	&	D	\\
1131+16	&	T	&	P$^a$	&	T	&	A	&	S	&	S	\\
1158-30	&	F	&	F	&	T	&	A	&	S	&	S	\\
1212-14	&	F	&	T	&	T	&	A	&	X	&	S	\\
1321+13	&	F	&	F	&	T	&	A	&	X	&	S	\\
1323-02	&	F	&	T	&	T	&	A	&	S	&	S	\\
1338-04	&	T	&	T	&	T	&	D	&	D	&	D	\\
1407+42	&	T	&	F	&	T	&	A	&	S	&	S	\\
1448+44	&	F	&	T	&	T	&	D	&	D	&	D	\\
2121-17	&	T	&	T	&	T	&	D	&	D	&	D	\\
\hline
\label{truth}
\end{tabular}
\begin{tablenotes}
    \item[a]$^a$ See \citet{rose} for full description.
     \end{tablenotes}
\end{table}

\subsubsection{Narrow-line Seyfert 1 galaxies}

\begin{figure}
\centering
\includegraphics[scale=0.26]{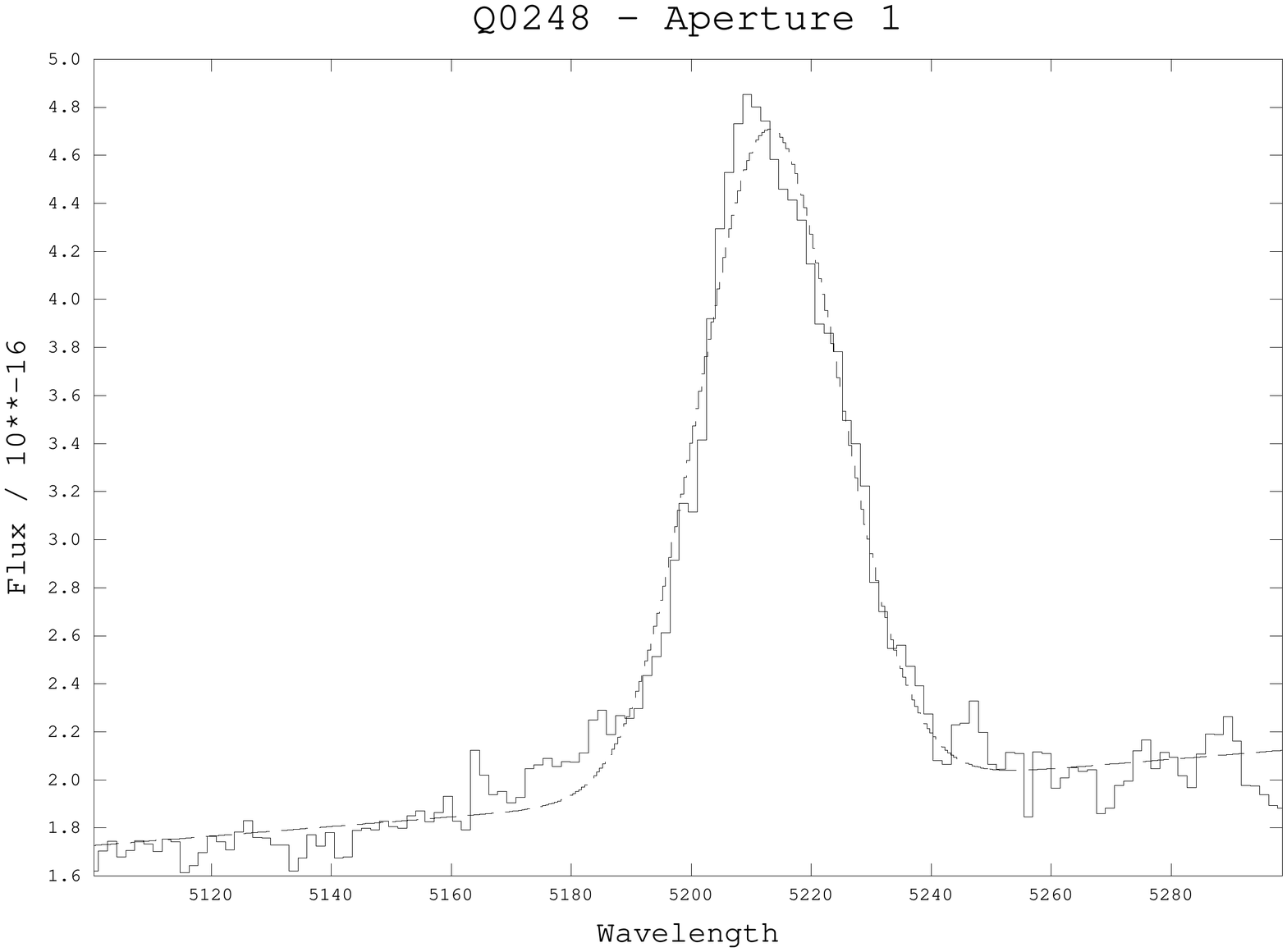}
\includegraphics[scale=0.26]{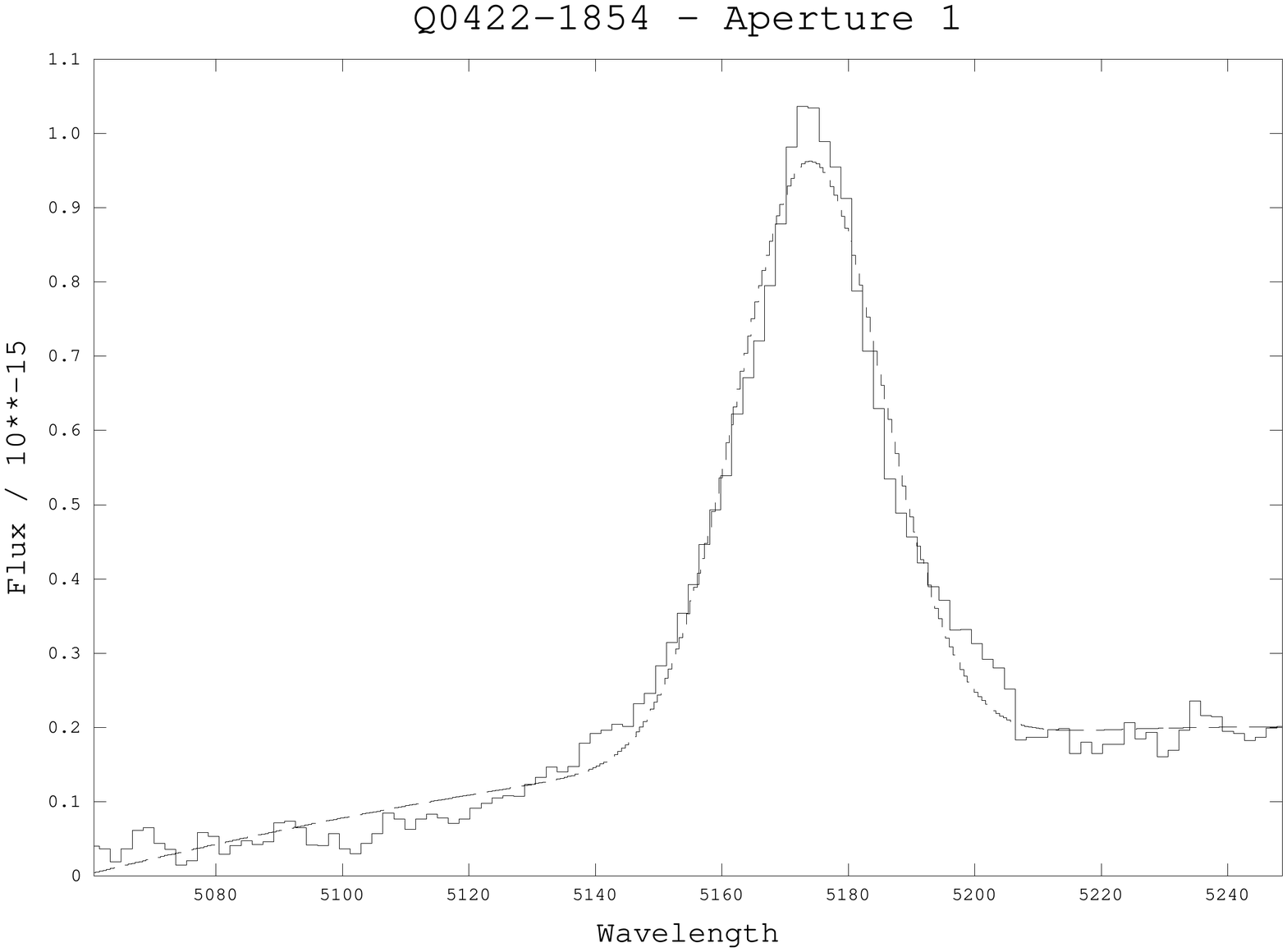}
\includegraphics[scale=0.26]{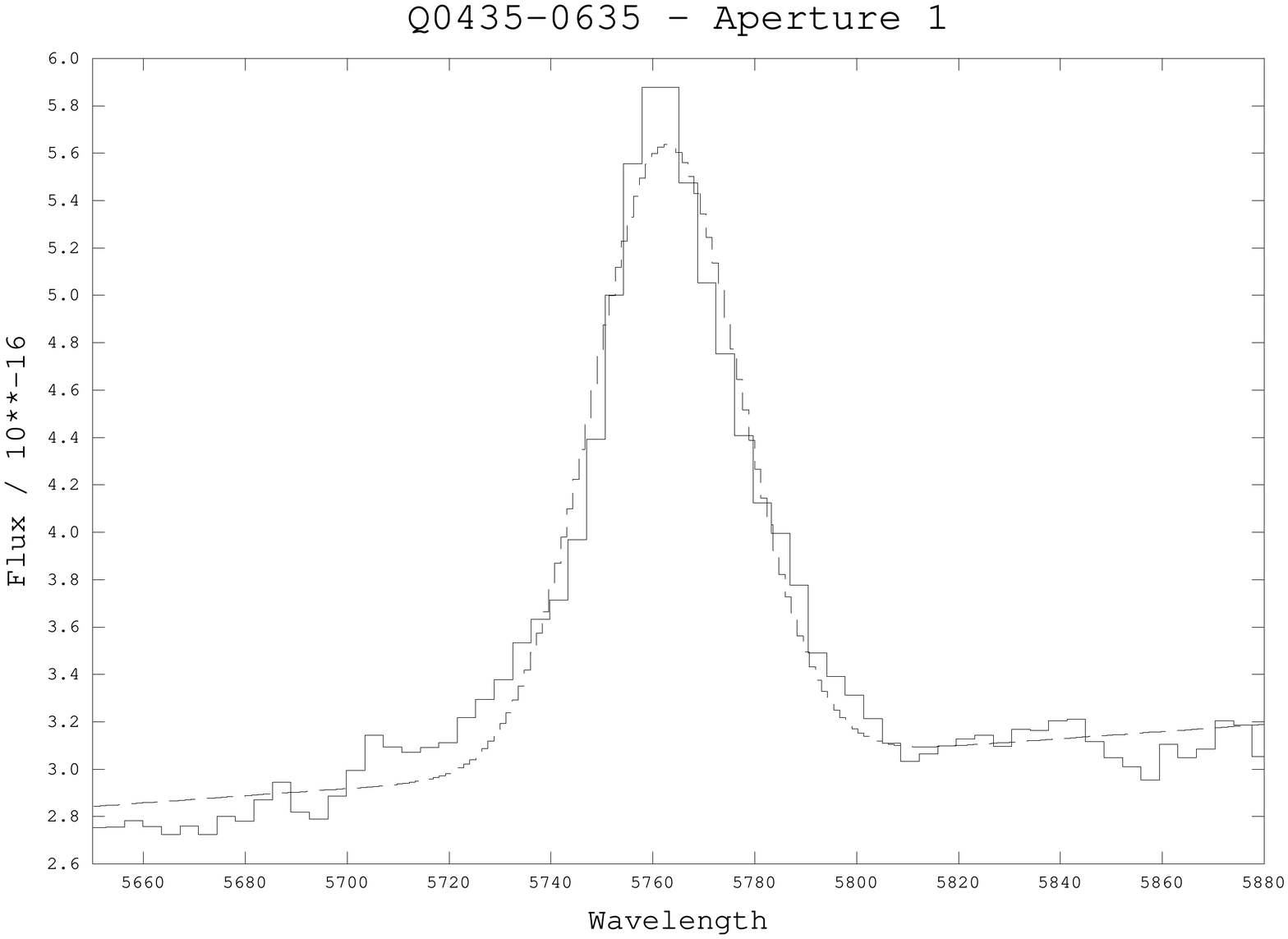}
\includegraphics[scale=0.26]{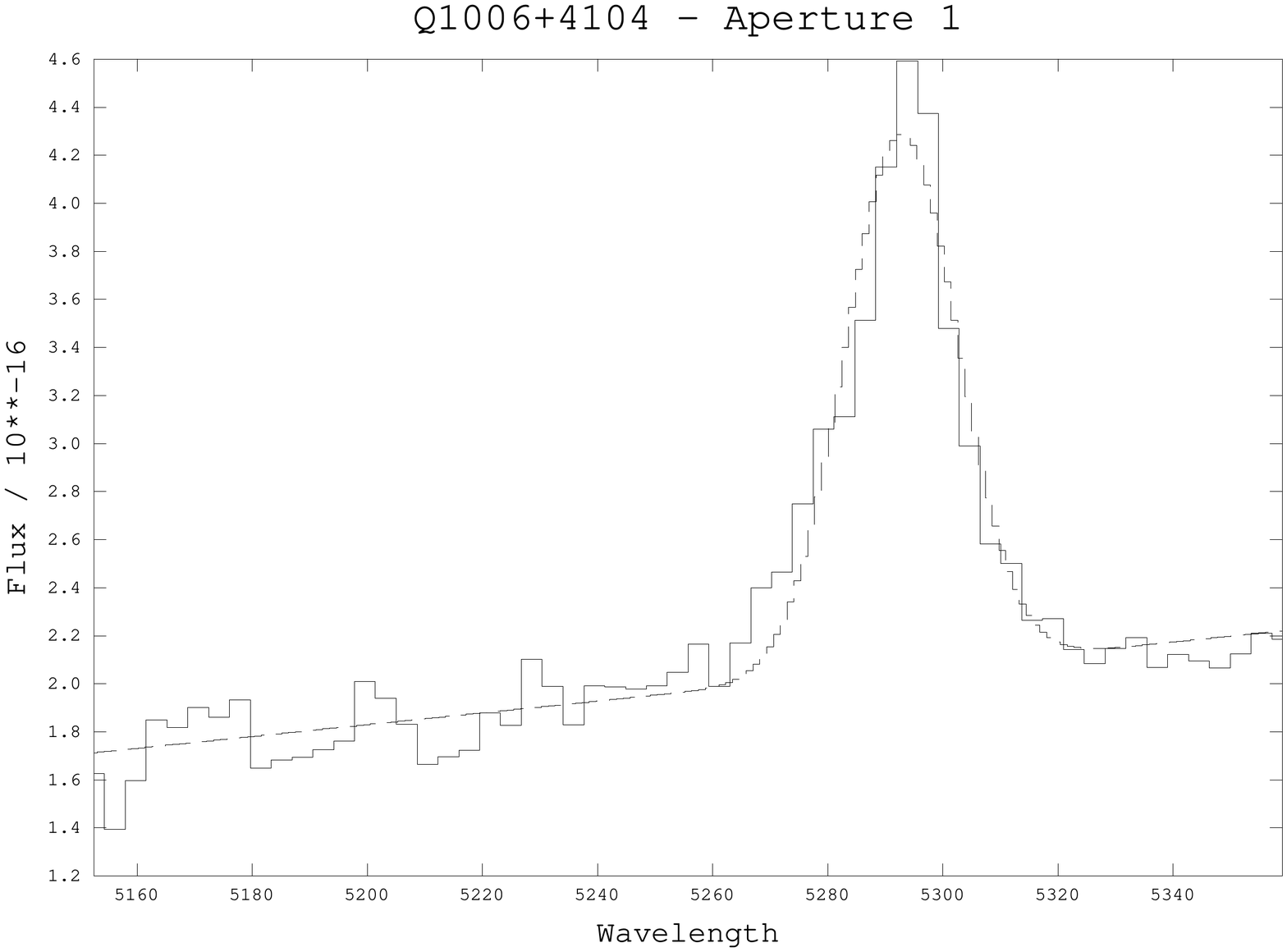}
\caption{Gaussian fits to the H$\beta$ emission lines of the NLS1 candidates in the 2MASS sample. From top to bottom: J0248+14, J0422-18, J0435-06 and J1006+41. The overall fits to the emission lines are adequate and therefore we use the velocity widths (FWHM) to characterise these objects.}
\label{nls1}
\end{figure}

Narrow-line Seyfert 1 galaxies (NLS1) are a sub-class of Seyfert galaxies with BLR Balmer emission lines that are relatively narrow (FWHM$_{H\beta}$ $<$ 2000 km s$^{-1}$: \citealt{o&p}; \citealt{veron}). There are several hypotheses surrounding the nature of these objects. One prevailing idea is that NLS1 are AGN with lower than typical black hole masses that are undergoing a phase of rapid accretion \citep{pounds95}. In this scenario the NLS1 fall below, but are growing towards, the M-$\sigma$ correlation \citep{f&m}; eventually, as they accrete more mass, they will appear as typical Seyfert objects \citep{mathur}. 

If the red 2MASS objects are young, recently triggered AGN, then we might expect there to be a relatively large fraction of NLS1 objects in the sample. We checked our sample for NLS1 based on single Gaussian fits to their H$\beta$ lines, to find objects with H$\beta$ or H$\alpha$ emission line widths that are below the NLS1 limit (FWHM $<$ 2000 km s$^{-1}$), but greater than the widths of the [OIII] lines. This is consistent with the fitting procedures of \citet{o&p} and \citet{goodrich}. We found that 4 objects with adequate single Gaussian fits to H$\beta$ have velocity widths consistent with them being NLS1: J0248+14, J0422-18, J0435-06 and J1006+41. These objects have H$\beta$ velocity widths (FWHM) of 1550$\pm$20 km s$^{-1}$, 1450$\pm$120 km s$^{-1}$, 1780$\pm$70 km s$^{-1}$ and 1170$\pm$70 km s$^{-1}$ respectively.

Figure \ref{nls1} shows the single Gaussian fits to the H$\beta$ emission lines in these objects. Apart from the line widths, there are two other key properties that are found to be associated with NLS1: [OIII] emission lines that are weak relative to the broad H$\beta$ lines ($[$OIII]/H$\beta_{total}$$<$3) and the presence of strong FeII emission (FeII$_{4570}$/H$\beta$$_{total}$$>$0.5: \citealt{o&p}; \citealt{goodrich}; \citealt{veron}). We find that all 4 objects with FWHM$_{H\beta}$$<$2000 km s$^{-1}$ in our sample also posses these typical NLS1 properties. 

Of the AGN in the 2MASS sample, only $\sim$16\% are NLS1 objects, or $\sim$24\% as a fraction of all the broad emission-line objects in our 2MASS sample. This is similar to the fraction of NLS1s found in other samples of broad-line AGN: $\sim$24\% for the unobscured type 1 AGN \citep{jina}; 11\% for the PG quasars \citep{hao05}; and $\sim$15\% of all QSOs in SDSS DR3 AGN catalogue (\citealt{zhou}). Therefore we conclude that the 2MASS sample does not contain an unusually high proportion of NLS1 objects.

\subsection{Balmer decrements and reddening}

One possible explanation for the red colours of the 2MASS objects at near-IR wavelengths is that their AGN suffer an unusual amount of dust extinction compared with other samples of AGN ( e.g. \citealt{glik1}). In order to investigate this possibility, we have calculated Balmer recombination line decrements for the 2MASS, PG/SDSS and \citealt{jinall} samples to determine whether the 2MASS objects have higher levels of dust extinction than $\lq$typical' blue quasars.

Ideally, Balmer decrements (F$_{H\alpha}$/F$_{H\beta}$) would be determined for the narrow and broad line components of each object separately. However, some objects with a detectable BLR have relatively narrow broad lines (e.g. J1006+41), which makes it difficult to separate the components at the resolution of our spectra (see $\oint$ 3.3). Therefore we have taken two approaches. For the first approach we used the measured total (broad and narrow) fluxes of the Balmer emission lines. The second approach was to separate the broad and narrow emission components of the blends, where the components could be confidently separated. We only used the total (broad + narrow) Balmer flux measurements for the \citet{jina} sample, because we were uncertain whether the broad and narrow emission components could be confidently separated in that sample. In addition, when we consider Balmer decrements for the total flux of the Balmer lines, we only use the data for the objects with a detectable BLR, and do not consider the Balmer decrements for the type 2 objects. Balmer decrements for the 2MASS objects are given in Table \ref{baltab}.   

\begin{figure}
\centering
\includegraphics[scale=0.45, trim=0cm 7cm 0cm 8cm]{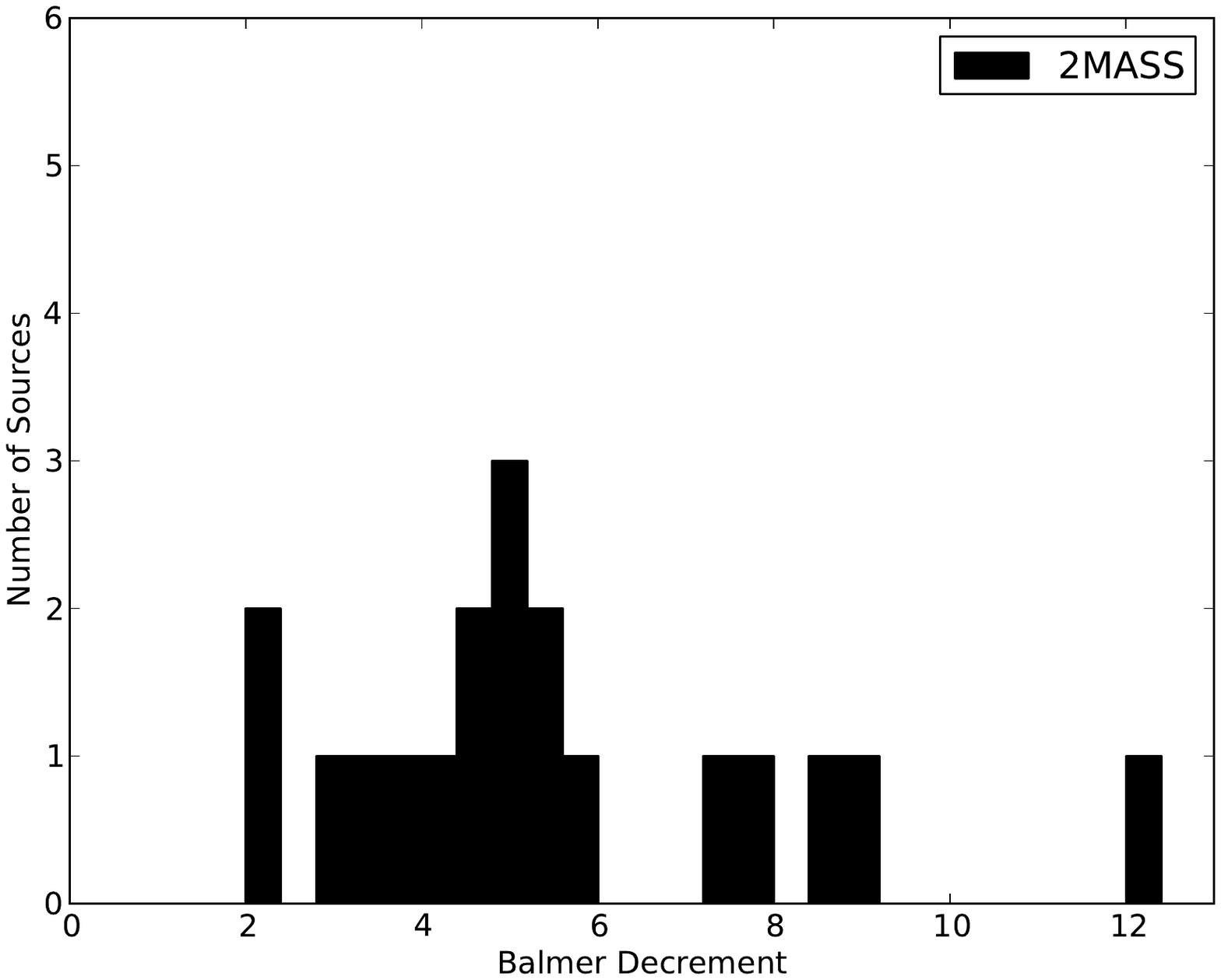}
\includegraphics[scale=0.45, trim=0cm 7cm 0cm 8cm]{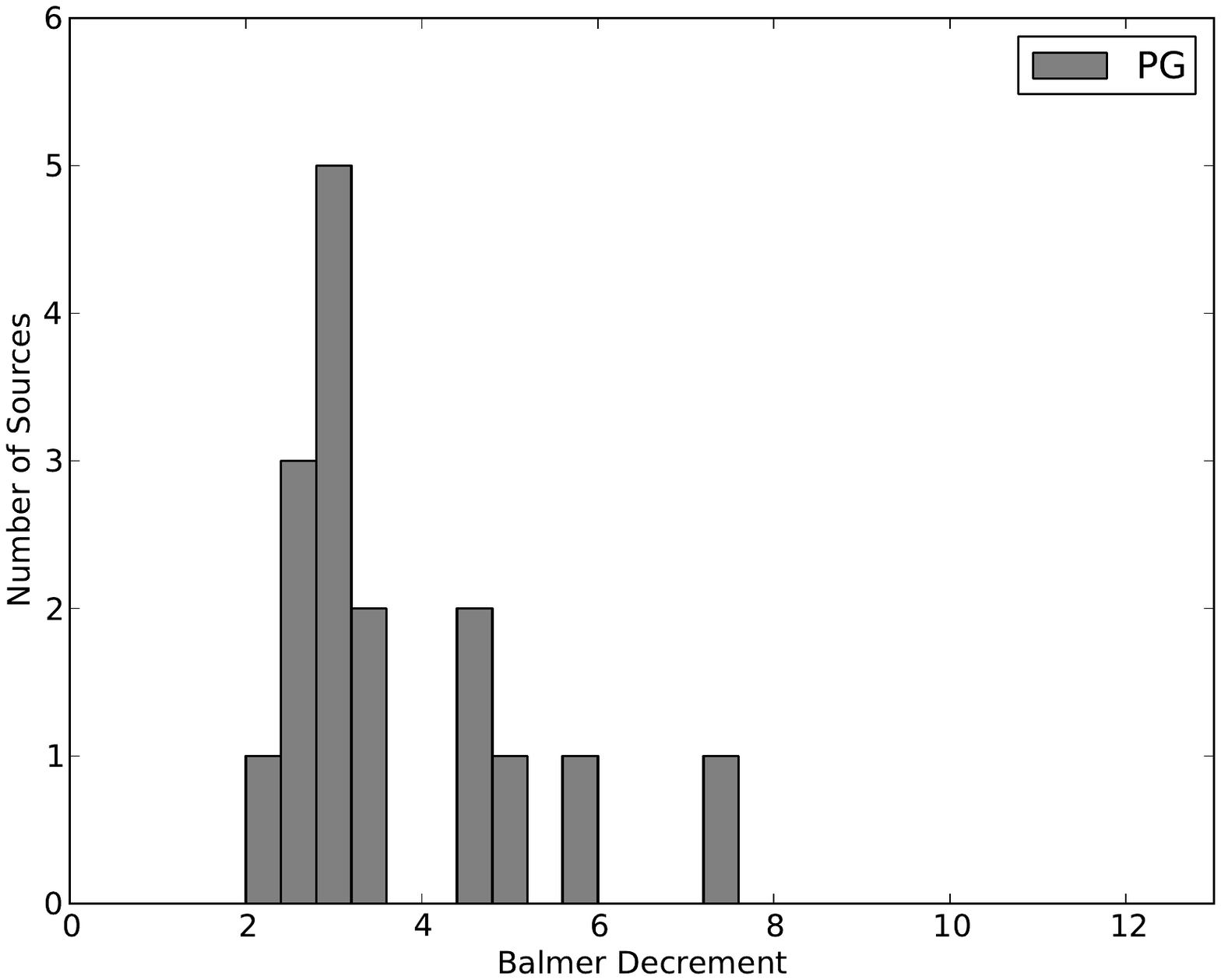}
\includegraphics[scale=0.45, trim=0cm 7cm 0cm 8cm]{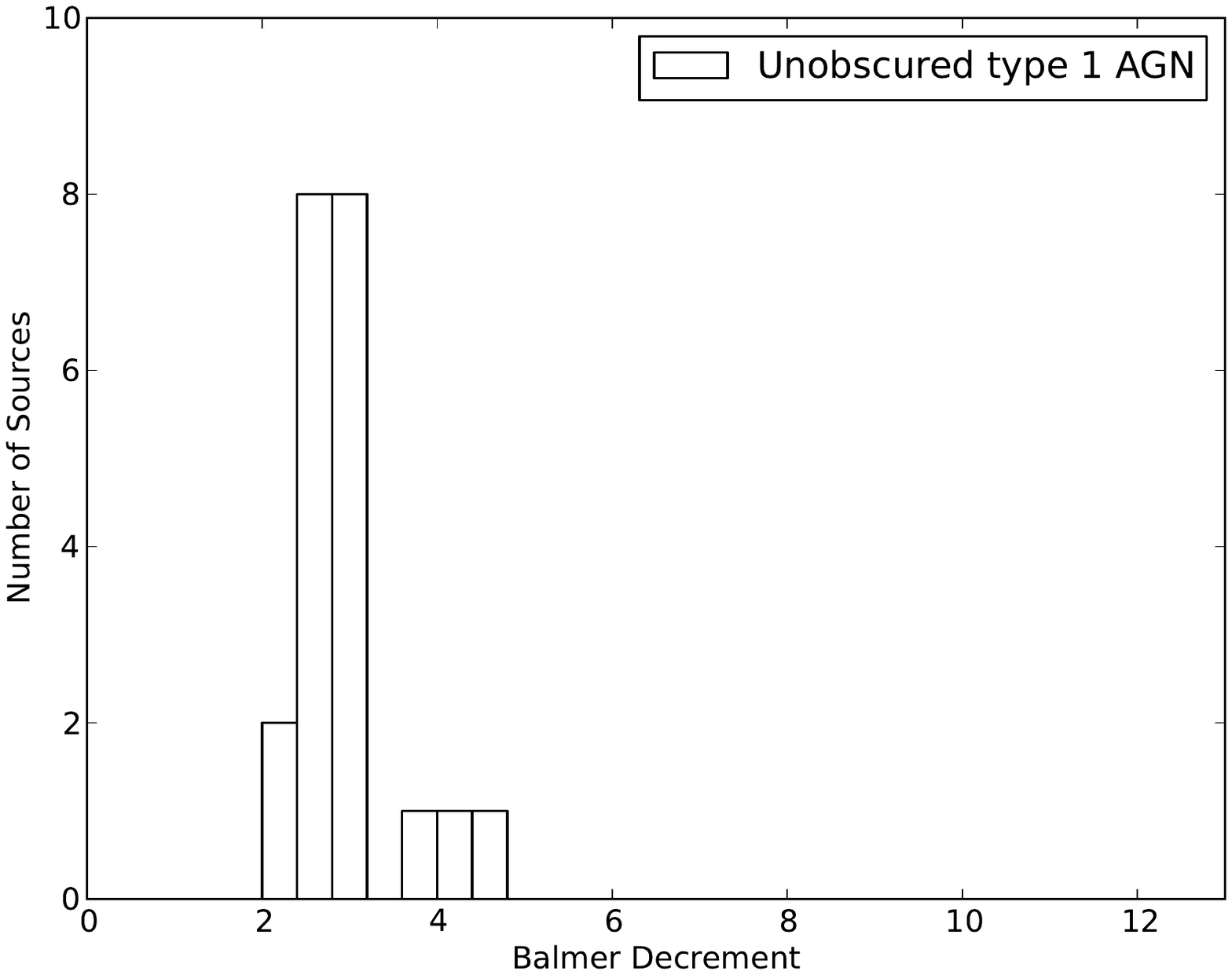}
\caption{Comparison of the total flux Balmer decrements (F$_{H\alpha}$/F$_{H\beta}$). The 2MASS population measurements (broad line objects only) are indicated in black (top), the PG quasar population are shown in grey (middle) and the unobscured type 1 AGN are given in white (bottom). Note that the Balmer decrements are only presented for those objects with broad Balmer emission.}
\label{baldecf}
\end{figure}

Figure \ref{baldecf} shows histograms of the Balmer decrements (F$_{H\alpha}$/F$_{H\beta}$) for the combined broad and narrow Balmer emission of the 2MASS sample (broad line objects only), the PG/SDSS sample and the unobscured type 1 AGN. 

It is clear from this figure that, on average, the 2MASS sample objects have higher F$_{H\alpha}$/F$_{H\beta}$ ratios than the PG quasars and unobscured type 1 AGN, implying that there is more dust extinction in the 2MASS sample. There is also a significant fraction of objects (5/19) with Balmer decrements that are higher than the most reddened PG quasars. This is reflected in the median F$_{H\alpha}$/F$_{H\beta}$ ratios: F$_{H\alpha}$/F$_{H\beta}$=4.9$\pm$0.5, 3.2$\pm$0.4, and 2.8$\pm$0.1 for the 2MASS, PG/SDSS and unobscured type 1 objects respectively.

To put this in context, the Balmer decrements for the 2MASS sample suggest reddening in the range 0.0 $\le$ E(B-V) $\le$ 1.2, with a median E(B-V)=0.52$\pm$0.05, assuming a Case B H$\alpha$/H$\beta$ ratio of 3.1 (\citealt{gaskell84}). However, although the 2MASS sample displays a significant degree of dust reddening, there are several objects ($\sim$25\% of all the broad line objects) which show very little (if any) significant BLR reddening. These objects include J0312+07, J0409+08, J0435-06, J1338-04 and J2121-17. This is interesting, because it highlights the diverse nature of the 2MASS sample. We further note that three of the objects with low Balmer decrements --- J0312+07, J0435-06 and J1338-04 --- have J-K$_S$ colours just below the J-K$_S$=2.0 limit. Also, J1040+59, an object with one of the highest Balmer decrements, has one of the reddest J-K$_S$ colours in the sample (J-K$_S$=3.02). However, overall we do not find a statistically significant correlation between J-K$_S$ colours and Balmer decrements for the 2MASS sample.

We have performed K-S tests (\citealt{peacock}; \citealt{ff}) to check the significance of any differences in the distribution of the Balmer decrements between the samples. We find that we can reject the null hypothesis that the 2MASS sample is drawn from the same parent population as the PG quasars and unobscured type 1 AGN at the 2$\sigma$ and $>$3$\sigma$ levels  of significance respectively (the corresponding p values are 0.007 and $<$0.001).

In the cases where it was possible to clearly separate the BLR and NLR components, we find that the Balmer decrements obtained for the BLR components alone are not significantly different from those derived from the total (BLR + NLR) fluxes (see Table \ref{baltab}). Therefore we do not consider the pure BLR Balmer decrements further in this paper.

When considering the range of reddening found for our 2MASS AGN, our results are consistent with the idea that dust reddening plays a significant role in causing the red near-IR colours of the 2MASS objects (see $\oint$ 6.1).

\begin{table}
\caption{The H$\alpha$/H$\beta$ Balmer decrements measured for the 2MASS sample. The second column presents the Balmer decrements for the total (BLR + NLR) fluxes. The third and fourth columns present the NLR and BLR Balmer decrements respectively.}
\begin{tabular}{lccc}
\hline
 & & F$_{H\alpha}$/F$_{H\beta}$ &\\
Name 	&	Total 	& NLR 	&	BLR 	\\
\hline
J0221+13	&	-	&	3.00	$\pm$	0.34	&	-			\\
J0248+14	&	7.21	$\pm$	0.64	&	-			&	-			\\
J0306-05$^a$	&	-	&	8.87	$\pm$	2.20			&	-			\\
J0312+07	&	5.20	$\pm$	0.45	&	-			&	-			\\
J0400+05	&	7.87	$\pm$	0.51	&	7.03	$\pm$	0.51	&	8.07	$\pm$	0.79	\\
J0409+07	&	2.38	$\pm$	0.46	&	5.00	$\pm$	0.46	&	1.67	$\pm$	0.08	\\
J0422-18	&	4.30	$\pm$	0.35	&	-			&	-			\\
J0435-06	&	3.45	$\pm$	0.43	&	-			&	-			\\
J0447-16	&	8.74	$\pm$	1.48	&	-			&	-			\\
J0504-19	&	-	&	3.83	$\pm$	0.82	&	-			\\
J0910+33	&	-	&	2.15	$\pm$	0.62	&	-			\\
J1006+41$^b$	&	5.96	$\pm$	0.28	&	-			&	{\bf $>$5.00}			\\
J1014+19	&	5.57	$\pm$	0.79	&	-			&	-			\\
J1040+59$^b$	&	12.14	$\pm$	0.57	&	-			&	{\bf $>$6.67}			\\
J1057-13	&	-	&	8.05	$\pm$	0.67	&	-			\\
J1127+24	&	4.89	$\pm$	0.46	&	3.25	$\pm$	0.46	&	5.38	$\pm$	0.60	\\
J1131+16$^c$	&	-	&	5.00	$\pm$	0.17	&	-			\\
J1158-30	&	-	&	10.98	$\pm$	0.48	&	-	\\
J1212-14	&	4.79	$\pm$	0.93	&	5.07	$\pm$	0.93	&	4.71	$\pm$	0.75	\\
J1321+13	&	-	&	5.67	$\pm$	0.52	&	-			\\
J1323-02	&	4.75	$\pm$	0.87	&	-			&	-			\\
J1338-04	&	3.75	$\pm$	0.69	&	-			&	-			\\
J1407+42	&	-	&	6.79	$\pm$	0.68	&	-			\\
J1448+44	&	4.90	$\pm$	1.21	&	-			&	-			\\
J2124-17	&	2.15	$\pm$	1.19	&	-			&	-			\\
\hline
\label{baltab}
\end{tabular}
\begin{tablenotes}
    \item[a]$^a$ Sky line features cut through the region of the H$\beta$ emission line. Results are based on an upper limit measurements for the H$\beta$ emission line, but are presented as a lower limit on the Balmer decrement.
    \item[b]$^b$ BLR Balmer decrement is based on an upper limit of the H$\beta$ flux and therefore is a lower limit of the Balmer decrement.
    \item[c]$^c$ The emission flux from H$\alpha$ is boosted by collisional excitation in this object \citep{rose}.
     \end{tablenotes}
     \end{table}

\subsubsection{Reddening in the NLR}

\begin{figure}
\centering
\includegraphics[scale=0.38, trim=0cm 4cm 0cm 0cm]{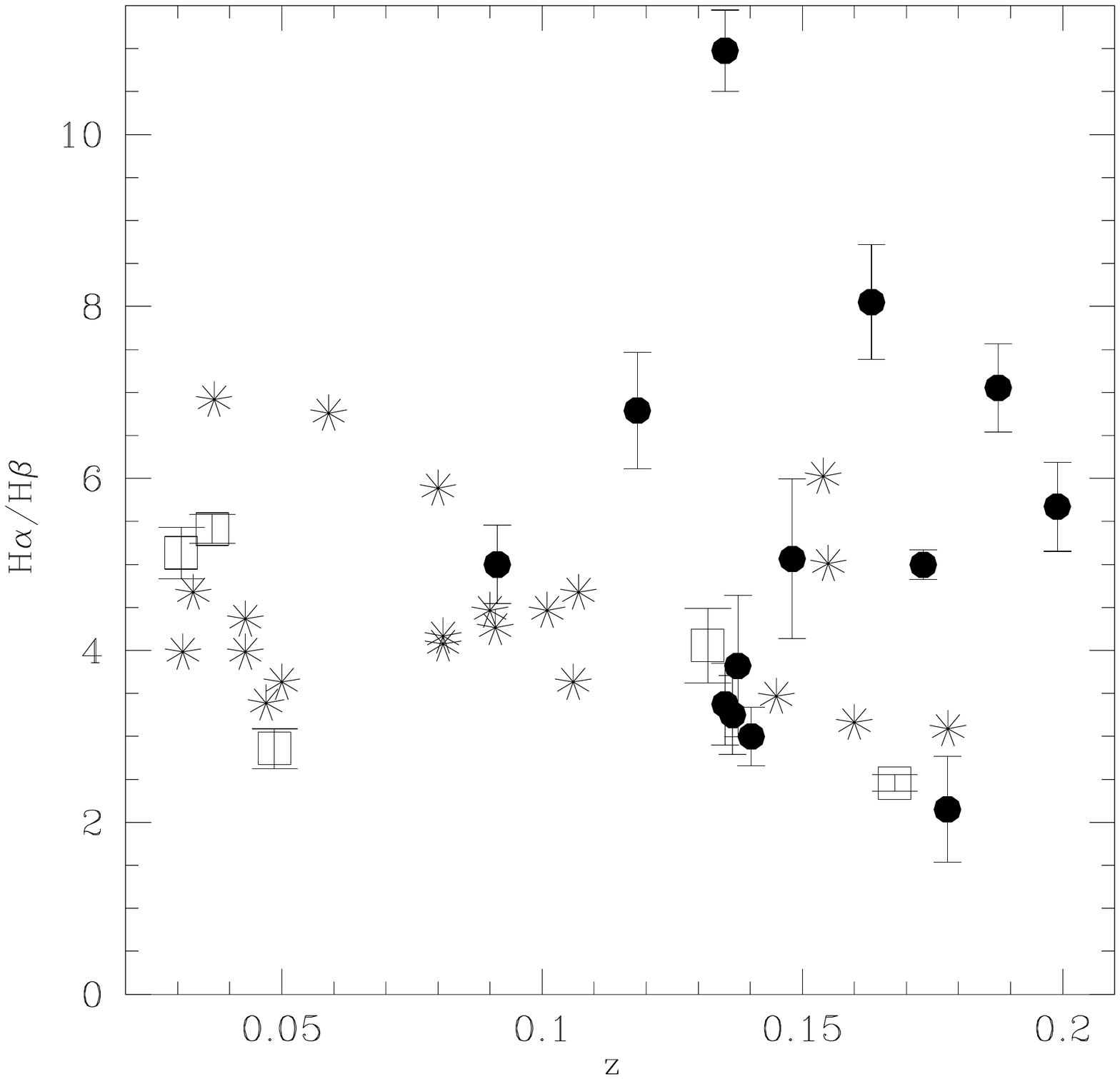}
\caption{The Balmer decrement (F$_{H\alpha}$/F$_{H\beta}$) versus redshift, z, for the NLRs of the 2MASS sample. The 2MASS population measurements are indicated by the filled circles, the PG/SDSS quasars are indicated by unfilled squares and the unobscured type 1 objects are indicated by asterisks.}
\label{baldecn}
\end{figure}

It is also interesting to consider the NLR Balmer decrements for the red 2MASS AGN in which the NLR and BLR components could be clearly separated. The results are presented in Figure \ref{baldecn}. We find that the NLR Balmer decrements cover a similar range to those measured for the BLR (and BLR + NLR combined), with some objects showing evidence for a high degree of reddening, but others showing no significant reddening. We have compared the distributions of the narrow emission line Balmer decrements for the 2MASS objects and unobscured type 1 AGN using a K-S test and find that we cannot reject the null hypothesis that these samples are drawn from the same parent population, because the test gives a high p value: 0.099. Note that small number statistics preclude a detailed comparison of the NLR Balmer decrements between the 2MASS and PG/SDSS samples. Also, in the case of the NLR it may not be valid to assume Case B recombination for the intrinsic reddened Balmer decrement: our detailed analysis of J1131+16 \citep{rose} demonstrates that collisional excitation of the H$\alpha$ emission line in the partially ionised zone can significantly enhance the H$\alpha$/H$\beta$ ratio if the gas densities are high (e.g. if part of the NLR emission originates in the circum-nuclear torus). 

\subsection{Outflows}

\begin{figure}
\centering
\includegraphics[scale=0.45, trim=0cm 7cm 0cm 8cm]{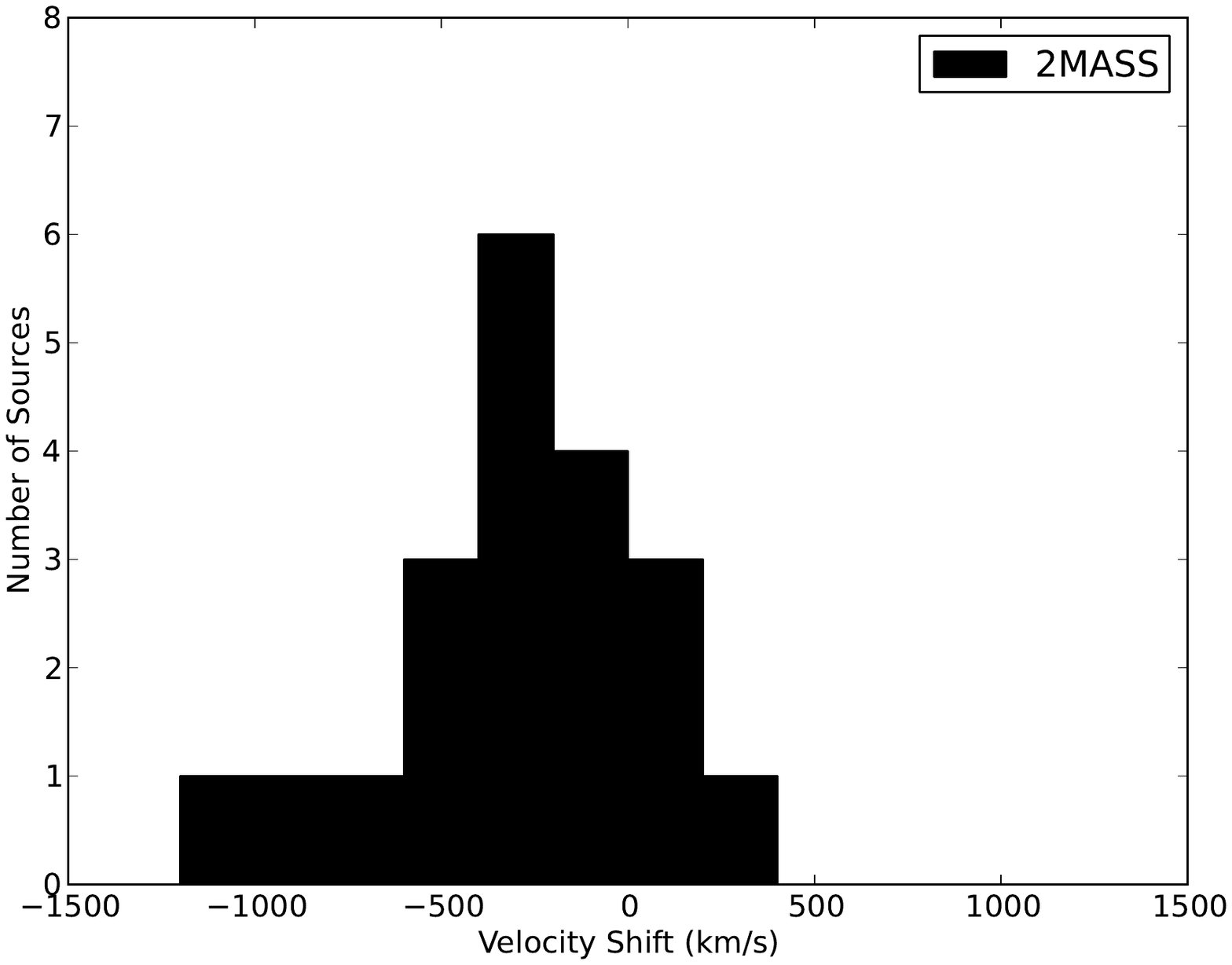}
\includegraphics[scale=0.45, trim=0cm 7cm 0cm 8cm]{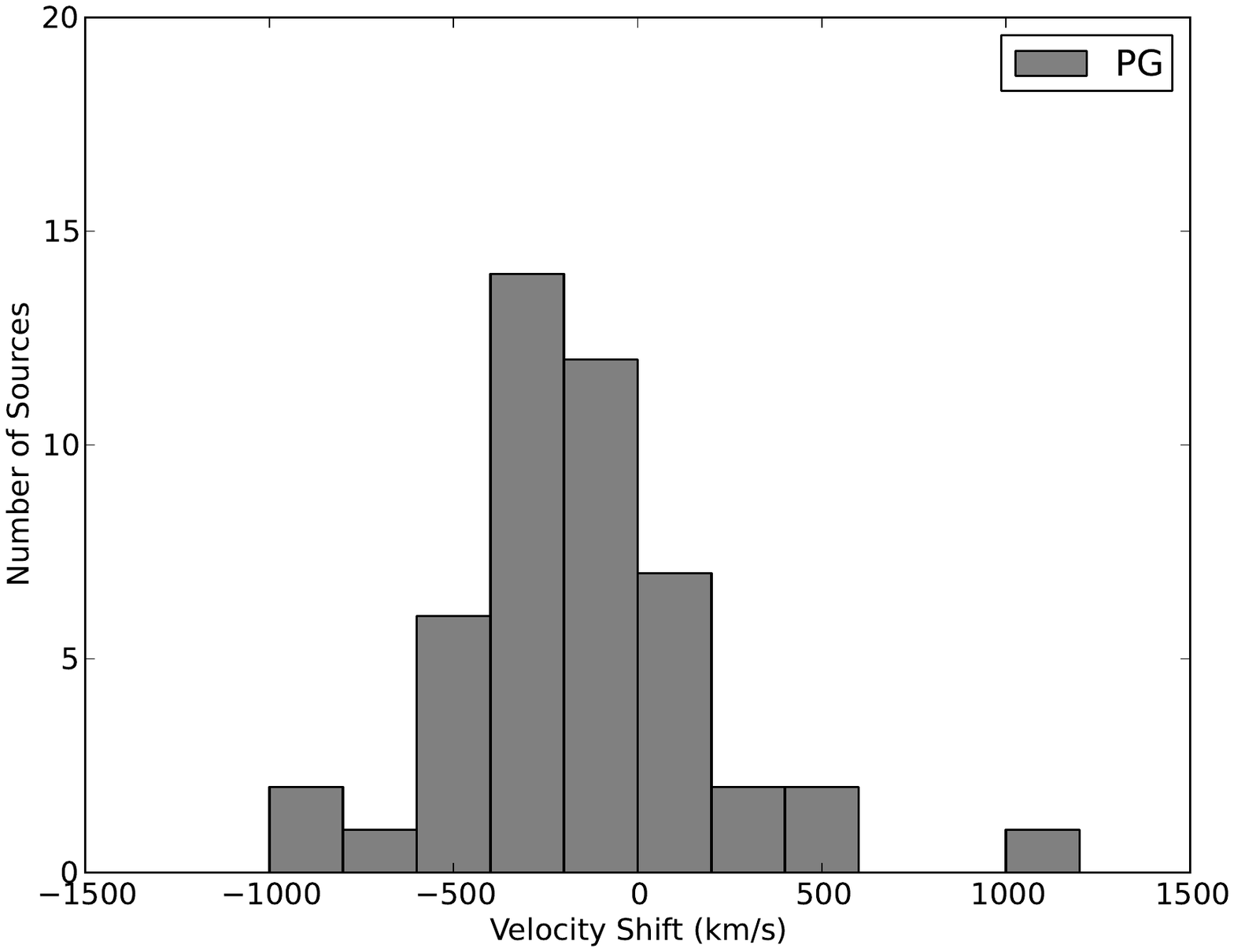}
\includegraphics[scale=0.45, trim=0cm 7cm 0cm 8cm]{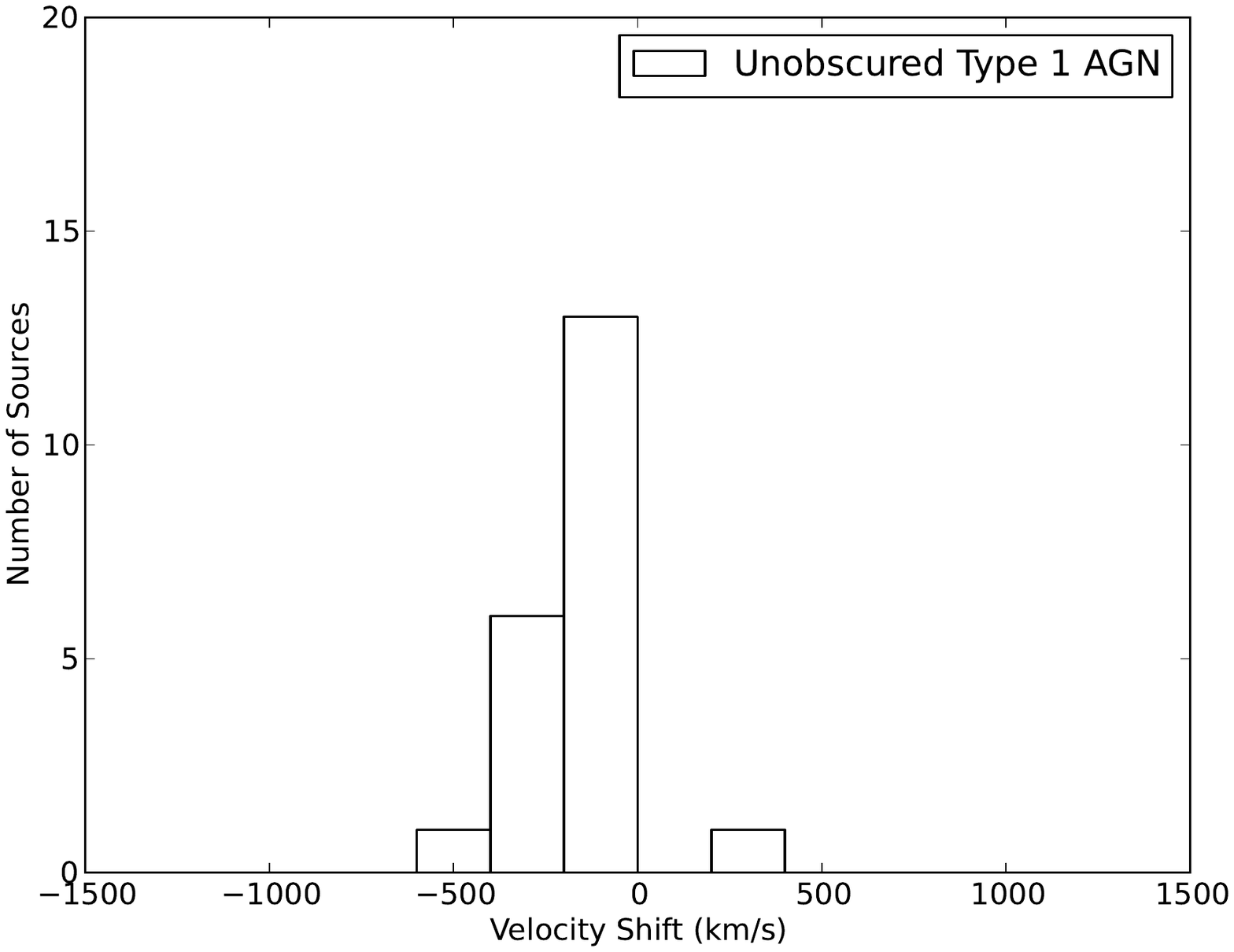}
\caption{Histogram of the shift of the broad [OIII]$\lambda$5007 components relative to the narrow components, as measured using {\sc dipso}. The 2MASS measurements are indicated by the black bars (top), the PG quasar population by the grey bars (middle) and the unobscured type 1 AGN by the open bars (bottom).}
\label{outflows}
\end{figure}

\begin{figure}
\centering
\includegraphics[scale=0.45, trim=0cm 7cm 0cm 8cm]{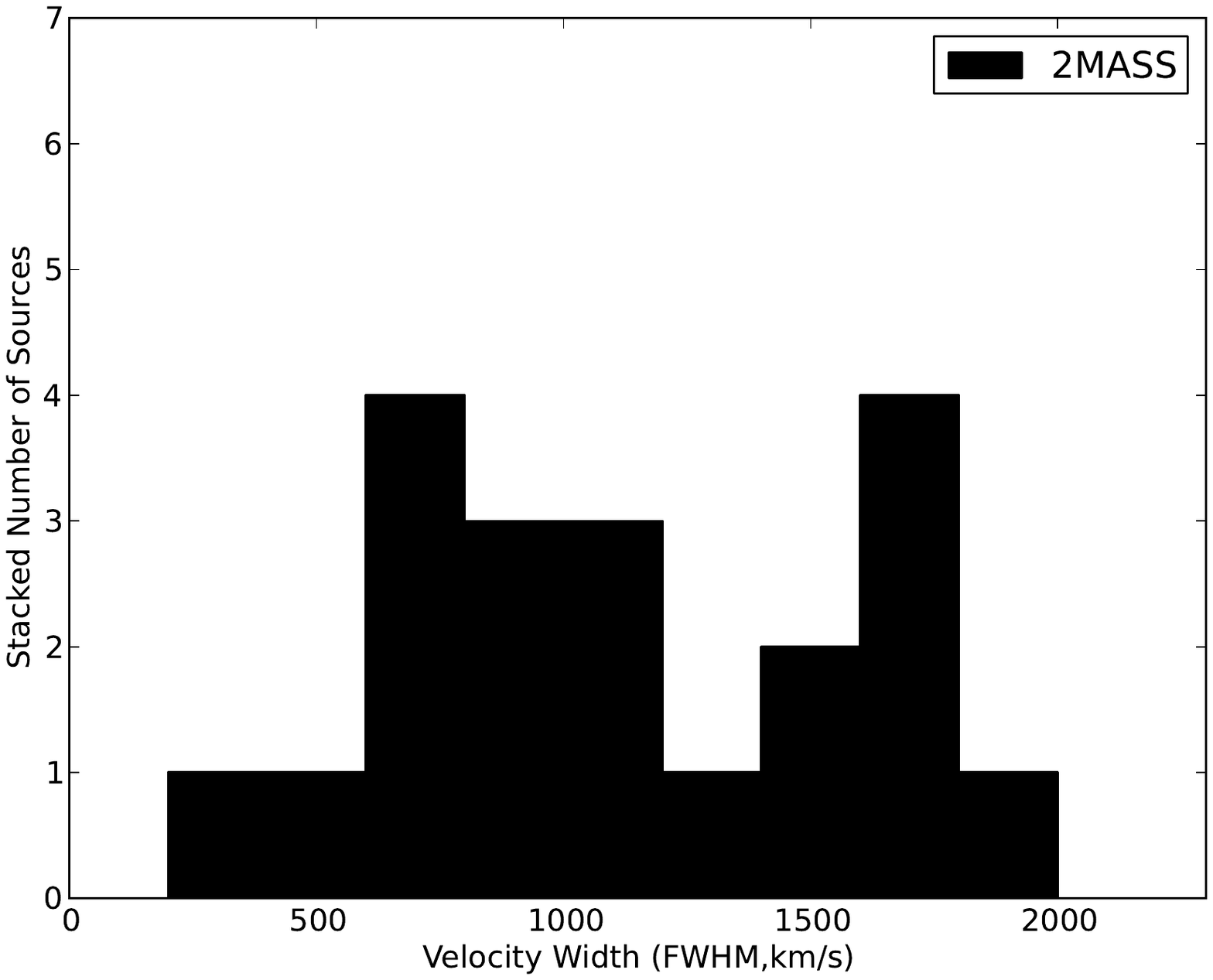}
\includegraphics[scale=0.45, trim=0cm 7cm 0cm 8cm]{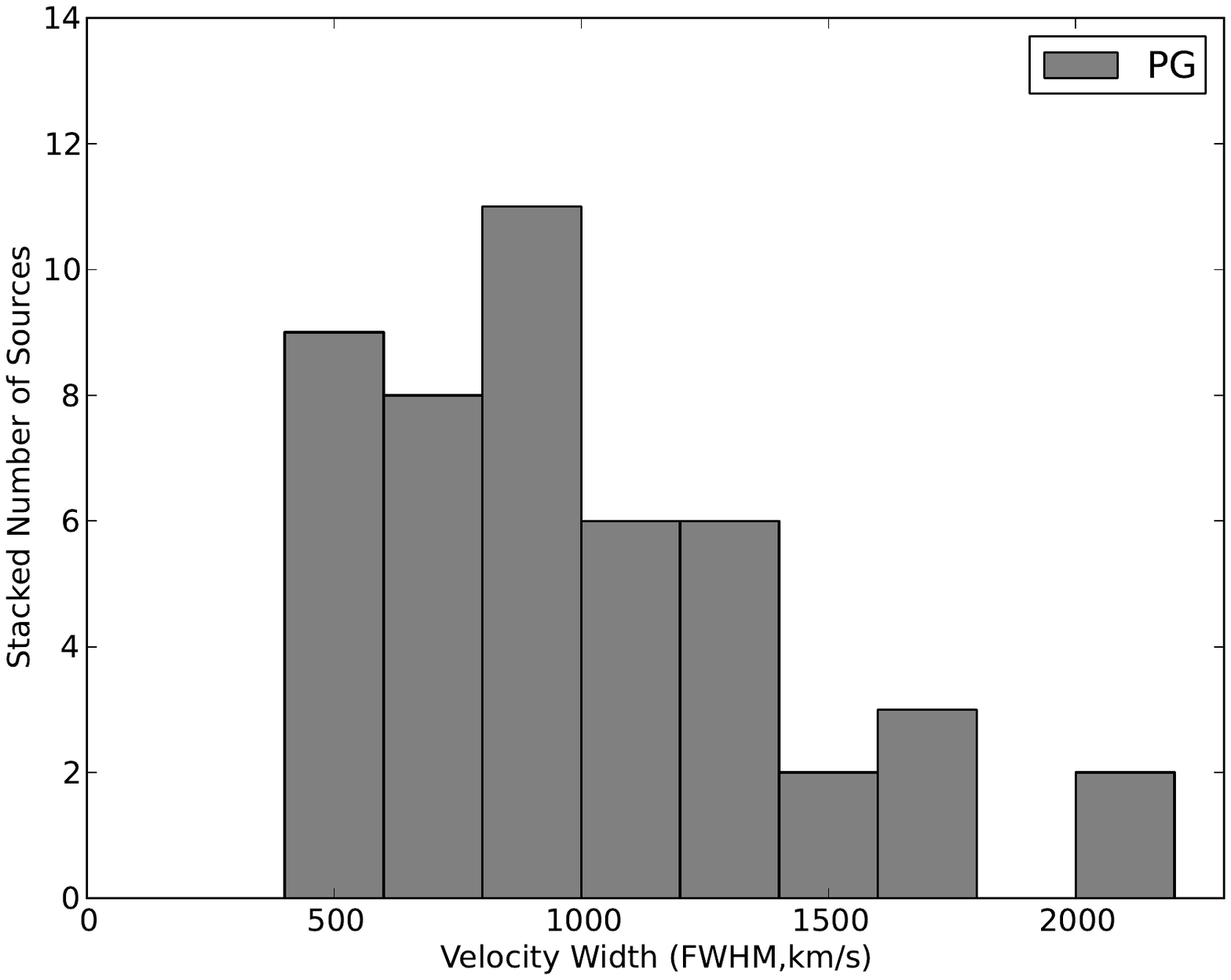}
\includegraphics[scale=0.45, trim=0cm 7cm 0cm 8cm]{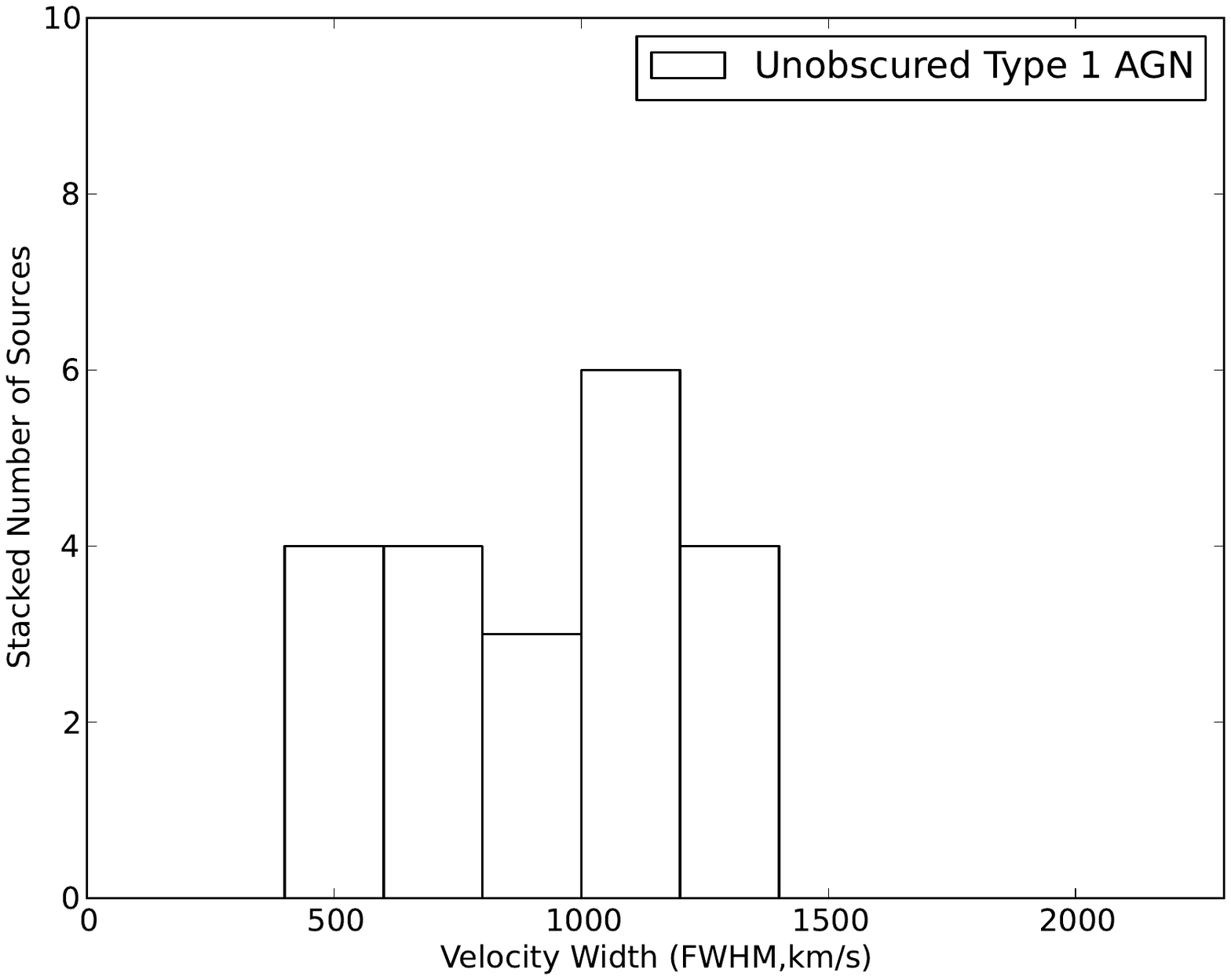}
\caption{Histogram of the instrumentally corrected velocity widths (FWHM, km s$^{-1}$) of the broad [OIII]$\lambda$5007 component, as measured using {\sc dipso}. The 2MASS measurements are indicated by the black bars (top), the PG quasar population by the grey bars (middle) and the unobscured type 1 AGN by the open bars (bottom).}
\label{fwhmhist}
\end{figure}

\begin{figure}
\centering
\includegraphics[scale=0.38, trim=0cm 4cm 0cm 0cm]{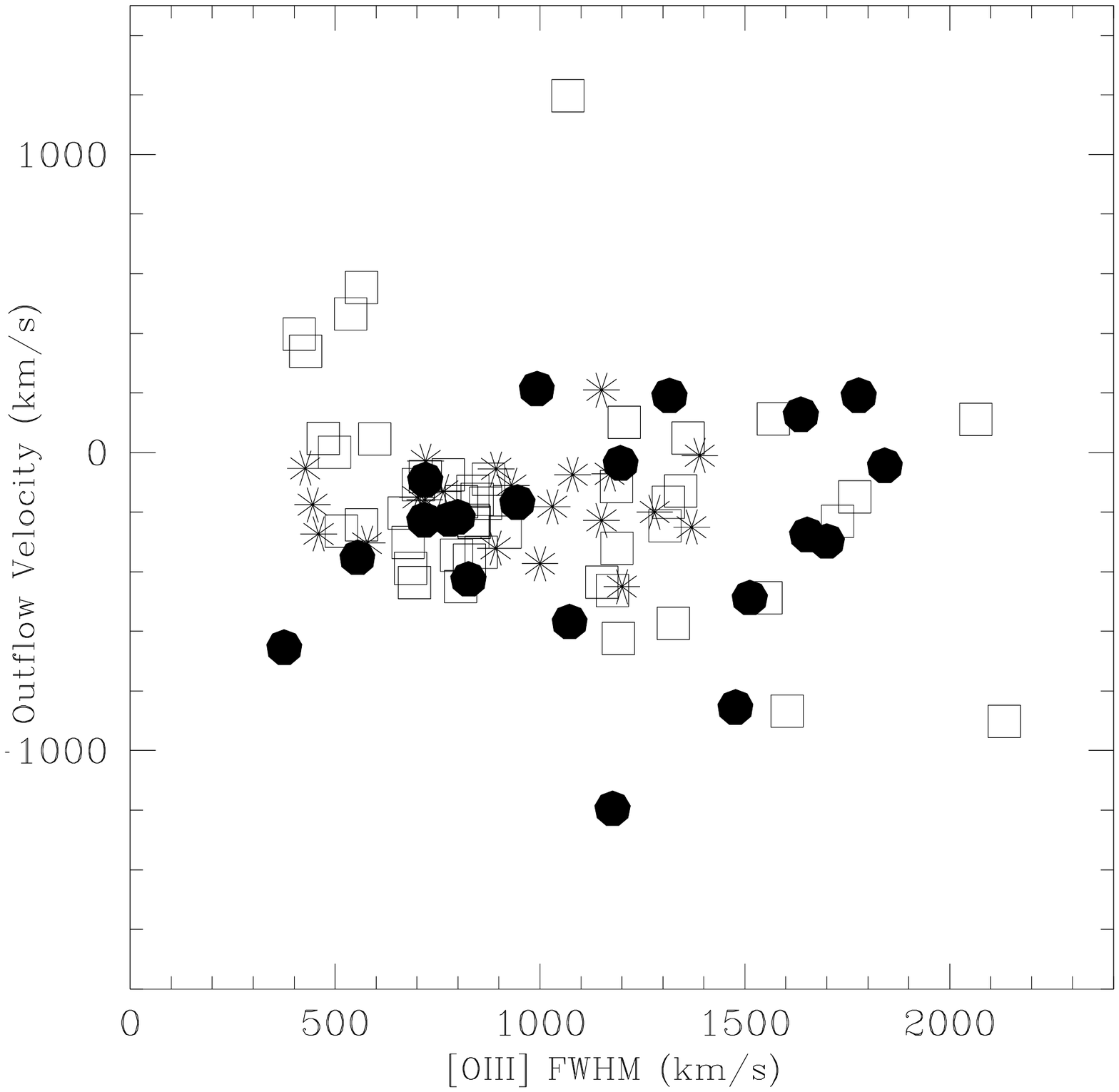}
\caption{Outflow velocity (km s$^{-1}$) plotted against the velocity width at FWHM (km s$^{-1}$) of the broad [OIII]$\lambda$5007 components as measured by {\sc dipso}. The 2MASS population measurements are indicated by the filled circles, the PG quasar population are indicated by the unfilled squares and the unobscured type 1 AGN are indicated by asterisks.}
\label{out}
\end{figure}

If the red 2MASS objects truly represent young AGN enshrouded in a cocoon of dust and gas, then we might expect them to show strong evidence for outflows in their spectra, when compared to other types of active galaxies. For example, in models of major galaxy mergers the AGN are triggered as gas and dust is driven by the tidal torques associated with the mergers into the nuclear regions of the galaxies; the AGN accrete at high Eddington rates, and then drive powerful outflows, blowing away the enshrouding dust and gas (\citealt{hopkins05}; \citealt{narayanan06}; \citealt{hopkins}). 

Figure \ref{outflows} shows a histogram of the rest-frame velocity shifts of the broad [OIII] components (where detected) relative to those of the narrower components, as calculated from the parameters given in Table \ref{models}. The 2MASS objects are plotted along with the PG quasars and the unobscured type 1 AGN for comparison. For the unobscured type 1 AGN, we used the online data tables from \citet{jina} to determine the outflow properties. For the 2MASS objects, PG quasars and unobscured type 1 AGN, we assume that the narrow component of the [OIII] emission represents the host galaxy rest frame emission, and the broader-line emission component the outflow \citep{holt08}. 

From Figure \ref{outflows} it is clear that there are no significant differences between the distributions of the velocity shifts of the samples and, the broad components are predominantly blue-shifted in all three samples, with only a few exceptions. We find the median velocity shifts of -240$\pm$80 km s$^{-1}$, \newline -230$\pm$50 km s$^{-1}$, and -160$\pm$30 km s$^{-1}$ for the 2MASS, PG and unobscured type I samples respectively. We have tested the significance of any difference between the distribution of velocity shifts of the samples using a 1D K-S test. Between the 2MASS objects and PG quasars we find a p value=0.731, and between the 2MASS objects and unobscured type 1 objects we find p value=0.230. Therefore we cannot confidently reject the null hypothesis that these samples are drawn from the same parent population. Note that, although there are no significant differences between the three samples, the most extreme object in all three samples in terms of its shift is a red 2MASS object: J1158-30. This object might be a good candidate for a young AGN.

In addition, in Figure \ref{fwhmhist} we plot a histogram of the rest-frame velocity widths (FWHM) of the broad [OIII] components. As with the velocity shifts, the widths (FWHM) of all three samples have a comparable range. Again, we have tested for any significant differences between the distribution of velocity widths (FWHM) of the samples using a 1D K-S test. Between the 2MASS objects and PG quasars we find a p value=0.385, and a p value=0.089 for the unobscured type 1 AGN. Therefore we cannot reject the null hypothesis that the samples are drawn from the same parent population when comparing the linewidths of the PG quasars and unobscured type 1 AGN with the 2MASS sample.

Figure \ref{out} shows the outflow velocity plotted against the velocity width (FWHM) of the broad [OIII]$\lambda$5007 component. We present this Figure to further illustrate that there are no significant differences between the [OIII] emission line kinematics of the samples. We have also tested for any significant difference between the populations using a 2D K-S test (\citealt{peacock}; \citealt{ff}). Again, we find no evidence for significant differences between the samples (p $>$ 0.05).

One important caveat when considering these results is that the emission line kinematics may be affected by the orientation of the AGN. One possibility is that our line of sight to the type 2 AGN is at a larger angle to the torus axis than type 1 AGN (as suggested by the orientation-based unified schemes). Therefore due to projection effects the measured blueshifts of the broad components might be lower in the type 2 than the type 1 objects. Alternatively, it is possible that the torus obscures the narrow line 
outflows from our direct line of sight in the type 2 objects. However, comparisons between the NLR kinematics of type 1 and 2 AGN show no clear differences (e.g. \citealt{whittle85}; \citealt{nelson95}).

\section{Black Hole and Accretion Properties}

Further insights into the nature of the red 2MASS objects can be obtained by studying the SMBH and accretion properties of the sample. Young, dust enshrouded objects are expected to have higher than average accretion rates, because there is more gas available for accretion (e.g. see \citealt{dimatteo}).

To determine the accretion properties, we must first determine the black hole masses of the objects. In this section we use the broad Balmer emission line properties to determine virial black hole masses and Eddington ratios for the 2MASS sample, the PG sample, and the unobscured type 1 AGN\footnote{For consistency, we use the H$\beta$ luminosities and velocity widths (FWHM) provided in \citealt{jina} to determine the black hole masses, bolometric and Eddington luminosities of the unobscured type 1 AGN sample (L$_{H\beta}$ and FWHM$_{H\beta}$ are flux weighted means of the intermediate and broad components of the H$\beta$ emission lines). Note that, the data used from \citealt{jina} is not corrected for extinction. This is reasonable as the sample is selected to be unreddened.}. Any significant differences in the Eddington ratios between the samples could indicate differences in their evolutionary stage.

To determine virial black hole masses, we have used the relationships between the Balmer emission and black hole mass from \citet{greene&ho}, which are based on correlations between the AGN bolometric luminosities (L$_{BOL}$), Balmer line luminosities, and line widths (FWHM) (see \citealt{greene&ho}, equations 6 and 7). We use equation 7 for most cases where we can use the broad H$\beta$ emission lines, minimising any degeneracies due to blending with other emission lines (e.g. [NII] with H$\alpha$). In cases where the H$\beta$ broad emission is not detectable, we used the H$\alpha$ broad emission with equation 6 instead. 

We fitted the broad H$\beta$ emission with the minimum number of Gaussian components required to obtain an adequate fit to the emission from the BLR, and took a flux weighted mean of the FWHM, consistent with the method of \citet{greene&ho}. Where possible, we corrected the BLR fluxes for extinction, and in cases where we could not confidently separate the broad and narrow emission, we used the total Balmer decrement measurements for the extinction correction. Note that, while our approach is consistent with that outlined in \citet{greene&ho} when measuring the broad H$\beta$ emission, problems with the separation of the broad and narrow emission in the objects where we could not separate the components may potentially have an affect on the line width measurements. Also, for the cases where we estimated an upper limit on the broad H$\beta$ flux, we used the lower limiting estimate of the extinction to find a lower limit on the SMBH mass, and therefore an upper limit on the Eddington ratio. The 2MASS objects have virial SMBH masses in the range 7.2$<$ log$_{10}$[M$_{BH}$]$<$9.2 M$_{\sun}$. 

As well as making virial SMBH estimates based on the broad emission line properties, for comparison we have also used the absolute R-band magnitudes (M$_{R}$) of the host galaxies of the 2MASS sample taken from \citet{hutchings03}. These estimates have the advantage of being free from any degeneracies caused by the blending of the emission lines from the BLR and the NLRs of the sample. In addition, masses can be estimated for all the sources, rather than just the sources with an observable BLR. To make these estimates we used the relation given in \citet{md}, which takes advantage of the correlation between black hole mass and host galaxy bulge luminosity, to determine the SMBH masses.

\begin{figure}
\centering
\includegraphics[scale=0.38, trim=0cm 4cm 0cm 0cm]{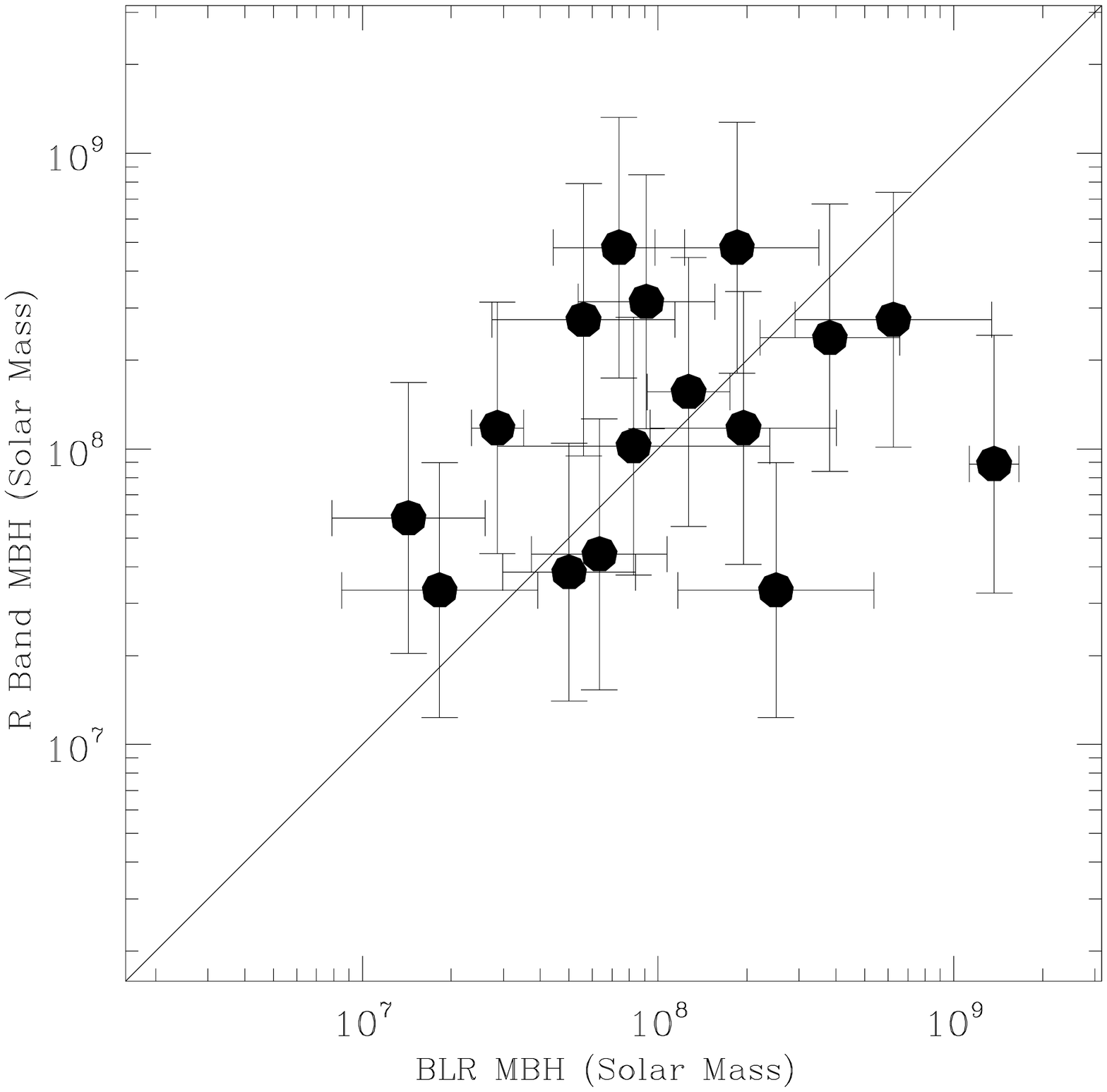}
\caption{A comparison of the SMBH mass estimates for the 2MASS AGN. We plot the mass estimates based on the the R-band magnitudes against the virial mass estimates based on the broad Balmer emission lines. In addition, we have plotted a line showing a one to one relationship for the black hole masses. The mass estimates obtained from the R-band magnitudes are higher on average than those obtained from the BLR measurements.}
\label{bhmcomp}
\end{figure}

Figure \ref{bhmcomp} shows the SMBH mass estimates from the R-band magnitudes plotted against the masses calculated using the properties of the BLR. In addition, we have plotted a line showing a one-to-one relationship between the black hole masses. While some objects show good agreement between the two estimates, there is a large scatter which is reflected in the uncertainty of the median ratio (M$_{Virial}^{SMBH}$/M$_{R-band}^{SMBH}$): 0.81$\pm$0.81. In cases where the virial mass estimates are lower than the R-band estimates, orientation/geometric and/or reddening effects could potentially lead to the Balmer fluxes, velocity widths and SMBH masses being underestimated. Alternatively the R-band measurements may be contaminated by light from the disks of the host galaxies, leading to over-estimates of the SMBH masses. Certainly, at least one object in our sample -- J1131+16 -- has a strong disk component \citep{rose}. However, in cases where the virial masses are higher, the flux of the BLR may be overestimated because of an inaccurate measurement in the broad H$\alpha$ flux due to blending with the narrow H$\alpha$+[NII] lines, which could lead to an overestimation of the level of extinction.

\begin{figure}
\centering
\includegraphics[scale=0.38, trim=0cm 4cm 0cm 0cm]{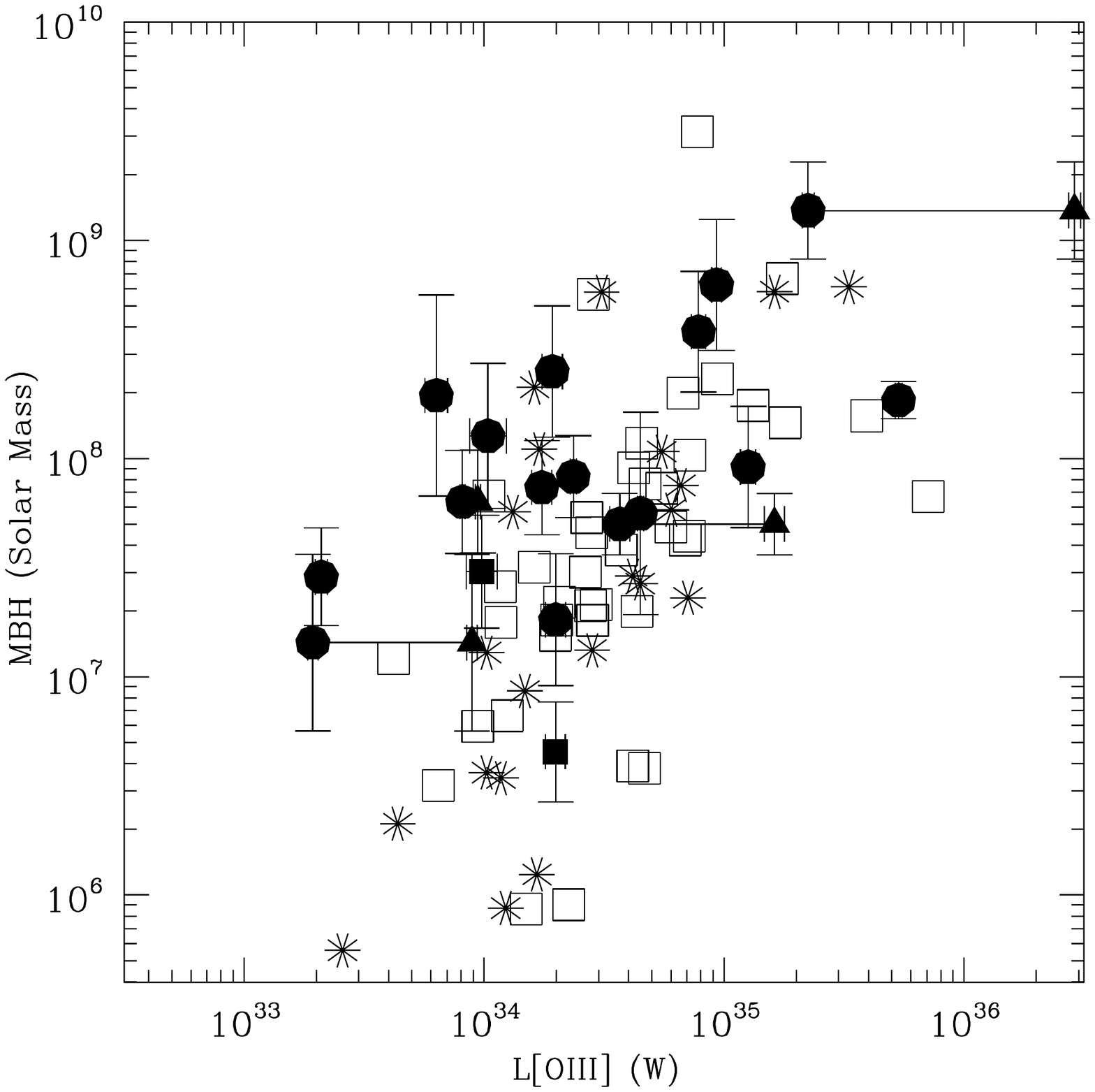}
\caption{Virial black hole mass (M$_{\sun}$) plotted against L$_{[OIII]}$ (W). The filled circles and triangles represent the 2MASS sample. Circles represent black hole masses estimated using H$\beta$ emission and filled squares are based on the H$\alpha$ emission. Where possible, we correct L$_{[OIII]}$ for reddening, these estimates are represented by filled triangles and are joined to their uncorrected counterparts by a straight line. The unfilled squares represent the black hole mass estimates of the PG sample and the asterisks indicate the type 1 AGN estimates.}
\label{bhmass}
\end{figure}

\begin{table*}
\caption{Virial SMBH masses and Eddington ratio results for the 2MASS sample. Values presented in brackets are for those objects where L$_{BOL}$ has been corrected for dust extinction.}
\begin{tabular}{lccc}
\hline
Name 	& M$_{R}$  &	Virial	 &	(L$_{BOL}$ /L$_{EDD}$)	\\
	& Log$_{10}$[M$_{SMBH}$] (M$\sun$) & Log$_{10}$[M$_{SMBH}$] (M$\sun$)	 &		\\
\hline
J0221+13	&	8.19	$\pm$	0.43	&	-			&	-			\\
J0248+14	&	8.07	$\pm$	0.44	&	7.46	$\pm$	0.09	&	0.020	$\pm$	0.005	\\
J0306-05	&	8.25	$\pm$	0.43	&	-			&	-			\\
J0312+07	&	8.07	$\pm$	0.44	&	8.29	$\pm$	0.32	&	0.009	$\pm$	0.001	\\
J0400+05$^a$	&	7.95	$\pm$	0.44	&	9.14	$\pm$	0.08	&	0.056(0.585)	$\pm$	0.009(0.125)	\\
J0409+07$^a$	&	7.58	$\pm$	0.46	&	7.70	$\pm$	0.05	&	0.204(0.901)	$\pm$	0.025(0.112)	\\
J0422-18	&	8.01	$\pm$	0.44	&	7.92	$\pm$	0.11	&	0.079	$\pm$	0.024	\\
J0435-06	&	8.68	$\pm$	0.41	&	7.87	$\pm$	0.26	&	0.065	$\pm$	0.003	\\
J0447-16	&	8.19	$\pm$	0.43	&	8.10	$\pm$	0.11	&	0.023	$\pm$	0.007	\\
J0504-19	&	8.01	$\pm$	0.44	&	-			&	-			\\
J0910+33	&	7.83	$\pm$	0.45	&	-			&	-			\\
J1006+41$^b$	&	7.52	$\pm$	0.46	&	7.26	$\pm$	0.10	&	0.302	$\pm$	0.105	\\
J1014+19	&	8.44	$\pm$	0.42	&	8.80	$\pm$	0.13	&	0.041	$\pm$	0.014	\\
J1040+59$^b$	&	7.52	$\pm$	0.46	&	8.40	$\pm$	0.11	&	0.021	$\pm$	0.007	\\
J1057-13	&	7.77	$\pm$	0.45	&	-			&	-			\\
J1127+24$^a$	&	7.64	$\pm$	0.45	&	7.80	$\pm$	0.10	&	0.036(0.041)	$\pm$	0.019(0.0010)	\\
J1131+16	&	8.31	$\pm$	0.43	&	-			&	-			\\
J1158-30	&	8.25	$\pm$	0.43	&	-			&	-			\\
J1212-14$^a$	&	7.77	$\pm$	0.45	&	7.16	$\pm$	0.18	&	0.037(0.172)	$\pm$	0.020(0.090)	\\
J1321+13	&	7.52	$\pm$	0.46	&	-			&	-			\\
J1323-02	&	8.38	$\pm$	0.42	&	8.58	$\pm$	0.16	&	0.057	$\pm$	0.026	\\
J1338-04	&	8.50	$\pm$	0.42	&	7.96	$\pm$	0.26	&	0.383	$\pm$	0.025	\\
J1407+42	&	7.95	$\pm$	0.44	&	-			&	-			\\
J1448+44	&	8.44	$\pm$	0.42	&	7.75	$\pm$	0.21	&	0.222	$\pm$	0.137	\\
J2124-17	&	8.68	$\pm$	0.41	&	8.27	$\pm$	0.04	&	0.802	$\pm$	0.068	\\
\hline
\label{rattab}
\end{tabular}
\begin{tablenotes}
    \item[a]$^a$ The NLR of this object has been corrected for dust extinction.
    \item[b]$^b$ M$_{SMBH}$ and therefore L$_{EDD}$ are calculated using broad H$\alpha$ emission as opposed to H$\beta$.
     \end{tablenotes}
\end{table*}

Figure \ref{bhmass} shows the virial black hole masses plotted against L$_{[OIII]}$ for the 2MASS and comparison samples, where the uncertainties are only plotted for the 2MASS objects for clarity. We do not plot the R-band black hole mass estimates because we do not have R-band magnitudes for the comparison samples, and we want to present the comparisons between mass estimates which have been determined via the same method. In addition, Table \ref{rattab} shows the individual SMBH masses of the 2MASS sample. Overall, from this figure it is clear that the black hole masses of the PG quasars (unfilled squares) and unobscured type 1 AGN (asterisks) are generally lower than the 2MASS objects (filled symbols) of equivalent [OIII] luminosity. Potentially, this difference can be explained in terms of extinction effects: in the cases of the few objects with virial mass estimates for which we have been able to correct L$_{[OIII]}$ for extinction in the NLR, the extinction correction moves the points onto the main correlation defined by the PG quasars and unobscured type 1 objects. 

Alternatively, the black hole masses of the 2MASS objects could be genuinely higher than those of the PG quasars and unobscured type 1 AGN of similar intrinsic power. In this context we note that the recent study of \citet{canalizo12} measured black hole masses and host galaxy properties for 29 red 2MASS AGN with 0.17 $<$ z $<$ 0.37, finding that the majority of red 2MASS AGN also have significantly more massive black hole masses than the masses of black holes in local AGN. However, we remain cautious in our interpretation of the derived black hole masses for our 2MASS sample because of the uncertainty surrounding the extinction corrections.

In terms of estimating the Eddington ratio (L$_{BOL}$/L$_{EDD}$), we used L$_{[OIII]}$ to calculate the bolometric luminosities of the AGN, rather than the AGN continuum, because the [OIII] emission line is well detected in all the objects, whereas the AGN continuum emission is subject to reddening effects and potential contamination by the host stellar continuum. L$_{BOL}$ was determined from L$_{[OIII]}$ using the relationship L$_{BOL}$=3500L$_{[OIII]}$ \citep{heckman04}. 

\begin{figure}
\centering
\includegraphics[scale=0.38, trim=0cm 4cm 0cm 0cm]{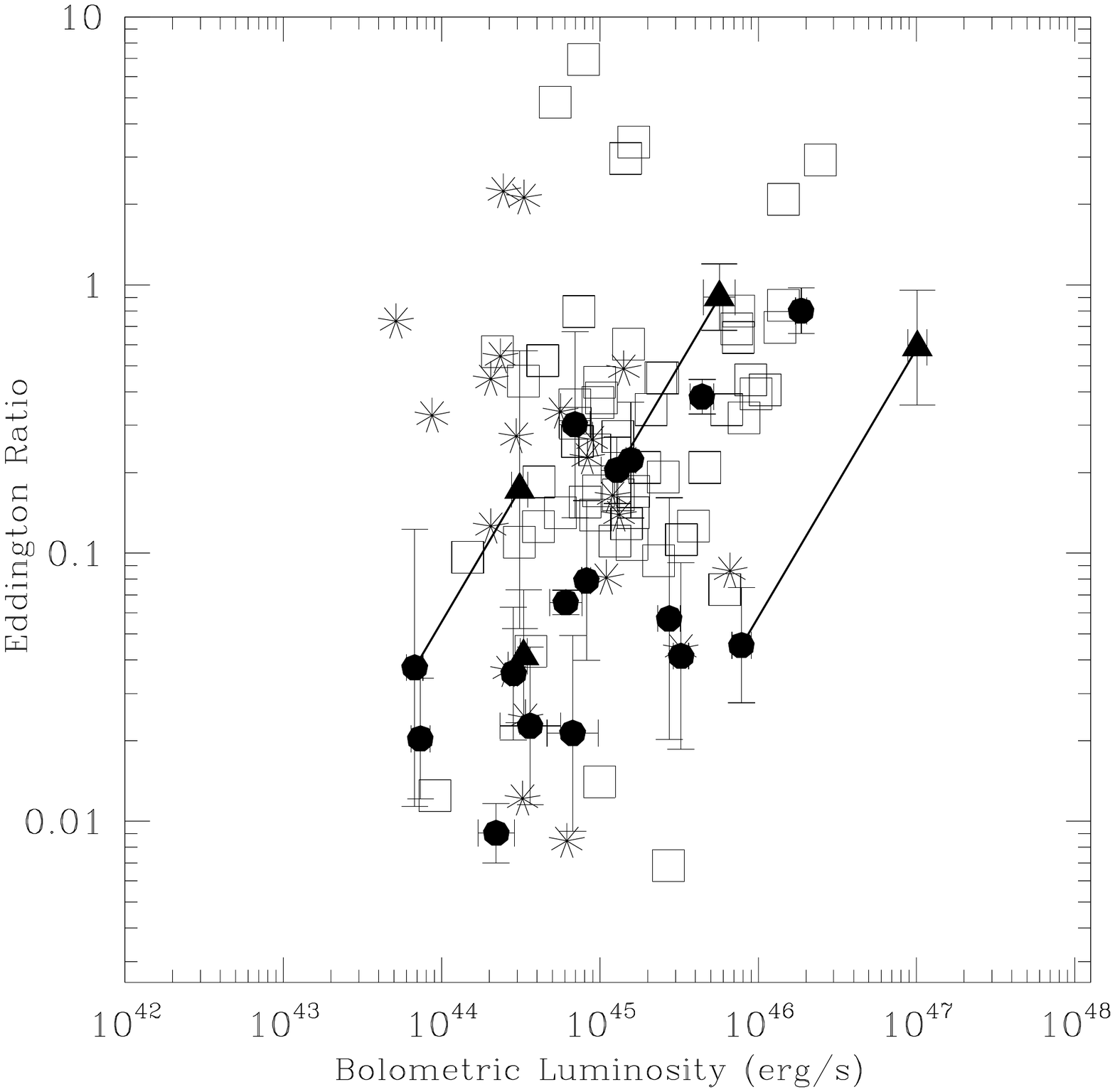}
\caption{Eddington ratio plotted against L$_{BOL}$ (erg s$^{-1}$) based on measurements of L$_{[OIII]}$ and virial black hole mass estimates. Filled circles represent the 2MASS objects where L$_{[OIII]}$ has not been corrected for dust extinction. For objects which could be corrected for dust extinction, we represent them with filled triangles and join them to their respective estimates where no correction has been made. The unfilled squares represent the PG quasars and the asterisks represent the unobscured type 1 objects.}
\label{loiiiedd}
\end{figure}

Figure \ref{loiiiedd} shows the Eddington ratios plotted against L$_{BOL}$. Once again the uncertainties are only plotted for the 2MASS objects for clarity. We present Eddington ratios for all possible objects in the 2MASS sample without correcting [OIII] for dust extinction (filled circles) and, where possible, we also show the Eddington ratios following the correction of the [OIII] luminosity for dust extinction (filled triangles), connecting both sets of estimates with a straight line. It is clear that, for the 4 objects in which we can measure reddening accurately, reddening has a significant effect on the Eddington ratio. There is an obvious, large scatter in Eddington ratios for all the samples: the 2MASS objects have ratios in the range: 0.009 $<$ L$_{BOL}$/L$_{EDD}$ $<$ 0.91. The median ratio for the 2MASS sample is 0.05$\pm$0.05, without correcting [OIII] for reddening and 0.08$\pm$0.04 in the reddening corrected case. In comparison, the PG quasar sample (unfilled squares) and the unobscured type 1 AGN sample (asterisks) have median ratios of 0.22$\pm$0.12 and 0.23$\pm$0.14 respectively; the large uncertainties reflect the large scatter in Eddington ratio. These ratios are all consistent within 2$\sigma$ of the uncertainties. Note that, if we could correct for NLR reddening in {\it all} the objects, the median ratio for the 2MASS objects would likely increase to a value closer to that measured for the PG quasar and unobscured type 1 AGN samples. Clearly, there is no evidence that the 2MASS objects have Eddington ratios that are significantly higher than those of the comparison samples.

\section{Discussion}

\subsection{Moderately Reddened Objects}

A key result from this study is that many of the red 2MASS AGN have moderate amounts of dust extinction when compared to the comparison samples (0.2 $<$ E(B-V) $<$ 1.2); the reddening values we measure are consistent with those determined for other samples of red 2MASS AGN (0.1 $<$ E(B-V) $<$ 3.2: Glikman et al. 2004, 2007, 2011 \& 2012; Shi et al. 2007; Urrutia et al. 2008, 2009 \& 2012; Georgakakis et al. 2009; Kuraszkiewicz et al. 2009a,b; Canalizo et al. 2012). Therefore it is possible that the 2MASS AGN have red J-K$_S$ colours because dust extinction reduces the J-band flux relative to the K-band flux such that the F$_K$/F$_J$ flux ratio is higher than a typical AGN. In Paper II we show that for an AGN with intrinsic near-IR colours that are typical of the PG quasars and the unobscured type 1 AGN (J-K$_S$$\sim$1.70), we require only a moderate level of reddening of E(B-V) $>$ 0.8 in order to give J-K$_S$ $>$ 2.0 --- this level of extinction is measured for 29\% of the 2MASS AGN with broad Balmer emission lines. 

Rather than the red near-IR colours being a result of dust extinction from a natal cocoon of dust, it is possible that the obscuring dust is located in the outer layers of the torus, and that we are observing these objects at an angle where our line of sight grazes the edge of the torus, leading to the observed red colours. However, such modest levels of reddening could also be produced by dust on a kpc-scale in the disks of the host galaxies.                                                                                                                                                                                                                                                 . 

Alternatively, rather than moderately extinguished AGN, it is possible that some of these objects are genuine type 2 objects in which the torus completely extinguishes the broad-line AGN in the optical and shorter wavelength near-IR bands. In this case, the level of extinction is expected to be much higher (A$_V$ $>$ 10) than that indicated by the Balmer decrements in the NLR because, while the SMBH and BLR are completely obscured at optical wavelengths by the torus, the NLR extends beyond the scale of the torus and is therefore relatively unaffected. In this case, due to the much higher levels of extinction caused by the torus, their J-K$_S$ colours should be much redder than is observed. Indeed, from our work in Paper II, we find that A$_v$ $>$ 5 would produce J-K$_S$ $>$ 3.00, a level only found for the type 2 2MASS object J1307+23 (excluded from the main study of this paper), but not for any of the other type 2 objects in our sample. However, the integrated near-IR colours of the heavily extinguished type 2 objects could be made bluer if the host galaxies make a significant contribution in the near-IR, because the colours of typical unreddened stellar populations are relatively blue at such wavelengths.

Finally, we note that dust extinction is not the only explanation for the red near-IR colours of the 2MASS objects. Some objects in our sample show relatively modest levels of extinction based on their BLR Balmer decrements, and have optical continuum shapes that appear similar to UV/optical selected AGN (e.g. J2124-17). Rather than moderate levels of dust extinction in such cases, a relatively large covering factor for the hot dust (T $\sim$ 1500K) could also result in red J-K$_S$ colours. Emission from hot dust in the torus becomes significant at wavelengths which coincide with the K-band. Therefore, a larger than average covering factor for this hot dust could increase the flux measured in the K-band relative to that of the J-band (likely dominated by accretion disk emission), thus producing the red J-K$_S$ colours measured for the 2MASS sample.  

All of these possibilities will be explored further in Paper II, where we will model the near- to mid-IR SEDs of the 2MASS objects with varying the level of extinction, dust covering factors and host galaxy contributions.  

\subsection{Young, dust enshrouded objects}

\begin{table*}
\caption{Basic properties of red AGN samples.}
\begin{tabular}{lcccc}
\hline
Sample 	&	Size 	&	Z	&	Data used in Study	&	Key Selection Criteria	\\
\hline
 & & & & \\
 This paper & 29 & 0.06 $<$ z $<$ 0.28 & Optical Spectra & J-K$_S$ $\ga$ 2.0,  \\
 & & & & selected from \citet{hutchings03}\\
 & & & &\\
 Glikman et al. 2004, 2007, & 120 & 0.1 $<$ z $<$ 2.5 & Optical \& IR Spectra & J-K$_S$ $\ge$ 1.7 \& R-K $\ge$ 4.0, \\
 2011 \& 2012  &  &  &  &  selected by cross-correlating the 2MASS \\
  & & & & PSC with the FIRST radio catalogue \\
   & & & &\\
 Georgakakis et al. 2009 & 10 & 0.29 $<$ z $<$ 3.1 & UV-FIR SED fitting & J-K$_S$ $\ge$ 1.5 \& R-K $\ge$ 5.0, \\
  &  &  &  &  selected by cross-correlating the 2MASS\\
 & & & & PSC with SDSS DR3 spectra\\
 & & & & \\
 Urrutia et al. 2008, 2009 & 13 & 0.1 $<$ z $<$ 1.0 & Optical Spectra & Selected from the  \\
 \& 2012& & & & \citet{glikman07} sample\\
 & & & & \\
 Shi et al. 2007 & 25 & 0.08 $<$ z $<$ 0.37 & IR Spectra & J-K$_S$ $\ga$ 2.0,  \\
 & & & & selected from \citet{cutri01}\\
 & & & &\\
Kuraszkiewicz & 44 & 0.07 $<$ z $<$ 0.37 & X-ray-FIR SED fitting,  & J-K$_S$ $>$ 2.0 and B-K$_S$ $>$ 4.3 \\
 et al. 2009a, & & & PCA analysis & \\
 & & & & \\
 Canalizo et al. 2012 & 9 & 0.13 $<$ z $<$ 0.37 & Optical/NIR spectroscopy,  & J-K$_S$ $>$ 2.0 and M$_K$ $\la$ -25 \\
 & & & HST images & \\
 & & & & \\
 Marble et al. 2003 & 29 & 0.13 $<$ z $<$ 0.6 &  HST images & J-K$_S$ $>$ 2.0 and M$_K$ $\la$ -23 \\
 & & & & \\
 Smith et al. 2002 & 20 & 0.06 $<$ z $<$ 0.6 &  Optical broadband   & J-K$_S$ $>$ 2.0 and M$_K$ $\la$ -23 \\
 & & & polarimetry & \\
\hline
\label{discussion}
\end{tabular}
\end{table*}

There have been several studies of red 2MASS AGN in the past that have focussed on different wavelength regions of the SEDs of the population (e.g. X-ray, UV, optical or IR). In this section we outline the basic properties and findings of these studies, and compare them with the results obtained in this paper, in order to determine the nature of our red, low-z, 2MASS AGN sample. The basic properties of the samples used for the previous studies are summarised in Table \ref{discussion}.

Many of these past studies conclude that the red 2MASS quasars are young, dust-enshrouded, transitional objects. This conclusion is
based on several lines of evidence, including a high rate of morphological disturbance of the host galaxies \citep{urrutia}, a relatively high rate of occurrence low ionization broad absorption lines (LoBALs; \citealt{urrutia09}), evidence for enhanced star formation rates compared to typical UV/optical selected AGN (\citealt{shi07}; \citealt{georgakakis}), and high Eddington ratios for 
the gas being accreted by the central super-massive black holes \citep{urrutia12}. 

In contrast, although the results from our local 2MASS AGN sample agree with the these previous studies in the sense that we determine the same moderate levels of reddening, we find no strong evidence that these are young quasars based on their [OIII] outflow properties, the incidence of NLS1, and the distribution of Eddington ratios. Indeed, the properties our 2MASS sample appear similar to those of the samples of UV/optical selected quasars in the local Universe.

The apparent discrepancy between the conclusions of our study and the previous studies that identify the red 2MASS objects as young, transition objects may be due
to different selection criteria.
In particular,  the Glikman et al. (2004, 2007, 2011, 2012), Georgakakis
(2009) and Urrutia et al. (2008, 2009, 2012) samples extend up to much higher
redshifts than our local sample, include optical colour selection criteria as well
as near-IR colour selection, and some require the objects to have broad lines;
in addition, Glikman
et al. (2004, 2007, 2011, 2012) require  detection of the objects 
in the FIRST radio survey. Such selection criteria favour 
more extreme objects that are
more likely to be genuinely young and dust enshrouded.

In this context it is notable that other studies of 2MASS AGN which are
based on lower redshift samples
and/or samples selected without additional optical colour criteria (the
bottom four entries in Table 8), agree with with the conclusions of our
investigation, albeit using different analysis techniques:

\begin{itemize}
\item [-] from X-ray to FIR SED studies, and PCA analysis of a population of red 2MASS AGN, \citet{kuraab} conclude that the red 2MASS AGN are not observed at
a younger evolutionary phase than other local samples; 

\item [-] using I-band HST images to study the host galaxies of red 2MASS AGN, and comparing their findings to the PG quasars and a sample of ULIRGs, \citet{marble03} find no indication that red 2MASS AGN represent young, dust enshrouded, transitional objects; 

\item [-] by comparing optical broad-band polarimetry results of red 2MASS AGN, PG quasars and BAL AGN, \citet{smith02} find that overall 2MASS AGN show relatively high levels of polarization (P $>$ 3\%), consistent with the idea that orientation/obscuration
effects are responsible for the red NIR colours.

\end{itemize}

These comparisons between the previous studies of the red 2MASS AGN emphasise the
sensitivity of the conclusions to the precise selection criteria employed for the particular samples. However,
it is seems unlikely that the majority of low redshift 2MASS AGN 
that are selected solely on the basis of near-IR colour   
[$(J-K) \ge 2$] are young, transitional objects.

\section{Conclusion and Future Work}

In this work we present a study of the optical spectra of a representative sample of 29 nearby (z $<$ 0.28) 2MASS-selected AGN with red near-IR colours (J-K$_S$ $\ga$ 2.0). We compare the results from this sample with those obtained for comparison samples of PG quasars and unobscured type 1 AGN. 
We find the following.
\begin{itemize}

\item Overall the local 2MASS AGN population show significantly higher reddening than samples of UV/optical selected AGN. However, while some objects are highly reddened (E(B-V)=1.2), others show relatively little (if any) reddening.

\item The velocity shifts and widths of the [OIII] lines of the red 2MASS AGN are not significantly larger than those of the comparison samples.

\item The Eddington ratios of the 2MASS AGN and the incidence of NLS1 type objects are comparable with those of `typical' AGN. 

\end{itemize}  

These results do not support the view that the local red 2MASS AGN represent young quasars in a transitional phase, where the AGN are emerging from their natal cocoons of gas and dust. However, they are consistent with the idea that these objects are reddened either by obscuring dust in the outer layers of the circum-nuclear torii, or by dust in the kpc-scale disks of the host galaxies.

In Paper II we will focus on the near-IR properties of this sample of red 2MASS AGN, and investigate the origin of their red J-K$_S$ colours. 

\section*{Acknowledgements}

We would like to thank the referee for useful comments and suggestions. M.R. acknowledges support in the form of an STFC PhD studentship. The authors thank T. Boroson for kindly supplying the spectra of the \citet{boroson} sample. The authors acknowledge the data analysis facilities provided by the
Starlink Project, which is run by CCLRC on behalf of PPARC. The William Herschel Telescope and its service programme are operated on the
island of La Palma by the Isaac Newton Group in the Spanish Observatorio del
Roque de los Muchachos of the Instituto de Astrof\'{i}sica de Canarias. This publication makes use of data products from the Two Micron All Sky Survey, which is a joint project of the University of Massachusetts and the Infrared Processing and Analysis Center/California Institute of Technology, funded by the National Aeronautics and Space Administration and the National Science Foundation. This research has made use of the NASA/IPAC Extragalactic Database (NED) which is operated by the Jet Propulsion Laboratory, California Institute of Technology, under contract with the National Aeronautics and Space Administration. Funding for SDSS-III has been provided by the Alfred P. Sloan Foundation, the Participating Institutions, the National Science Foundation, and the U.S. Department of Energy Office of Science.

\appendix

\section{Individual Properties}

\textbf{J0221+13} {is classified as an intermediate type AGN (type 1.8) in both \citet{smith02} and \citet{kuraab}, however our classification is consistent with the findings of \citet{hutchings03}: a type 2 AGN. Such a difference can be explained by the difference in S/N between the spectra used to identify the AGN type.}

\noindent \textbf{J0306-05:} while we find no evidence of broad line emission in J0306-05, \citet{hutchings03} classify this object as a type 1 AGN.

\noindent \textbf{J0312+07} is a type 1 Seyfert galaxy with J-K$_S$ = 1.98. A companion object is detected on the 2D spectrum at a distance of $\sim$400$\pm$80 kpc from the nucleus of J0312+07. Interestingly, the continuum of this companion appears redder than the continuum of J0312+07. It is possible that the companion is interacting with the host galaxy of J0312+07, implying that J0312+07 may be a young AGN triggered in a recent merger. By measuring the difference in redshift in the H$\alpha$ emission between J0312+07 and the companion galaxy, we have found that the companion is shifted by -240$\pm$20 km s$^{-1}$ relative to the rest-frame of J0312+07. Therefore, we can reasonably argue that J0312+07 is in an interacting system.

\begin{figure}
\centering
\includegraphics[scale=0.3, trim=0cm 4cm 0cm 0cm]{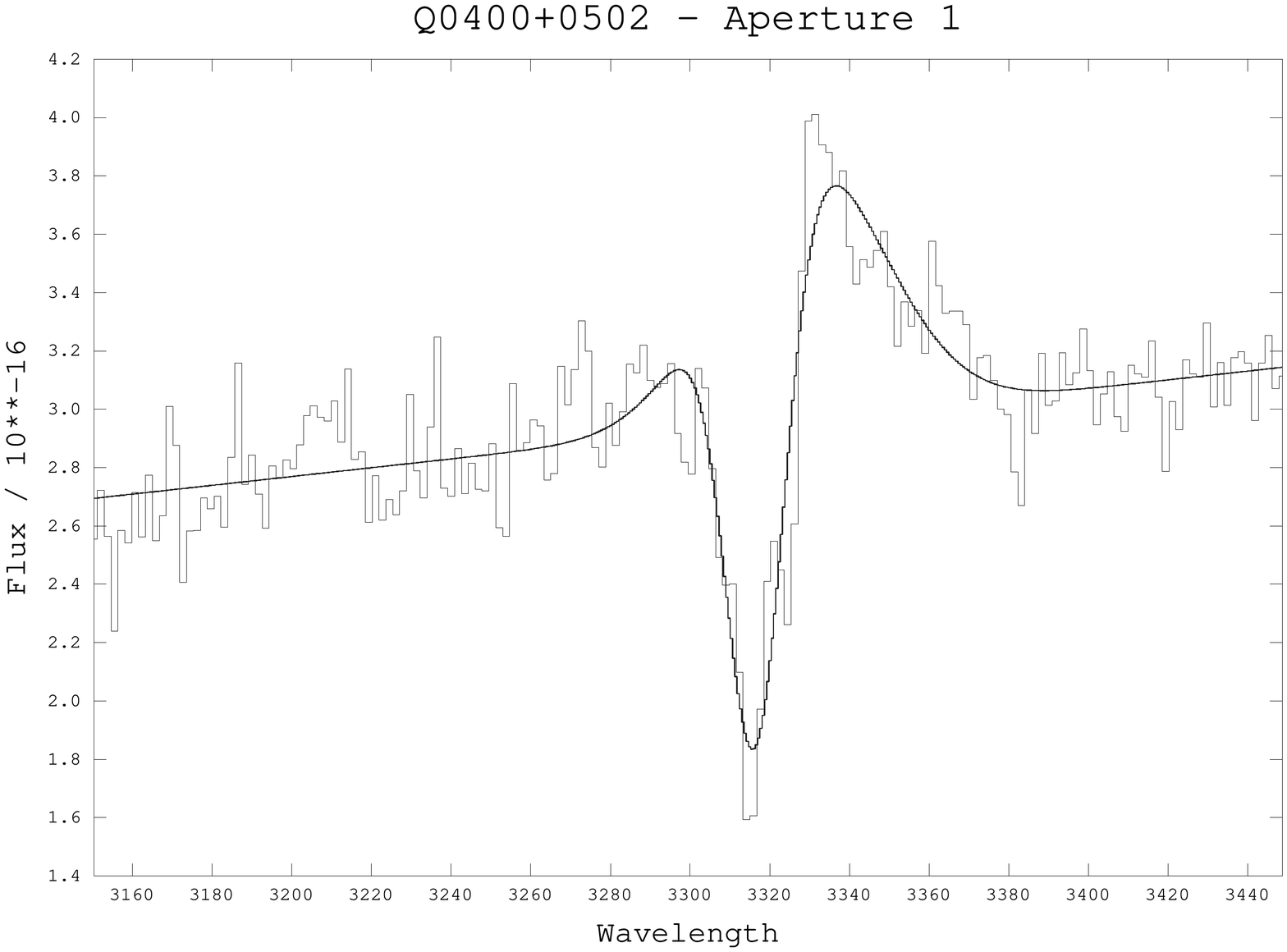}
\caption{The Gaussian fit to the MgII]$\lambda\lambda$2798,2802 absorption and emission features. Note that both the MgII] absorption and emission features overlap, making it difficult to obtain accurate measurements of the absorption/emission line properties.}
\label{lobal}
\end{figure}

\noindent \textbf{J0400+05} is a type 1 quasar with J-K$_S$ = 2.00. J0400+05 is the only object in the 2MASS sample with a detectable absorption feature for MgII]$\lambda\lambda$2796,2802. Figure \ref{lobal} shows the fit to the MgII] absorption feature. From this fit we deduce that the absorption feature is blueshifted by 1200$\pm$160 km s$^{-1}$ relative to the rest-frame of the 0400+05 narrow line emission, and has a velocity width of 1510$\pm$120 km s$^{-1}$ (FWHM). These properties fall short of the LoBAL properties found for the objects in the \citet{urrutia09} and \citet{glikman12} samples because BAL objects are defined to by blueshifted by $\sim$3000--25000 km s$^{-1}$ and have FWHM $>$ 2000 km s$^{−1}$ \citep{weymann}.

As well as absorption lines, the continuum of this object has an unusual shape which includes an unidentified, broad emission feature between 4200 and 5200 \AA\ in the rest-frame (see figure \ref{specs}). In addition, the BLR Balmer emission in J0400+05 is significantly blue shifted by -2870$\pm$160 km s$^{-1}$ relative to the NLR emission. Such shifts in the BLR are predicted for the recoils (`kicks') that result from the merger of two black holes \citep{shields08}. The magnitude of the kick can be large: \citet{dain08} have computed kicks of up to 3300 km s$^{-1}$, but the most extreme kicks require black hole binaries of equal mass and spin, where the spins are extremely high. We further note that J0400+05 has the highest virial black hole mass calculated for the 2MASS sample. We note that \citet{hutchings03} does not detect the presence of the shifted broad line emission in J0400+05; they classify this object as a type 2 AGN.

\noindent \textbf{J1006+41:} \citet{hutchings03} does not detect the presence of broad line emission in J1006+41, whereas we detect broad line emission in H$\alpha$ and FeII.

\noindent \textbf{J1040+59} is a type 1 Seyfert galaxy with J-K$_S$ = 3.02 --- far higher than the median near-IR colour of the sample. In addition to its unusually red J-K$_S$ colour, the optical continuum of this object rises more steeply to the red than any other object in the sample. Its shape is reminiscent of the object PKS 1549-79 which also has extremely red colours (J-K$_S$=2.82; see \citealt{bellamy}; \citealt{holt06}). PKS 1549-79 is believed to be a quasar which is enshrouded in a cocoon of natal gas and dust which is supplied by a recent merger (\citealt{tadhunter01} \citealt{holt06}). J1040+59 also features in the \citet{glikman07} sample. Consistent with our results, while \citet{glikman07} found that J1040+59, and many of their objects, are sufﬁciently extinguished such that the broad Balmer emission lines are not present at optical wavelengths, broad Paschen lines are detected in the near-IR. 

\noindent \textbf{J1057-13:} \citet{hutchings03} classify J1057-13 as a LINER/HII type object. However, based on the BPT diagrams (see Figures \ref{diag1}-\ref{diag3}), and presence of FHILs, we find clear evidence that the emission-line properties of J1057-13 are consistent with those of a Seyfert type AGN.

\noindent \textbf{J1131+16} is a type 2 quasar with J-K$_S$ = 2.15. This object has an unusually rich spectrum of FHILs, which are strong compared to the [OIII]$\lambda$5007 line, and has been studied in detail by \citet{rose}. We believe that the unusual strengths of the FHILs in this object are due to a specific viewing angle of the far wall of the torus, coupled with a lack of dust on larger scales that might otherwise obscure our view of the torus. The spectrum of the hot dust emission ($\sim$1500 K) is expected to peak at wavelengths similar to the K-Band. The enhanced flux in this band could be the explanation for the red J-K$_S$ colour observed for this object.

\noindent \textbf{J1158-30:} we note that \citet{hutchings03} classify J1158-30 as a type 1 AGN, but we see no evidence of broad line emission in our spectra.

\noindent \textbf{J1307+23} is a type 2 Seyfert galaxy/LINER with J-K$_S$ = 3.31 (the reddest J-K$_S$ colour in the sample) and has one of the largest redshifts in the sample (z = 0.274). It is possible that this object is a genuine type 2 AGN in which the AGN is subject to extremely high levels of extinction because the torus is viewed close to edge on, but due to the larger scale of the NLR, shows very little evidence for dust extinction in the narrow Balmer decrements. However, both \citet{smith02} and \citet{hutchings03} report the presence of broad line emission in J1307+23, inconsistent with our analysis of its spectrum. On the other hand, we note that, like us, \citet{kuraab} classify this object as a type 2 object.

\noindent \textbf{J1321+13:} like J1057-13, \citet{hutchings03} classify J1321+13 as a LINER/HII type object. Again, based on the BPT diagrams (see Figures \ref{diag1}-\ref{diag3}), and the detection of FHILs, we find that the emission-line properties of J1321+13 are consistent with those of a Seyfert type AGN.

\noindent \textbf{J1637+25} is classified as an intermediate type AGN in both \citealt{smith02} (type 1.9) and \citealt{kuraab} (type 1.5), however our classification is consistent with the finding of \citet{hutchings03} that this object is a type 2 AGN.

\noindent \textbf{J2124-17} is a type 1 quasar with J-K$_S$ = 2.37. Based on the visual inspection of the optical spectrum, this object appears to have an optical SED that is typical of UV/optical selected type 1 quasars. Another notable property of this object is that it has the strongest FeII emission of all of the red 2MASS sample. However, a single Gaussian fit to the H$\beta$ emission line gives a velocity width FWHM of 2040$\pm$30 km s$^{-1}$---which is higher than expected for a NLS1 (FWHM $<$ 2000 km s$^{-1}$).

Another interesting property of this object is that it has been previously identified as an ULIRG \citep{k&s}. Indeed, this object, along with a further 25 ULIRGs, is studied in more detail in \citet{javierboth}. 

\noindent \textbf{ A note on the comparison of classifications with other work.} Most of the observed differences between AGN classifications noted in this section can be explained a by differences in S/N between the studies, or by differences in the method used to fit the H$\alpha$+[NII] blend of the observations, particularly for objects with weak broad lines, or classified as LINER/intermediate types. 

\label{lastpage}

\end{document}